\documentclass[apj,revtex4,numberedappendix]{emulateapj}
\usepackage{graphics}
\usepackage{subfigure}
\usepackage{hyperref}
\usepackage{graphicx}
\usepackage{pdflscape}
\usepackage{rotating}
\usepackage{color}

\usepackage{amsmath}
\usepackage{natbib}
\usepackage{float}
\newcommand{\CH}[1]{\colhead{#1}}

\newcommand\ii{{\sc ii}}
\newcommand\iii{{\sc iii}}
\newcommand\iv{{\sc iv}}
\newcommand\W{{$\lambda$}}

\begin{document}

\shortauthors{Berg et al.}
\title{Carbon and Oxygen Abundances in Low Metallicity Dwarf Galaxies\footnote{
Based on observations made with the NASA/ESA Hubble Space Telescope,
obtained from the Data Archive at the Space Telescope Science Institute, which
is operated by the Association of Universities for Research in Astronomy, Inc.,
under NASA contract NAS 5-26555.}\footnote{Some observations reported here 
were obtained at the MMT Observatory, a joint facility of the University of Arizona 
and the Smithsonian Institution.}}

\author{Danielle A. Berg\altaffilmark{1}, 
	    Evan D. Skillman\altaffilmark{2}, 
	    Richard B.C. Henry\altaffilmark{3}, 
	    Dawn K. Erb\altaffilmark{1},
	    Leticia Carigi\altaffilmark{4}}

\altaffiltext{1}{Center for Gravitation, Cosmology and Astrophysics, Department of Physics, University of Wisconsin Milwaukee,
3135 N Maryland Ave., Milwaukee, WI 53211; bergda@uwm.edu; erbd@uwm.edu}
\altaffiltext{2}{Minnesota Institute for Astrophysics, University of Minnesota, 116 Church St. SE, Minneapolis, MN 55455;  skillman@astro.umn.edu}
\altaffiltext{3}{University of Oklahoma; rhenry@ou.edu}
\altaffiltext{4}{Instituto de Astronom\'{i}a, Universidad Nacional Aut\'{o}noma de M\'{e}xico, Apartado Postal 70-264, Ciudad Universitaria, M\'{e}xico DF 04510, M\'{e}xico; carigi@astro.unam.mx}

\begin{abstract}

The study of carbon and oxygen abundances yields information on the time evolution
and nucleosynthetic origins of these elements, yet remains relatively unexplored.
At low metallicities (12+log(O/H) $<$ 8.0), nebular carbon measurements
are limited to rest-frame UV collisionally excited emission lines.
Therefore, we present UV spectrophotometry of 12 nearby, low-metallicity, 
high-ionization \ion{H}{2} regions in dwarf galaxies obtained with
the Cosmic Origins Spectrograph on the Hubble Space Telescope.
We present the first analysis of the 
C/O ratio in local galaxies based solely on simultaneous 
significant detections of the UV O$^{+2}$ and C$^{+2}$ collisionally 
excited lines in seven of our targets and five objects from the literature,
to create a final sample of 12 significant detections. 
Our sample is complemented by optical SDSS spectra,
from which we measured the nebular physical conditions 
and oxygen abundances using the direct method.

At low metallicity (12+log(O/H) $<$ 8.0), no clear trend is evident in C/O
vs. O/H for the present sample given the large dispersion observed.
When combined with recombination line observations at higher values of O/H,
a general trend of increasing C/O with increasing O/H is also viable, 
but with some significant outliers.
Additionally, we find the C/N ratio appears to be constant (but with significant scatter) 
over a large range in oxygen abundance, indicating carbon is 
predominantly produced by similar nucleosynthetic mechanisms as nitrogen.
If true, and our current understanding of nitrogen production is correct,
this would indicate that primary production of carbon (a flat trend)  
dominates at low metallicity, but quasi-secondary production (an increasing 
trend) becomes prominent at higher metallicities. 
A larger sample will be needed to determine the true nature and dispersion of the relation.

\end{abstract}



\section{INTRODUCTION}\label{sec1}


A key tracer of galaxy evolution is the change in the 
chemical composition over time.
The metallicity of a galaxy is a sensitive observational diagnostic 
of its past star formation history (SFH) and present-day evolutionary 
state because metallicity increases monotonically with each successive 
generation of massive stars (MS) \citep{wheeler89}. 
Oxygen abundance is an important tracer of metallicity because it is the most 
abundant element in the Universe after hydrogen and helium and is convenient to 
observe since its emission lines are ubiquitous in the rest-frame optical regime. 
After oxygen, carbon and nitrogen play the largest role in the chemical evolution of galaxies.  
Because C and O are important sources of interior opacity in stars, 
knowledge of the time evolution of CNO abundances is necessary to 
properly model stellar isochrones. 
Furthermore, because C and O emission lines originate principally in 
star forming regions, they trace the physical conditions in the gas from 
which the current generation of mass stars is forming. 
Equally significant, this interstellar gas carries the signature of the interplay between 
SF, gas accretion, and supernova-driven feedback across cosmic time.

The C content in metal-poor star-forming dwarf galaxies was first investigated in detail by 
\citet{garnett95} using \textit{Hubble Space Telescope} (HST) \textit{Faint Object Spectrograph} 
(FOS) observations, who found that C/O increases with increasing O/H.
This trend was supported by further UV studies of irregular and spiral galaxies
\citep{garnett97,kobulnicky97,kobulnicky98,garnett99,izotov99},
implying the apparent secondary\footnotemark[4] behavior of C is largely produced by 
either the time delay in the release of C by low- and intermediate-mass 
stars relative to O or metallicity-dependent yields in massive stars 
\citep{garnett99,henry00,carigi00,chiappini03}.

\footnotetext[4]{If a particular isotope is produced from the original H and He in a star, 
the production is said to be ``primary" and its relative abundance remains constant.
If instead the isotope is a daughter element of heavier elements initially present in the star,
the production is called ``secondary" and is linearly dependent on the initial abundance
of the parent heavier elements (metallicity). }

The C/O ratio is fairly well studied for stars in the Galactic disk \citep[e.g.,][]{gustafsson99,bensby06}
and the Galactic halo \citep[e.g.,][]{akerman04,spite05,fabbian09}, 
both of which have been found to be consistent with \ion{H}{2} region recombination line (RL)
abundances at moderate to high metallicities \citep[12+log(O/H) $\gtrsim$ 8.0; e.g.,][]{esteban14}.
However, the monotonic increase in C/O with oxygen abundance in metal-poor 
galaxies reported by \citet{garnett95} contrasts with the trend seen in Galactic stars,
where the C/O ratio is roughly constant over the same range in oxygen abundance. 
At high redshifts (z $\gtrsim$ 1), UV C and O lines become observable in the optical,
opening a window to ground-based C/O studies using emission lines from 
lensed galaxies, composite spectra of large samples, or deep 
spectra of low metallicity objects in which the C and O lines are strong
\citep[e.g.,][]{shapley03,erb10,james14,stark14}, 
as well as absorption lines from metal-poor damped Ly$\alpha$ systems (DLAs).
In particular, C/O observations of DLAs appear to follow the 
trend of increasing C/O abundance with decreasing metallicity at very low oxygen 
abundance seen for Milky Way disk stars \citep[e.g.,][]{pettini08,cooke11}.

\subsection{The Need For Improved UV C/O Measurements} \label{sec1.2}
Carbon, as measured by its bright collisionally excited lines (CELs), 
has been historically difficult to observe in extragalactic \ion{H}{2} regions. 
Carbon has no strong transitions of its important ionization states in the optical, 
and there are no IR transitions for C$^{+2}$, the main ionization state of C, in \ion{H}{2} regions. 
Other methods have been explored as means to measure C abundances, 
such as optical RLs of \ion{C}{2} 
\citep{esteban02,peimbert05,lopez-sanchez07,garcia-rojas07,esteban09,esteban14}. 
However, the RLs become very faint at low metallicities; 
in fact all of the C/O measurements from RLs are at oxygen 
abundances (corrected for temperature fluctuations) above 
12+log(O/H) = 8.0 \citep[see][and references therein]{esteban14}.
Stars provide another way to determine abundances, but this is done with the most luminous stars, 
which in external galaxies tend to be evolved giants and super giants, 
whose surfaces are complicated by the effects of internal mixing \citep{venn95}.

Little advancement in our understanding has occurred since the insight gained 
into C/O abundances from observations obtained with FOS on HST in the 1990's.
In particular, the dispersion in this relationship is unmeasured due to the 
extremely small sample size.
This is largely because few instruments, past and present, have the requisite 
wavelength coverage and sensitivity, resulting in a paucity of nebular C/O 
measurements in local galaxies, especially at low metallicities. 
Further, the current body of work is hampered by significant uncertainties due 
to marginal detections and the difficulties of combining space-based UV
and ground-based optical observations \citep[see discussion in][]{garnett95}.
We can overcome many of these uncertainties and challenges by observing 
the UV CELs from the dominant ions of C and O in ionized nebulae
(\ion{C}{3}] $\lambda\lambda1907,1909$\footnotemark[5] 
and \ion{O}{3}] $\lambda\lambda1660,1666$)
with the Cosmic Origins Spectrograph \citep[COS;][]{green12} on HST.

\footnotetext[5]{The UV C$^{+2}$ doublet, commonly notated as \ion{C}{3}], is actually a combination of 
the forbidden magnetic quadrupole [\ion{C}{3}] $\lambda$1907 emission line and the semi-forbidden 
electrodipole \ion{C}{3}] $\lambda$1909 emission line. 
For ease of notation, we will adopt the simplified, albeit incorrect, form of \ion{C}{3}] $\lambda\lambda$1907,1909.} 

We obtained HST spectroscopic measurements of the UV CELs 
(\ion{O}{3}] $\lambda\lambda$1660,1666 and \ion{C}{3}] $\lambda\lambda$1907,1909) 
in extragalactic \ion{H}{2} regions in a sample of targets spanning the metallicity range of dwarf galaxies. 
We describe our sample selection in Section~\ref{sec2} and details of the UV HST 
and complimentary optical observations in Sections~\ref{sec3.1} and \ref{sec3.2} respectively.
With the intention of minimizing sources of discrepancies, 
all data used in this paper were analyzed in a uniform manner as outlined in Section~\ref{sec4}.
We define criteria for a high-quality C/O sample in \S~\ref{sec5.1}
From these data we investigate the relative C and N abundances in Sections~\ref{sec5.2}
and ~\ref{sec5.3}, comparing to RL studies in Section~\ref{sec5.4}.
In Section~\ref{sec6} we discuss potential sources of carbon.
We compare to stellar abundances in \S~\ref{sec6.1},
discuss potential sources of dispersion in \S~\ref{sec6.2}, 
and compare to chemical evolution models in \S~\ref{sec6.3}.
A summary of our results are given in Section~\ref{sec7}.
The Appendix contains three sections: first, we include the details
of the supplemental optical spectra; second, we test the spectrophotometry
of the SDSS spectrum for an outlier from our sample by comparing 
to a follow-up spectrum from the MMT; third, we discuss strategy for future studies.


\section{SAMPLE SELECTION} \label{sec2}


The purpose of this study is to obtain new gas-phase UV observations of 
C$^{+2}$ and O$^{+2}$ in a sample of extragalactic \ion{H}{2} regions to aid 
our understanding of the C/O relationship with nebular oxygen abundance.
The energies required for ionization to C$^{+2}$ and O$^{+2}$ are 24.8 eV and 
35.1 eV respectively, and thus these observations are limited to high ionization \ion{H}{2} regions.
Further, the collisionally excited C and O transitions of interest have high 
excitation energies (6-8 eV), and so are best observed in low metallicity 
environments where nebular electron temperature ($T_e$) is high. 

Using the \ion{C}{3}] $\lambda\lambda1907,1909$/\ion{O}{3}] 
$\lambda1666$ line ratio is a robust way to investigate the C/O 
relationship for the following reasons \citep[c.f.,][]{garnett95}:
(1) The C/O ratio exhibits minimal uncertainty due to reddening,
as the interstellar extinction curve is nearly flat over the wavelength
range of interest, $1600-2000$ \AA.
(2) The UV \ion{O}{3}] and \ion{C}{3}] lines have similar excitation and ionization
potentials and so their ratio has little dependence on the physical conditions of 
the gas, i.e., nebular $T_e$ and ionization structure.
(3) Measuring the O$^{+2}$ and C$^{+2}$ lines simultaneously eliminates the 
aperture matching and positioning uncertainties that can arise when combining
different observational setups. 

In order to establish the C/O relationship with O/H in the sparsely measured 
metal poor regime, we have chosen objects with high emission line surface 
brightnesses and low metallicity which span a large range in O abundance. 
With the completion of the Sloan Digital Sky Survey Data Release 
12\footnotemark[6] \citep[SDSS-III DR12;][]{eisenstein11,alam15}, many 
new low-metallicity nearby dwarf galaxies have been identified
\citep[e.g.,][]{guseva09,ekta10,izotov12}.
We culled potential targets from the SDSS observations deemed optimal by several studies:
\begin{enumerate}
\item \citet{izotov12} presented 42 low-metallicity emission-line galaxies 
selected from 7th data release of the SDSS \citep{abazajian09}, including 17 of the most 
metal-deficient emission-line galaxies known in the local universe at the time. 
\item \citet{ekta10} presented 59 blue compact galaxies (BCGs) selected from the 3rd 
data release of the SDSS \citep{abazajian05} and 32 dwarf irregular galaxies selected from the literature. 
These 91 galaxies were selected to have accurate gas-phase oxygen abundance determinations. 
\item \citet{guseva09} presented 44 emission-line galaxies, selected mostly 
from the 6th data release of the SDSS \citep{adelman-mccarthy08} as metal-deficient galaxy candidates.
\item The MPA-JHU group\footnotemark[7] created an emission line data base 
for the 818,333 unique star-forming targets in the 7th data release of the SDSS.
\end{enumerate}

\footnotetext[6]{\url{http://www.sdss.org/dr12/}} 
\footnotetext[7]{Data catalogues are available from \url{http://www.mpa-garching.mpg.de/SDSS/}.
The Max Plank institute for Astrophysics/John Hopkins University(MPA/JHU) SDSS data base was produced by a collaboration 
of researchers(currently or formerly) from the MPA and the JHU. The team is made up of Stephane Charlot (IAP), 
Guinevere Kauffmann and Simon White (MPA),Tim Heckman (JHU), Christy Tremonti (U. Wisconsin-Madison $-$ formerlyJHU) 
and Jarle Brinchmann (Leiden University $-$ formerly MPA).} 

From these literature sources we set the following criteria for our sample:
\begin{enumerate}
\item \textit{12+log(O/H) $\leq 8.2$:} 
In order to sample the low-metallicity galaxy population, we considered only 
targets with direct gas-phase oxygen abundances of 12+log(O/H) $\leq 8.2$
based on [\ion{O}{3}] $\lambda4363$ detections at strengths of 3$\sigma$ or greater.
\item \textit{$z < 0.26$:} 
The G140L grating is the only grating on COS that allows simultaneous 
observations of \ion{O}{3}] $\lambda\lambda1660,1666$ and 
\ion{C}{3}] $\lambda\lambda1907,1909$ in nearby galaxies. 
Limited wavelength coverage ($< 2405$ \AA) and rapidly declining red-ward 
throughput required targets with redshifts of $z < 0.26$. 
\item \textit{$D_{25} \lesssim 5\arcsec$:}
Through visual inspection of SDSS photometry \citep{gunn98} 
using the Catalog Archive Server (CAS) database, 
we selected candidate targets which have compact morphologies
in the sense that the diameter of their optical light profiles $\lesssim 5\arcsec$.
This step is important to optimize the sample observations for the 2.5\arcsec\ COS aperture.
\item \textit{E(B$-$V) $< 0.1$:}
Only targets with small reddening values were selected,
which reduces uncertainties associated with the reddening correction of the UV lines
and the relative amount of C and O locked up in dust grains.
\item \textit{$m_{FUV} \leq 19.5$ AB:}
To ensure targets were bright enough in the FUV to enable continuum 
detections with COS, we required targets to have \textit{Galaxy Evolution 
Explorer} (GALEX) photometry \citep[GR6;][]{bianchi14} values of $m_{FUV} \leq 19.5$ AB.
\item \textit{EW(5007) $>$ 50 \AA:}
To improve upon previous studies which largely lack significant detections
of \ion{O}{3}] $\lambda1666$, we selected galaxies with large [\ion{O}{3}] 
$\lambda5007$ equivalent widths.
\item \textit{Predicted F(\ion{O}{3}]$\lambda1666$) S/N $> 5$:}
The GALEX FUV flux within the 2.5\arcsec\ HST/COS aperture was
used alongside the optical [\ion{O}{3}] line fluxes to select targets for 
which the COS exposure time calculator (ETC) predicted S/N $> 5$ in 
the \ion{O}{3}] $\lambda1666$ line, which is the weakest of the desired UV lines. 
\end{enumerate}

As a result of the literature search just described, our final sample contained 12 
nearby (0.003 $< z <$ 0.135), UV-bright ($m_{FUV} \leq 19.5$ AB), 
compact (D $< 5\arcsec$), low-metallicity (12+log(O/H) $\leq 8.2$) dwarf galaxies.
Basic properties of our final sample are listed in Table~\ref{tbl1}.
Assuming solar $M_B$ = 5.47, luminosities are estimated by converting the 
SDSS $u-$ and $g-$band c-model magnitudes
to apparent B-band magnitudes, using the conversion from \citet{jester05},
before determining the absolute B-band magnitude given the SDSS redshift. 
Median total stellar masses \citep[following][]{kauffmann03b}, average total 
star formation rates \citep[SFR; based on][]{brinchmann04}, and
average total specific SFRs (sSFR) are taken from the MPA-JHU catalogue.
Given our selection of nearby, compact, bright targets, our sample has
very low-masses and high sSFRs. 
Figure~1 displays our sample targets.


\begin{deluxetable*}{lccccccccccc}
\tablewidth{0pt}
\tablecaption{Bright, Compact, Nearby Dwarf Sample}
\tablehead{
\CH{Target} & \CH{R.A.}	  & \CH{Dec.}		& \CH{$z$} & \CH{$M_B$} & \CH{log $L_B$}	& \CH{log $M_\star$} & \CH{log SFR }		& \CH{log sSFR}	& \CH{AB FUV} & \CH{12+log(O/H)} & \CH{$t_{exp}$}  	\\
\CH{   	} & \CH{(J2000)} & \CH{(J2000)}	& \CH{}	 & \CH{(mag)}	& \CH{(L$_\odot$)}	& \CH{(M$_\odot$)}	 & \CH{(M$_\odot$/yr)}	& \CH{(yr$^{-1}$)}	& \CH{(mag)} 	& \CH{(dex)} 		& \CH{(s)}	 }
\startdata
J141454		& 14:14:54.14	& -02:08:22.90	& 0.005	& -13.4	& 7.57		& 6.61	& -1.91	& -8.58	& 19.32	& 7.28	& 4746	\\	
J082555		& 08:25:55.52	& +35:32:31.9	& 0.003	& -12.6	& 7.21		& 6.04	& -1.98	& -8.11	& 18.80	& 7.42	& 2180	\\	
J104457		& 10:44:57.79	& +03:53:13.1	& 0.013	& -15.9	& 8.56		& 6.80	& -0.82	& -7.82	& 18.28	& 7.44	& 2066	\\	
J120122		& 12:01:22.31	& +02:11:08.3	& 0.003	& -12.4	& 7.16		& 6.09	& -1.97	& -8.14	& 18.64	& 7.50	& 2056	\\	
J085103		& 08:51:03.64	& +84:16:13.8	& 0.006	& -9.6	& 6.03		& 6.08	& -2.30	& -8.46	& 17.01	& 7.66	& 2154	\\	
J124159		& 12:41:59.34	& -03:40:02.4	& 0.009	& -14.5	& 7.97		& 6.59	& -1.40	& -8.05	& 19.44	& 7.74	& 1847	\\	
J115441		& 11:54:41.22	& +46:36:36.2	& 0.004	& -13.8	& 7.71		& 6.14	& -2.02	& -8.22	& 17.98	& 7.75	& 2229	\\	
J122622		& 12:26:22.71	& -01:15:12.2	& 0.007	& -12.4	& 7.15		& 7.21	& -1.10	& -8.37	& 17.67	& 7.77	& 1944	\\	
J122436		& 12:24:36.71	& +37:24:36.5	& 0.040	& -18.0	& 9.41		& 7.86	& -0.06	& -8.01	& 18.41	& 7.78	& 2136	\\	
J124827		& 12:48:27.79	& +48:23:03.3	& 0.030	& \nodata	& \nodata		& 7.47	& -0.49	& -8.05	& 19.44	& 7.80	& 2044	\\	
J025426		& 02:54:26.12	& -00:41:22.6	& 0.015	& -16.4	& 8.77		& 7.65	& -0.67	& -8.40	& 18.54	& 8.06	& 2048	\\	
J095137		& 09:51:37.47	& +48:39:41.2	& 0.135	& -14.8	& 8.13		& 9.34	& 0.49	& -8.93	& 19.10	& 8.20	& 2168	\\	
\enddata 
\tablecomments{Selected target sample. All objects are bright, compact,
    nearby dwarf galaxies with low metallicities as measured by their ground-based optical spectra. 
    The first four columns give the target name used in this work, location, and redshift.
    Column 5 gives the galaxy B-band AB magnitude, estimated using the conversion of \citet{jester05}
    with the SDSS c-model u and g magnitudes. 
    Luminosity in Column 6 is determined from Column 5 assuming $M_{B,\odot}$ = 5.47.
    Note that no SDSS magnitudes are available for J124827. 
    Columns 7$-$9 list the median total stellar masses, average star formation rates, and average specific star formation rates from the MPA-JHU database.
    Column 10 lists the FUV magnitudes for these objects from GALEX.
    Column 11 lists the range of metal-poor direct oxygen abundances of our sample taken from the literature.
    Column 12 gives the total HST/COS science exposure times.}
\label{tbl1}
\end{deluxetable*}


\begin{figure*}\label{fig1}
   \begin{center}
   \resizebox{!}{50mm}{ 
      \begin{tabular}{cccc}
             \resizebox{40mm}{!}{\includegraphics{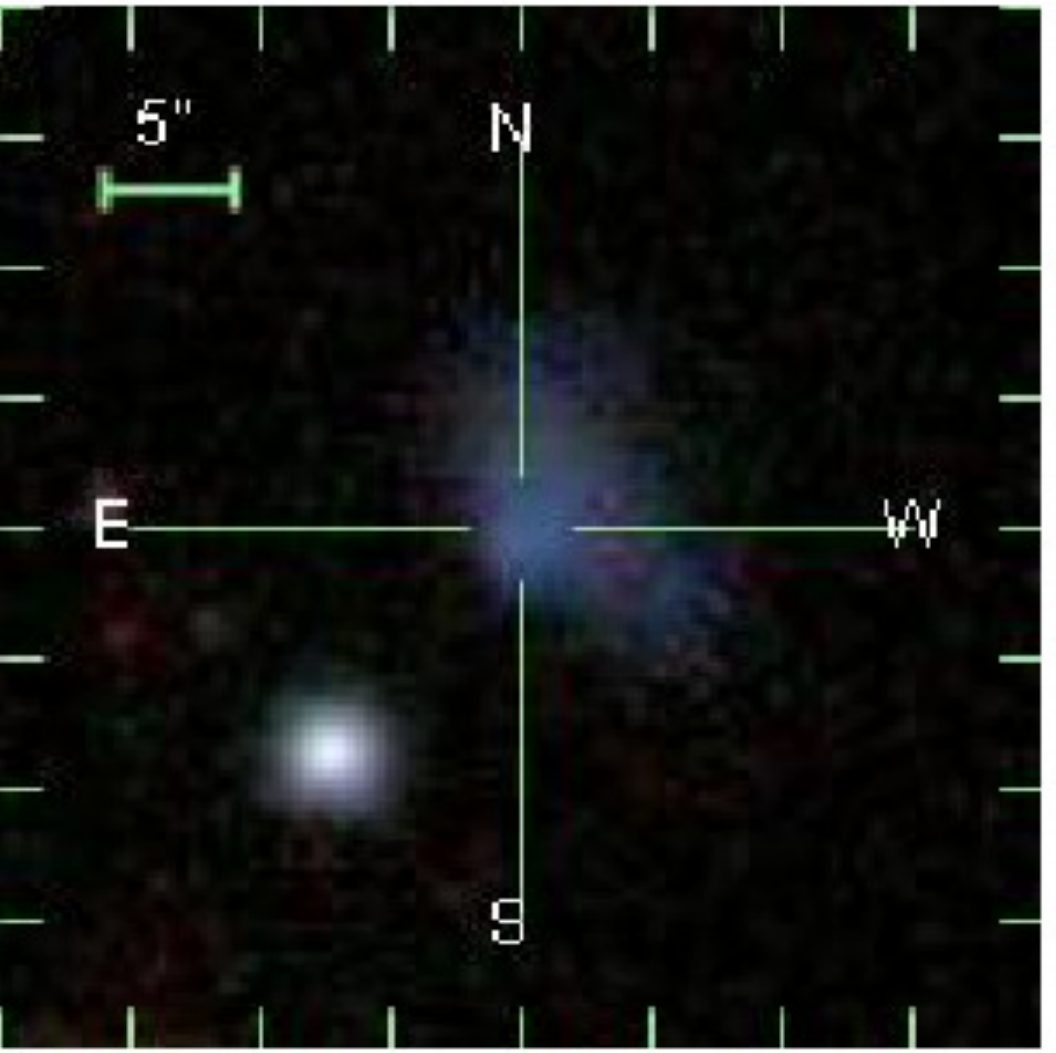}}     		& \resizebox{40mm}{!}{\includegraphics{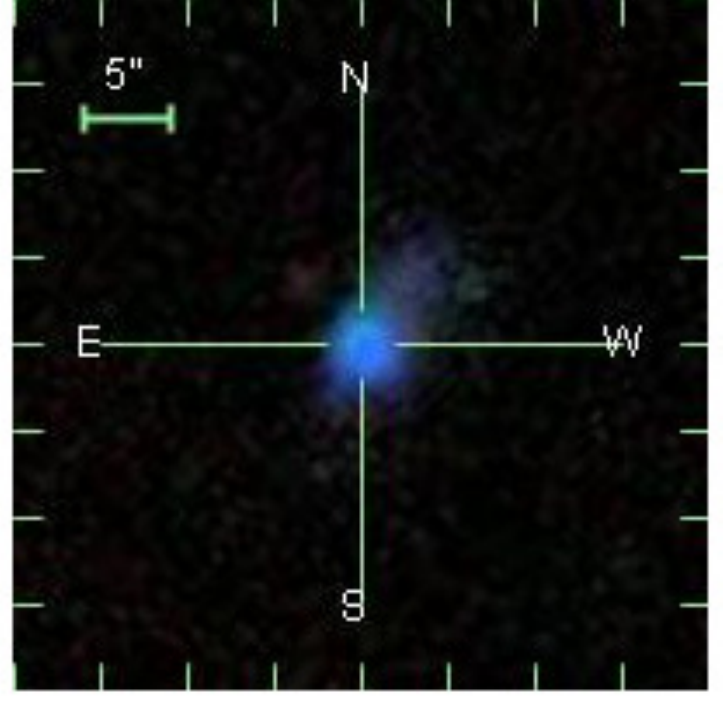}}     
      	& \resizebox{40mm}{!}{\includegraphics{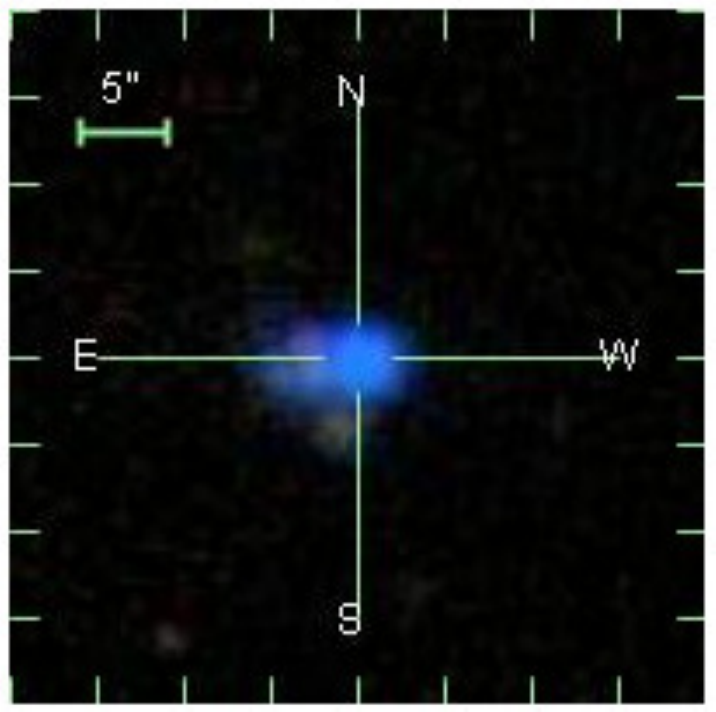}}   		& \resizebox{40mm}{!}{\includegraphics{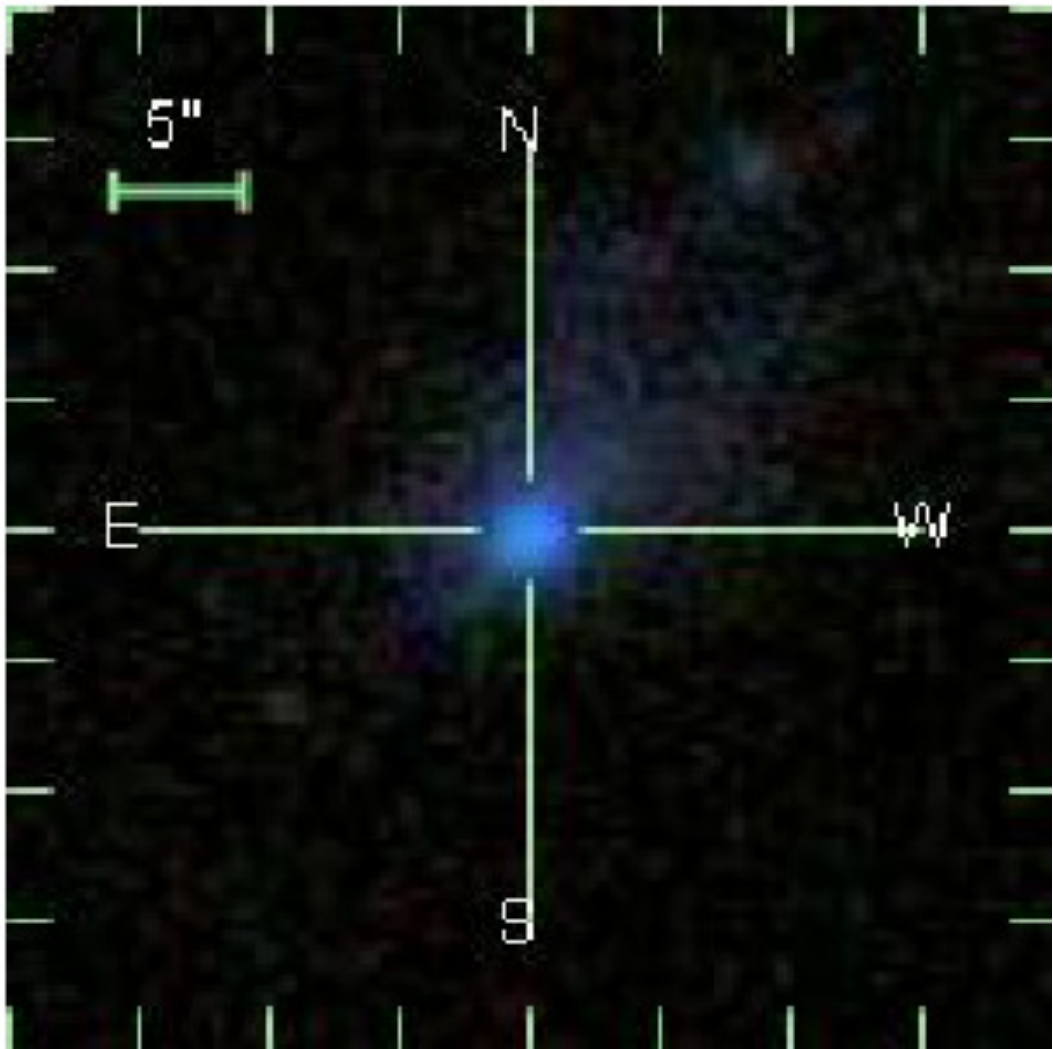}} \\
	{J141454.14-020822.9}	& {J082555.52+353231.9}	& {J104457.79+035313.1} & {J120122.31+021108.3} \\
	   \resizebox{40mm}{!}{\includegraphics{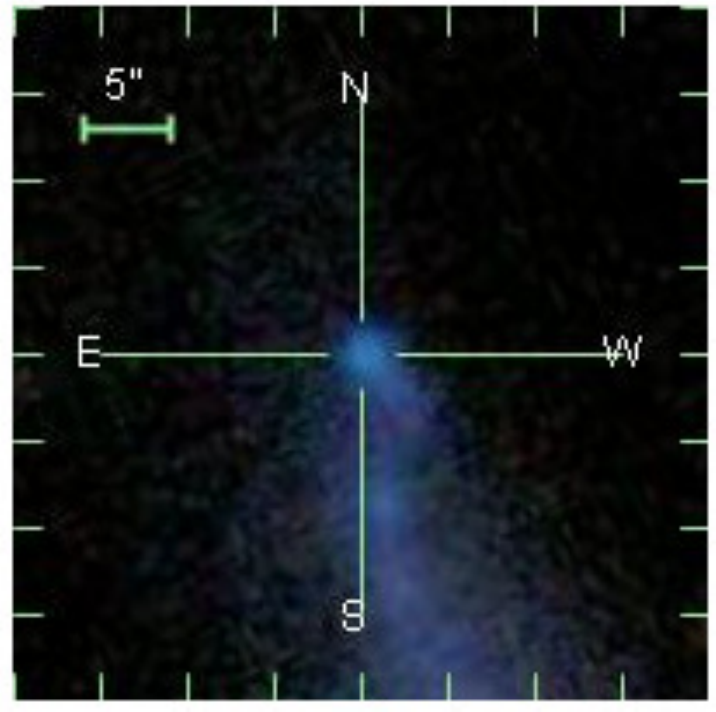}} 			& \resizebox{40mm}{!}{\includegraphics{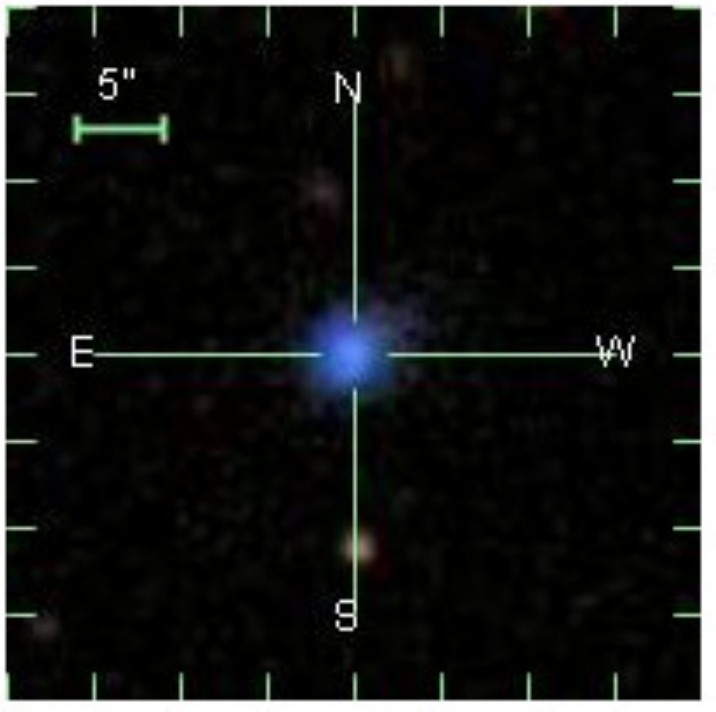}} 
	& \resizebox{40mm}{!}{\includegraphics{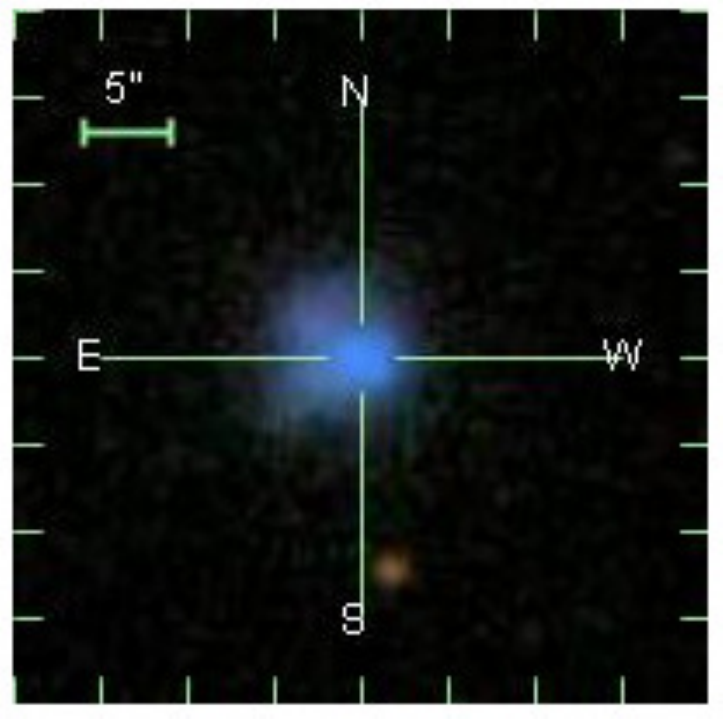}} 			& \resizebox{40mm}{!}{\includegraphics{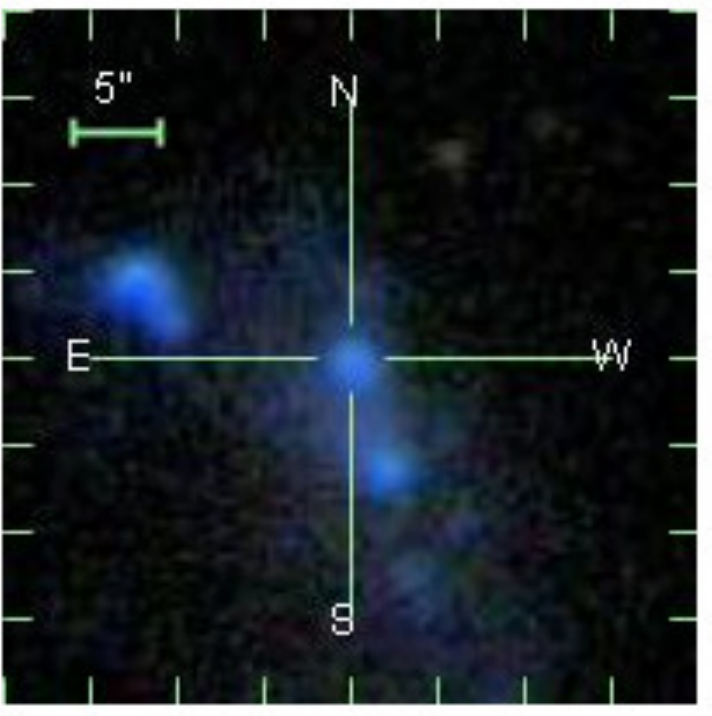}} \\  
	{J085103.64+841613.8}	& {J124159.34-034002.4} & {J115441.22+463636.2} & {J122622.71-011512.2} \\
   	   \resizebox{40mm}{!}{\includegraphics{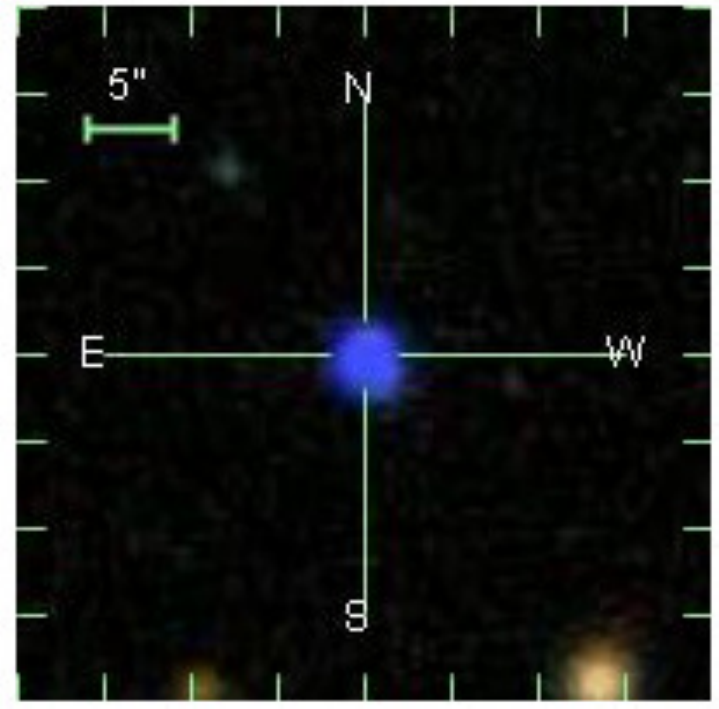}} 			& \resizebox{40mm}{!}{\includegraphics{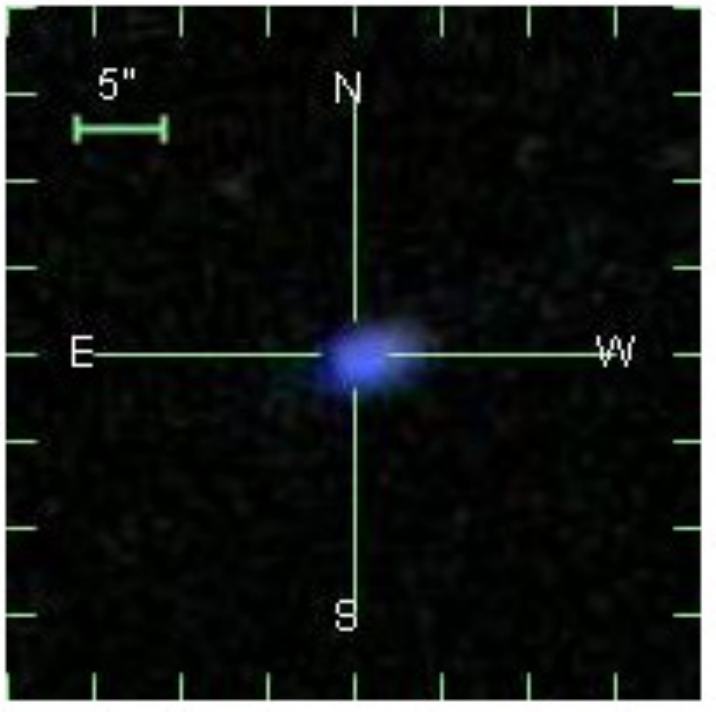}} 	
      	& \resizebox{40mm}{!}{\includegraphics{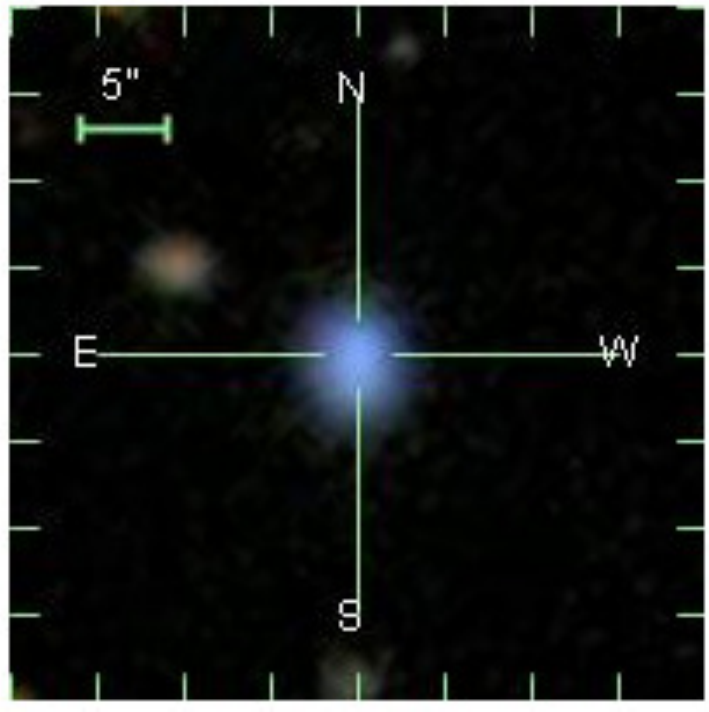}} 			& \resizebox{40mm}{!}{\includegraphics{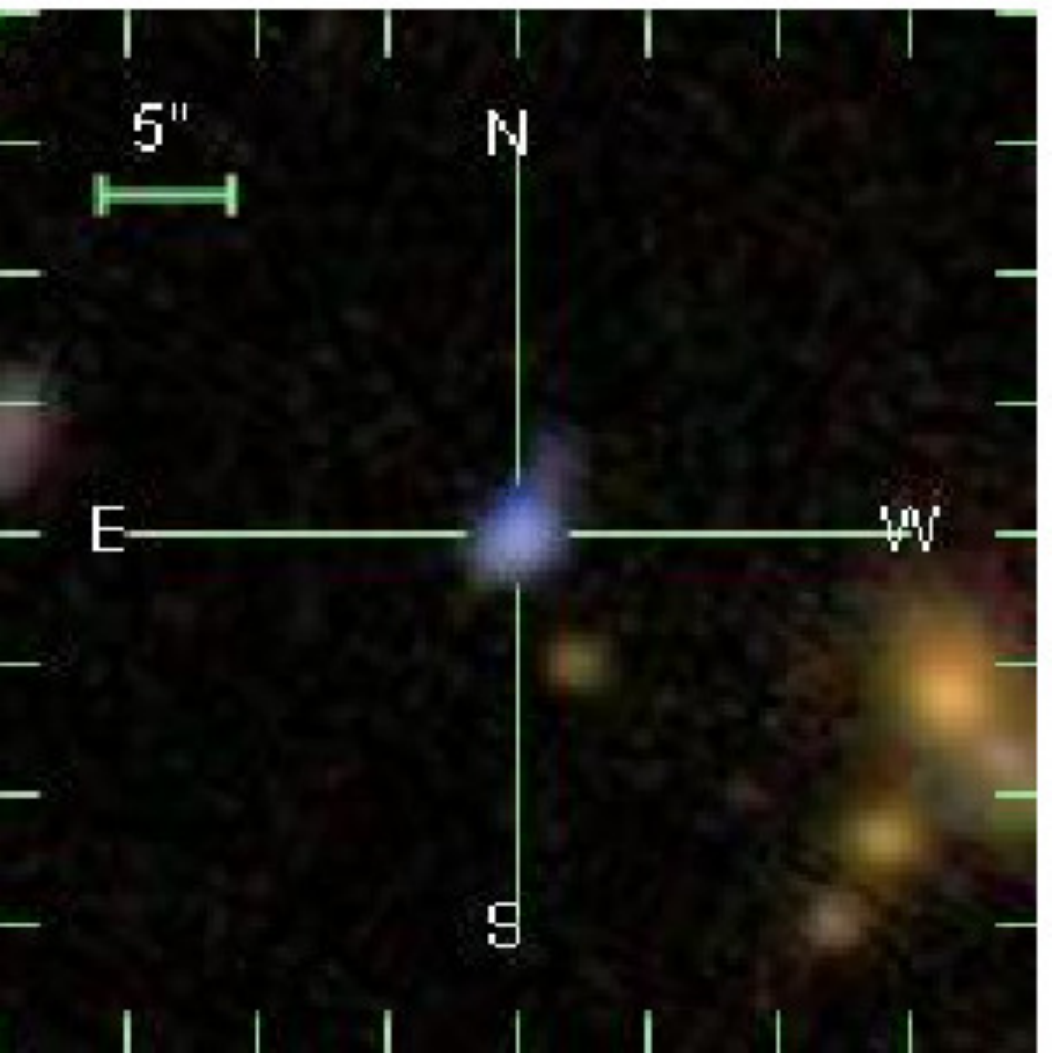}} \\
	{J122436.71+372436.5} & {J124827.79+482303.3} & {J025426.12-004122.6} & {J095137.47+483941.2} \\
  	\end{tabular}    }
	\caption{SDSS images of the 12 targets in our sample. Notice the bright blue appearance and compact morphologies
	signifying ongoing star formation and high surface brightnesses which allow for maximum flux through the 2.5\arcsec\ COS aperture. }
   \end{center} 
\end{figure*}     


\section{SPECTROSCOPIC OBSERVATIONS AND DATA REDUCTION}\label{sec3}


\subsection{HST/COS FUV Spectra} \label{sec3.1}

As discussed in \citet[][hereafter G95]{garnett95}, the approach described in 
Section~\ref{sec2} to simultaneously observe C and O CELs in the UV 
has many advantages over studies which combine space-based UV 
data with ground-based optical data in their C/O abundance calculations.
This method is currently limited to observations 
from COS as it is the only space-based UV spectrograph with the 
necessary sensitivity and wavelength coverage.
We therefore acquired 13 orbits of UV spectra of our targets with the 
COS on the HST as part of HST-GO13312 during cycle 21. 
To maximize our exposure times, we  
chose to use a spectroscopic peak search acquisition.
As shown in the SDSS images in Figure~1, the central blue bulge of 
emission in each of our targets falls within a 5\arcsec\ diameter aperture.
\citet{henry15} obtained COS spectroscopic observations of 10 Green Pea 
galaxies \citep{cardamone09}, which have similarly compact sizes and 
morphologies in the SDSS as our sample, and demonstrated that most 
of the UV continuum emission fell within a central 1\arcsec\ aperture.
Assuming good initial pointing coordinates and successful completion 
of the acquisition, the COS 2.5\arcsec\ fiber should be well aligned 
with our targets and capture most of the UV emission.

COS FUV observations were taken in the TIME-TAG mode using the 2.5\arcsec\ 
PSA aperture and the G140L grating at a central wavelength of 1280 \AA.
In this configuration, segment A has a wavelength range of 1282$-$2148 \AA\footnotemark[8],
allowing the simultaneous observation of the \ion{O}{3}] $\lambda\lambda1660,1666$
and \ion{C}{3}] $\lambda\lambda1907,1909$ emission lines. 
We used the FP-POS=ALL setting, which takes 4 images offset from one 
another in the dispersion direction, increasing the cumulative S/N and 
mitigating the effects of fixed pattern noise. 
The 4 positions allow a flat to be created and cosmic rays to be eliminated.
Each target was observed for the maximum time allotted in a single orbit as 
determined by the object orbit visibility, except for J141454 which was prioritized 
to be observed for 2 orbits to extend the sample to lower metallicities.
All data from GO13312 were processed with CALCOS version 2.21\footnotemark[9].

\footnotetext[8]{The G140L grating on COS is characterized as having wavelength coverage out to 2405 \AA.
However, our experience with this setup indicates a range of usefulness out to only 2000 \AA.}
\footnotetext[9]{\url{http://www.stsci.edu/hst/cos/pipeline/CALCOSReleaseNotes/notes/}}

In order to gain signal-to-noise we chose to bin the spectra in the dispersion direction.
For the G140L grating, six pixels (80.3 m\AA/pix) span a resolution element 
of roughly 0.55 \AA\ at $\lambda1660$.
By measuring individual airglow emission lines in our spectra, we found a typical
FWHM $\approx$ 3 \AA, allowing us to re-bin our spectra by the six pixels of a 
resolution element while maintaining six resolution elements per FWHM.

\subsection{Supporting SDSS Optical Spectra} \label{sec3.2}

Each of the targets in our sample has been previously observed as part of the SDSS DR7.
We used the publicly available SDSS data \citep{york00}, 
which have been reduced with the SDSS pipeline \citep{bolton12}.
Preliminary emission line fluxes from the MPA-JHU data catalog were used
to select these targets such that they had significant [\ion{O}{3}] $\lambda4363$ 
auroral line detections.
However, to ensure uniformity we have remeasured the SDSS emission lines, as
described below, and used the most recent atomic data for the subsequent analysis.
From the optical spectral line measurements we obtain the physical parameters 
needed to analyze the UV spectra: 
interstellar reddening, electron temperature and density, direct oxygen abundance, 
and degree of ionization of the nebular gas.
The details of these calculations are given in Sections~\ref{sec3.3} and \ref{sec4}, 
with the resulting parameters listed in Tables~\ref{tbl2} and \ref{tbl3}.

\subsection{Nebular Emission Line Measurements} \label{sec3.3}

Emission line strengths for both the COS and SDSS spectra were measured 
using the {\tt SPLOT} routine within IRAF\footnotemark[10]. 
Groups of nearby lines were fit simultaneously, constrained by a single Gaussian 
FWHM and a single line center offset from the vacuum wavelengths (i.e., redshift).
Special attention was paid to the Balmer lines, 
which can be located in troughs of significant underlying stellar absorption. 
In the cases where Balmer absorption was clearly visible, 
the bluer Balmer lines (H$\delta$ and H$\gamma$) were fit simultaneously with 
multiple components such that the absorption was fit by a broad, negative Lorentzian 
profile and the emission was fit by a narrow, positive Gaussian profile. 
Note that our sample is composed of high ionization star-formation regions that 
display only weak Balmer absorption, consistent with the hard radiation 
fields from main sequence O and B stars, such that the absorption component
is negligible for the stronger H$\beta$ and H$\alpha$ emission lines.  
To ensure that noise spikes are not fit, only emission lines with a 
strength of 3$\sigma$ or greater are used in the subsequent abundance analysis. 

\footnotetext[10]{IRAF is distributed by the National Optical Astronomical Observatories.}

The errors of the flux measurements were approximated using
\begin{equation}
	\sigma_{\lambda} \approx \sqrt{(2 \times \sqrt{N} \times \mbox{rms})^2 + (0.01 \times F_{\lambda})^2},
	\label{eq1}
\end{equation}
where N is the number of pixels spanning the Gaussian profile fit to the narrow emission lines. 
The rms noise in the continuum was taken to be the average of the rms on each side of an emission line. 
The two terms in Equation~\ref{eq1} approximate the errors from continuum subtraction and flux calibration.
For weak lines, such as the UV CELs, the rms term determines the approximate uncertainty. 
In the case of strong Balmer, [\ion{O}{3}], and other lines, 
the error is dominated by the inherent uncertainty in the flux calibration and
accounted for by adding the 1\% uncertainty of standard star calibrations in CALSPEC \citep{bohlin10}.

The COS and SDSS spectra were de-reddened using Balmer 
line ratios and the application of the  \citet{cardelli89} reddening 
law, parametrized by $A_{V}=3.1\ E(B-V)$.
An initial estimate of the electron temperature was determined 
from the ratio of the [\ion{O}{3}] $\lambda4363$ auroral line to the 
[\ion{O}{3}] $\lambda\lambda4959,5007$ nebular lines and used as an 
input to determine the reddening.
We iterate on this process, using the de-reddened [\ion{O}{3}] line 
ratio to determine the new electron temperature, until the change 
in temperature is less than 10 K. 
The final reddening estimate is an error weighted average of the 
individual reddening values determined from the H$\alpha$/H$\beta$, 
H$\gamma$/H$\beta$, and H$\delta$/H$\beta$ ratios. 

All of our targets have low extinction in the range of E(B-V) of $\sim0.05$ to 0.19.
Note that the C and O abundances presented here have not been corrected 
for the fraction of atoms embedded in dust.
\citet{peimbert10} have estimated that the depletion of O ranges between 
roughly 0.08$-$0.12 dex, and has a positive correlation with O/H abundance.
C is also expected to be depleted in dust, 
mainly in polycyclic aromatic hydrocarbons and graphite.
The estimates of the amount of C locked up in dust grains in the local 
interstellar medium shows a relatively large variation depending on the 
abundance determination methods applied \citep[see, e.g.,][]{jenkins14}.
For the low abundance targets presented here, and their corresponding 
small extinctions, the depletion onto dust grains is likely small.

Reddening corrected line intensities measured for the seven objects with 
significantly detected \ion{O}{3}] and \ion{C}{3}] are reported in Table~\ref{tbl2}.
Figure~\ref{fig2} shows the rest-frame (corrected by SDSS redshift) 
\ion{C}{4}, \ion{O}{3}], and \ion{C}{3}] emission line regions of 
the COS spectra for each of the seven targets in our sample with 
significant C and O detections.
Note that the spectra of J122426 and J124827 were noticeably 
noisier than those of the other targets, and so have been displayed 
with a box-car smoothing of 3 pixels for visual aid.
The other five targets in our sample did not have detectable C or O emission lines.
We discuss the cause of these non-detections in Appendix~\ref{A3} and use them 
to improve our selection criteria for a future sample.

\subsection{Diagnostic Diagrams} \label{sec3.4}
\begin{figure*}
\plotone{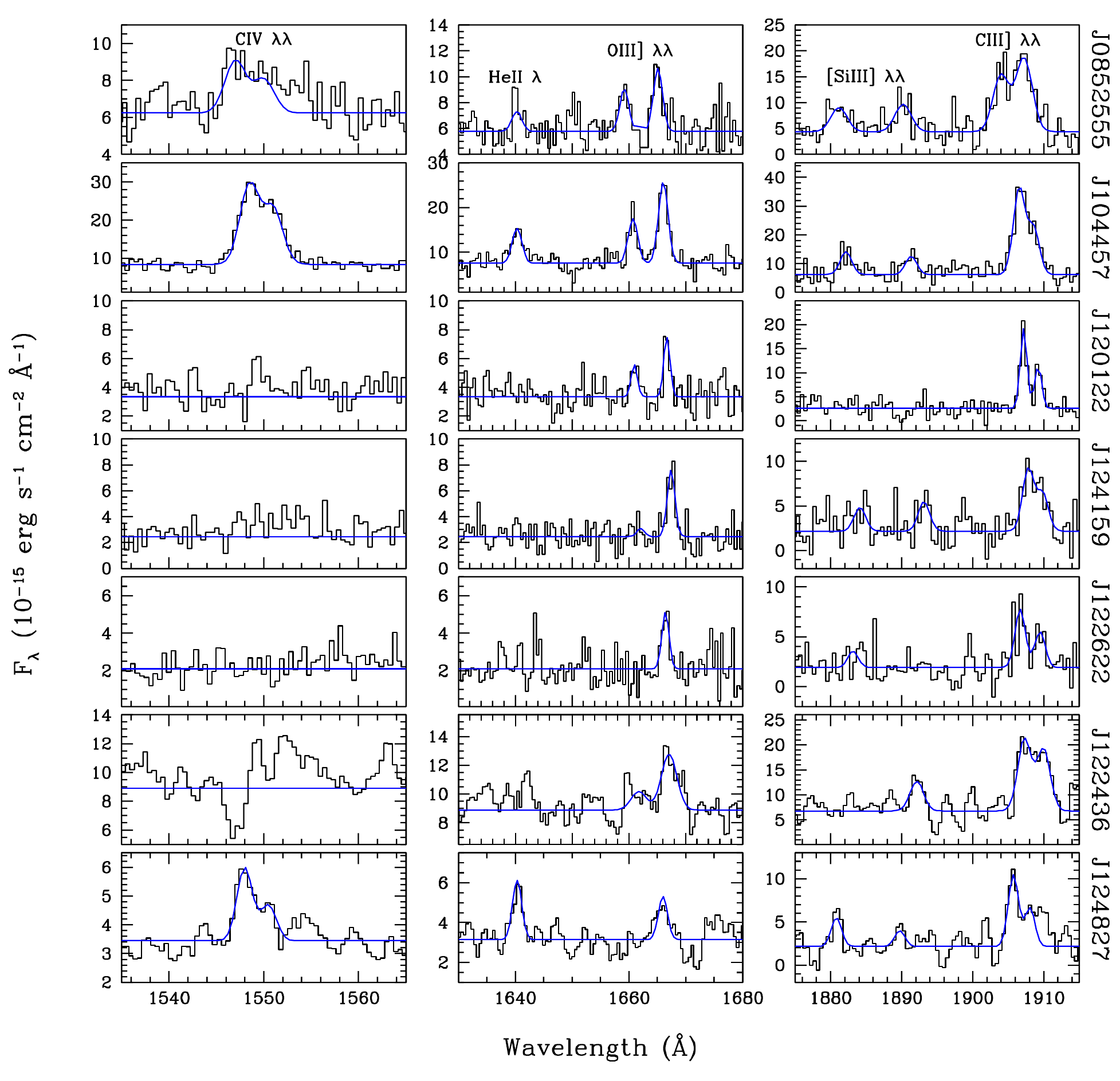}
\caption{HST/COS rest-frame emission line spectra for the seven dwarf 
galaxies in our sample with significant C and O detections.
The blue line represents the best fit to the emission lines as described in Section~\ref{sec3.3}.
Note that the spectra in the last two rows are noisy and so are displayed
with a box-car smoothing of 3. }
\label{fig2}
\end{figure*}

Several of the targets in our sample exhibit strong high-ionization emission lines. 
As shown in Figure~\ref{fig2}, significant \ion{C}{4} $\lambda\lambda1548,1550$
and \ion{He}{2} $\lambda1640$ emission is present in 3 of our 7 targets with C/O detections.
With ionization potentials of 24.6 eV and 47.9 eV to reach \ion{He}{2} and \ion{C}{4} respectively,
emission line features of these species are more commonly observed in high energy 
objects, such as AGN.
\citet{hainline11} used a composite AGN spectrum to show that narrow-lined AGN 
have very strong \ion{C}{4} emission on average (EW(\ion{C}{4}) = 16.3 \AA), which
is typically much stronger than \ion{C}{3}] (average \ion{C}{4}/\ion{C}{3}]
flux ratio $\sim7.5$; \cite{alexandroff13}).
In contrast, we measure much smaller ratios of \ion{C}{4}/\ion{C}{3}] ($< 1.0$) 
for our sample, similar to \citet{stark14} who observed \ion{C}{4}/\ion{C}{3}] 
in lensed extreme emission line galaxies.
Additionally, the \ion{He}{2} $\lambda1640$ emission observed for our sample 
appears to be narrow, indicative of a nebular origin. 
\citet{brinchmann08} have investigated the origin of nebular \ion{He}{2},
arguing that O stars are the main sources of the \ion{He}{2} ionizing photons
in metal poor systems with 12 + log(O/H) $< 8.0$.

\begin{figure}
\plotone{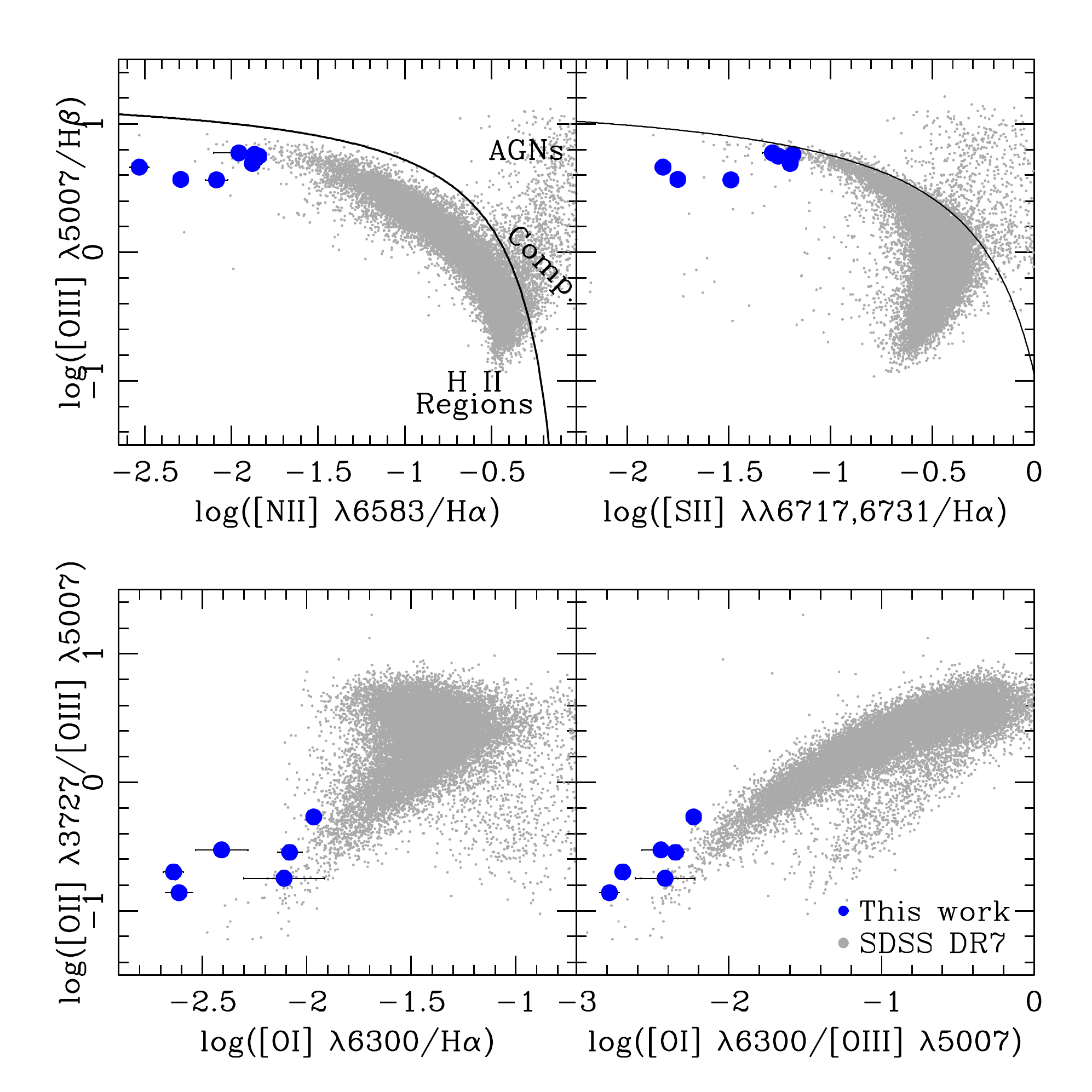}
\caption{SDSS emission line ratios for the seven dwarf galaxies in our sample
with significant C and O detections.
A low-mass subset of the SDSS DR7 is plotted in grey as a comparison sample. 
The solid lines are the theoretical starburst limits from \citet{kewley06}.}
\label{fig3}
\end{figure}

To verify that our target observations originate from photoionized \ion{H}{2} regions,
we plot four of the standard BPT emission line diagnostic diagrams 
\citep{baldwin81} in Figure~\ref{fig3}.
Line measurements for the current sample are plotted as blue points
in comparison to the grey locus of SDSS DR7 low-mass \citep[M$_{\star} < 10^9$ M$_{\odot}$, which are expected 
to be relatively low-metallicity following the mass-metallicity relationship; e.g.,][]{tremonti04,berg12}, 
star-forming galaxies taken from the MPA-JHU database.
In the top two panels, the solid lines are the theoretical starburst limits from \citet{kewley06}.
Based on these plots, our sample exhibits the expected properties of photoionized regions,
although they are clearly [\ion{N}{2}] and [\ion{S}{2}] deficient outliers with respect to average star-forming galaxies.
In the bottom panels of Figure~\ref{fig3} we find no indication of contributions from shock excitation,
which can manifest as strong [\ion{O}{1}] emission.


\begin{deluxetable*}{lccccccc}
\tabletypesize{\scriptsize}
\tablewidth{0pt}
\setlength{\tabcolsep}{3pt}
\tablecaption{ Emission-Line Intensities for HST/COS Observations of Nearby Compact Dwarf Galaxies}
\tablewidth{0pt}
\tablehead{
\CH{} & \multicolumn{7}{c}{$I(\lambda)/I(\mbox{H}\beta)$} }
\startdata
{Ion}                       			& {J082555}      	& {J104457}      	& {J120122}     		& {J124159}      	& {J122622}      	& {J122436}      	 	& {J124827}		\\
\hline
{C~\iv $\lambda$1548.19}		& 0.55$\pm$0.02	& 2.07$\pm$0.07	& \nodata			& \nodata			& \nodata			& \nodata			& 0.82$\pm$0.24	\\
{C~\iv $\lambda$1550.77}		& 0.42$\pm$0.01	& 1.52$\pm$0.06	& \nodata			& \nodata			& \nodata			& \nodata			& 0.40$\pm$0.24	\\
{He~II~$\lambda$1640.42} 	& 0.38$\pm$0.01	& 0.70$\pm$0.05	& \nodata			& \nodata			& \nodata			& \nodata			& 0.96$\pm$0.24	\\
{O~\iii] $\lambda$1660.81}	& 0.49$\pm$0.02	& 0.85$\pm$0.05	& 0.57$\pm$0.20	& 0.20$\pm$0.20	& \nodata			& 0.53$\pm$0.11	& \nodata			\\
{O~\iii] $\lambda$1666.15}	& 0.83$\pm$0.02	& 1.49$\pm$0.06	& 1.05$\pm$0.20	& 1.72$\pm$0.28	& 0.010$\pm$0.001	& 1.28$\pm$0.11	& 0.77$\pm$0.22	\\
{Si~\iii] $\lambda$1883.00}	& 1.16$\pm$0.03	& 0.75$\pm$0.06	& \nodata			& 1.10$\pm$0.25	& \nodata			& \nodata			& \nodata			\\
{Si~\iii] $\lambda$1892.03}	& 1.25$\pm$0.04	& 0.66$\pm$0.05	& \nodata			& 1.38$\pm$0.25	& \nodata			& 1.29$\pm$0.04	& \nodata			\\
{C~\iii] $\lambda$1906.68}	& 2.63$\pm$0.08	& 2.83$\pm$0.08	& 3.89$\pm$0.22	& 2.94$\pm$0.32	& 0.022$\pm$0.001	& 3.38$\pm$0.13	& 2.12$\pm$0.18	\\
{C~\iii] $\lambda$1908.73}	& 3.44$\pm$0.10	& 1.12$\pm$0.06	& 2.04$\pm$0.20	& 1.80$\pm$0.27	& 0.012$\pm$0.001	& 2.96$\pm$0.12	& 1.05$\pm$0.16	\\
\hline
{[O~\ii]~$\lambda$3727}   		& \nodata			& \nodata			& \nodata			& \nodata			& \nodata			& 0.973$\pm$0.029	& 0.779$\pm$0.025 	\\
{[Ne~\iii]~$\lambda$3868}		& 0.276$\pm$0.008	& \nodata			& \nodata			& \nodata			& 0.485$\pm$0.014	& \nodata			& \nodata			\\
{H$\delta$ $\lambda$4101} 	& 0.266$\pm$0.008	& 0.295$\pm$0.009	& 0.302$\pm$0.010	& 0.265$\pm$0.007	& 0.265$\pm$0.007	& 0.257$\pm$0.008	& 0.239$\pm$0.008	\\ 	   		
{H$\gamma$ $\lambda$4340}	& 0.476$\pm$0.013	& 0.506$\pm$0.015	& 0.513$\pm$0.017	& 0.495$\pm$0.016	& 0.493$\pm$0.014 	& 0.458$\pm$0.014	& 0.470$\pm$0.018	\\
{[O~\iii]~$\lambda$4363}  		& 0.116$\pm$0.003	& 0.146$\pm$0.004	& 0.103$\pm$0.006  & 0.105$\pm$0.012	& 0.116$\pm$0.003	& 0.111$\pm$0.004	& 0.129$\pm$0.014	\\	 	
{H$\beta$ $\lambda$4861}  	& 1.000$\pm$0.028	& 1.000$\pm$0.031	& 1.000$\pm$0.037	& 1.000$\pm$0.028	& 1.000$\pm$0.028	& 1.000$\pm$0.031	& 1.000$\pm$0.036	\\	
{[O~\iii]~$\lambda$4958}  		& 1.232$\pm$0.035	& 1.470$\pm$0.004	& 1.204$\pm$0.041	& 1.608$\pm$0.086	& 1.899$\pm$0.054	& 1.840$\pm$0.054	& 1.990$\pm$0.064	\\
{[O~\iii]~$\lambda$5006}  		& 3.620$\pm$0.103	& 4.525$\pm$0.134	& 3.577$\pm$0.103	& 4.811$\pm$0.136	& 5.688$\pm$0.161	& 5.554$\pm$0.166	& 5.915$\pm$0.198	\\	
{[N~\ii]~$\lambda$6548}   		& 0.005$\pm$0.001	& 0.003$\pm$0.001	& \nodata			& 0.014$\pm$0.004	& 0.015$\pm$0.001	& 0.019$\pm$0.003	& 0.010$\pm$0.012	\\
{H$\alpha$ $\lambda$6562} 	& 2.757$\pm$0.080	& 2.748$\pm$0.082	& 2.779$\pm$0.092	& 2.793$\pm$0.076	& 2.791$\pm$0.079	& 2.791$\pm$0.085	& 2.775$\pm$0.094	\\
{[N~\ii]~$\lambda$6583}   		& 0.014$\pm$0.001	& 0.008$\pm$0.001	& 0.023$\pm$0.004	& 0.037$\pm$0.004	& 0.038$\pm$0.001	& 0.040$\pm$0.003	& 0.031$\pm$0.012	\\
{[S~\ii]~$\lambda$6716}   		& 0.026$\pm$0.001	& 0.022$\pm$0.001	& 0.052$\pm$0.004	& 0.098$\pm$0.005	& 0.106$\pm$0.003	& 0.087$\pm$0.004	& 0.084$\pm$0.012	\\
{[S~\ii]~$\lambda$6730}   		& 0.022$\pm$0.001	& 0.018$\pm$0.001	& 0.037$\pm$0.004	& 0.076$\pm$0.005	& 0.074$\pm$0.002	& 0.066$\pm$0.004	& 0.059$\pm$0.012	\\			
{[O~\ii]~$\lambda$7319}   		& 0.006$\pm$0.001	& 0.005$\pm$0.001	& 0.005$\pm$0.003	& 0.015$\pm$0.001	& 0.019$\pm$0.005	& 0.015$\pm$0.003	& 0.008$\pm$0.002	\\
{[O~\ii]~$\lambda$7330}   		& 0.006$\pm$0.001	& 0.005$\pm$0.001	& 0.011$\pm$0.003	& 0.011$\pm$0.001	& 0.032$\pm$0.001	& 0.013$\pm$0.003	& 0.011$\pm$0.002	\\
{[S~\iii]~$\lambda$9068}  		& 0.045$\pm$0.001	& 0.040$\pm$0.001	& 0.063$\pm$0.003	& 0.082$\pm$0.002	& 0.103$\pm$0.005	& \nodata			& \nodata			\\
\hline
{E(B$-$V)}				& 0.160$\pm$0.009	& 0.120$\pm$0.010	& 0.190$\pm$0.010	& 0.170$\pm$0.008	& 0.120$\pm$0.009	& 0.080$\pm$0.010	& 0.050$\pm$0.012	\\
{F$_{H\beta}$}				& 230.8			& 413.7			& 114.3			& 98.3			& 8131			& 138.4			& 78.1			\\
\enddata
\tablecomments{
The first column lists the vacuum wavelengths of the observed ions for wavelengths 
of $\lambda < 2000$ \AA, and approximate air wavelengths for optical emission lines.
The flux values for each object listed are reddening corrected intensity ratios relative to H$\beta$.
The last two rows are extinction and the H$\beta$ raw fluxes, in units of 
$10^{-16}$ erg s$^{-1}$ cm$^{-2}$, measured from the SDSS spectra. }
\label{tbl2}
\end{deluxetable*}


\section{Chemical Abundances} \label{sec4}

In order to minimize sources of uncertainty,
we compute the chemical abundances for our sample, as well as the supplemented 
literature sources (see \S~\ref{sec5.1}), in a uniform, consistent manner.
With the exception of the C/O ratio, 
all physical conditions and abundances are calculated from the optical spectra.
The C/O abundances are determined from the UV O$^{+2}$ and C$^{+2}$ CELs
for the reasons given in Sections~\ref{sec1} and ~\ref{sec2}.


\subsection{Temperature and Density} \label{sec4.1}

A simple \ion{H}{2} region can be modeled by three separate volumes: 
a low-, intermediate-, and high-ionization zone.
Accurate \ion{H}{2} region abundance determinations require 
reliable electron temperature measurements for each volume.
This is typically done by observing a temperature-sensitive auroral-to-strong-line ratio. 
The [\ion{O}{3}] I($\lambda\lambda$4959,5007)/I($\lambda$4363) ratio 
is expected to reflect the temperature in the high ionization zone.
We use this ratio to determine electron temperatures using the reddening 
corrected line-ratios from the SDSS spectra and updated atomic data following \citet{berg15}, 
assuming the ions are well-approximated by a 5-level atom\footnotemark[11].
We note that the electron temperature can also be determined from the 
[\ion{O}{3}] $\lambda5007$/\ion{O}{3}]$\lambda1666$ ratio, as is commonly done 
in high redshift targets where the intrinsically faint optical auroral line is often undetected.
However, for nearby targets, this requires combining space- and ground-based observations,
potentially introducing flux matching issues and mismatched aperture effects.
In fact, for the seven targets presented here, on average the [\ion{O}{3}] $\lambda5007$/\ion{O}{3}]$\lambda1666$
electron temperatures are $\sim3000$ K lower than the [\ion{O}{3}] $\lambda5007$/$\lambda4363$ temperatures.
Further observations are needed to understand this difference.
For simplicity, we determine the high ionization zone electron 
temperature from the optical [\ion{O}{3}] ratio only.

\footnotetext[11]{\url{https://github.com/moustakas/impro};
IMPRO is an assortment of routines written in IDL by J. Moustakas for reducing 
and analyzing multiwavelength imaging and spectroscopy of galaxies.}

Once a direct electron temperature is determined, the physical conditions 
of the other zones are needed to complete the \ion{H}{2} region picture.
Following \citet[hereafter G92;][]{garnett92}, photoionization models 
can be used to relate the direct temperatures of different ionization zones: 
\begin{align}
       \mbox{T[S~\iii]} & =  0.83\times \mbox{T[O~\iii]} + 1700\mbox{ K} \label{eqn:G92-1} \\
       \mbox{T[N~\ii]} & =  0.70\times \mbox{T[O~\iii]} + 3000\mbox{ K,} \label{eqn:G92-2}
\end{align}
where Equation~\ref{eqn:G92-2} was adopted from \citet{campbell86}, based on models from \citet{stasinska82}.
These relationships are valid for temperatures typical 
of \ion{H}{2} regions: $T_e$ = 2,000$-$18,000\,K.
We adopt the [\ion{O}{3}] temperature for the high ionization zone, and use the G92 
relationships to determine the low- and intermediate-ionization zone temperatures. 
We note that several recent studies have investigated the validity of the G92 temperature 
relationships with respect to updated atomic data, larger samples, and higher quality 
observations \citep[e.g.,][]{kennicutt03a,binette12,berg15} and found significant differences.
However, by the nature of our target selection, our sample is very high excitation,
and so there are minimal contributions from the low-ionization lines 
([\ion{O}{2}] $\lambda3727$ and [\ion{O}{2}] $\lambda\lambda7320,7330$),
and thus little dependence on the low-ionization zone temperatures
for the oxygen abundances.

The [\ion{S}{2}] $\lambda\lambda$6717,6731 ratio is used to determine the electron densities.
Ideally, we would compare with the density determinations from the \ion{C}{3}] 
$\lambda\lambda$1906,1909 line ratio, but, five of our seven targets have
ratios outside of the calibration range of the \ion{C}{3}] density diagnostic 
and the other two have highly discrepant \ion{C}{3}] densities.
Despite the large uncertainties on the \ion{C}{3}] ratios due to the rapidly declining 
throughput of the G140L grating at those wavelengths, the ratios of J104457 and 
J122622 fall significantly outside of the theoretically predicted range ($> 3\sigma$).
This optical vs.\ UV density discrepancy problem has been noted by several studies of high-redshift
lensed galaxies \citep[e.g.,][]{hainline09,quider09,christensen12,bayliss14},
and may be the result of different zones of origination for [\ion{S}{2}] and \ion{C}{3}] 
\citep{james14}.
Additionally, the \ion{C}{3}] ratio is an insensitive measure of electron density
below values of 10$^3$ cm$^{-3}$, in particular, for values of $n_e\sim100$ cm$^{-3}$
that are typical of \ion{H}{2} regions in the local universe. 
Opportunely, both C$^{+2}$ and O$^{+2}$ have high critical densities 
($> 10^5$/cm$^3$) and so are insensitive to electron density.
The electron temperature and density determinations are listed in Table~\ref{tbl3}.

\subsection{Ionic And Total Abundances} \label{sec4.2}

Ionic abundances relative to hydrogen are calculated using:
\begin{equation}
	{\frac{N(X^{i})}{N(H^{+})}\ } = {\frac{I_{\lambda(i)}}{I_{H\beta}}\ } {\frac{j_{H\beta}}{j_{\lambda(i)}}\ }.
	\label{eq:Nfrac}
\end{equation}
The emissivity coefficients, $j_{\lambda(i)}$, which are functions of both 
temperature and density, are determined using a 5-level atom approximation 
with the updated atomic data reported in \citet{berg15}. 

Total oxygen abundances (O/H) are calculated from the simple 
sum of O$^{+}$/H$^{+}$ and O$^{+2}$/H$^{+}$, as contributions
from O$^{+3}$ (requiring an ionization potential of 54.9 eV) 
are typically negligible in \ion{H}{2} regions.
Traditionally, O$^{+}$/H$^{+}$ is determined from the optical [\ion{O}{2}] $\lambda3727$ 
blended line. However, due to the limited SDSS blue wavelength coverage and the
redshift of our sample, [\ion{O}{2}] $\lambda3727$ is not detected in 5 of our targets. 
In these cases, O$^{+}$/H$^{+}$ is determined from the 
optical red [\ion{O}{2}] $\lambda\lambda7320,7330$ doublet \citep[e.g.,][]{kniazev04}.

Other abundance determinations require ionization correction 
factors (ICF) to account for unobserved ionic species. 
For nitrogen, we adopt the convention of N/O = N$^{+}$/O$^{+}$ 
\citep{peimbert67}, which is valid at a precision of about 10\% \citep{nava06}.
Multiple ions of sulfur are measured in 5 of our 7 targets.
However, unlike the simple ionization structure of nitrogen,
both S$^{+2}$ and S$^{+3}$ lie in the O$^{+2}$ zone.
To correct for the unobserved S$^{+3}$ state, we employ the ICF from \citet{thuan95}: 

\footnotesize{
\begin{align}
        \mbox{ICF(S)} & =  \frac{\mbox{S}}{\mbox{S}^{+} + \mbox{S}^{+2}} \nonumber \\
        			       & = \big[0.013 + x\{5.10 + x[-12.78 + x(14.77 - 6.11x)]\}\big]^{-1}, \nonumber \\
\end{align}}
\normalsize 
where $x =$ O$^{+}$/O.
Ionic and total O, N, and S determined from the optical spectra are listed for our 7 
galaxies in Table~\ref{tbl3}.

\subsubsection{C/O Abundance} \label{sec4.2.1}

In the simplest case, C/O can be determined from the C$^{+2}$/O$^{+2}$ ratio alone.
Since O$^{+2}$ has a higher ionization potential than C$^{+2}$ (54.9 eV versus 47.9 eV, respectively), 
regions ionized by a hard ionizing spectrum may have a significant amount of carbon in the C$^{+3}$ form, 
causing the C$^{+2}$/O$^{+2}$ ionic abundance ratio to underestimate the true C/O abundance. 
The metallicity dependence of the stellar continua, the stellar mass-$T_{eff}$
relation \citep{maeder90}, and stellar mass \citep{terlevich85} will also systematically 
affect the relative ionization fractions of these species.  
To correct for this effect, we apply the ICF described by G95:
\begin{align}
	{\frac{\mbox{C}}{\mbox{O} }} & = {\frac{\mbox{C}^{+2}}{\mbox{O}^{+2}}\ }\times \Bigg[{\frac{X(\mbox{C}^{+2})}{X(\mbox{O}^{+2})}}\Bigg]^{-1} \nonumber \\
			     			    & = {\frac{\mbox{C}^{+2}}{\mbox{O}^{+2}}\ }\times{\mbox{ICF}},	
\end{align}
where X(C$^{+2}$) and X(O$^{+2}$) are the C$^{+2}$ and O$^{+2}$ volume fractions, respectively.

\begin{figure} 
\begin{center}
	\includegraphics[scale=0.9,trim=10mm 7mm 50mm 15mm,clip]{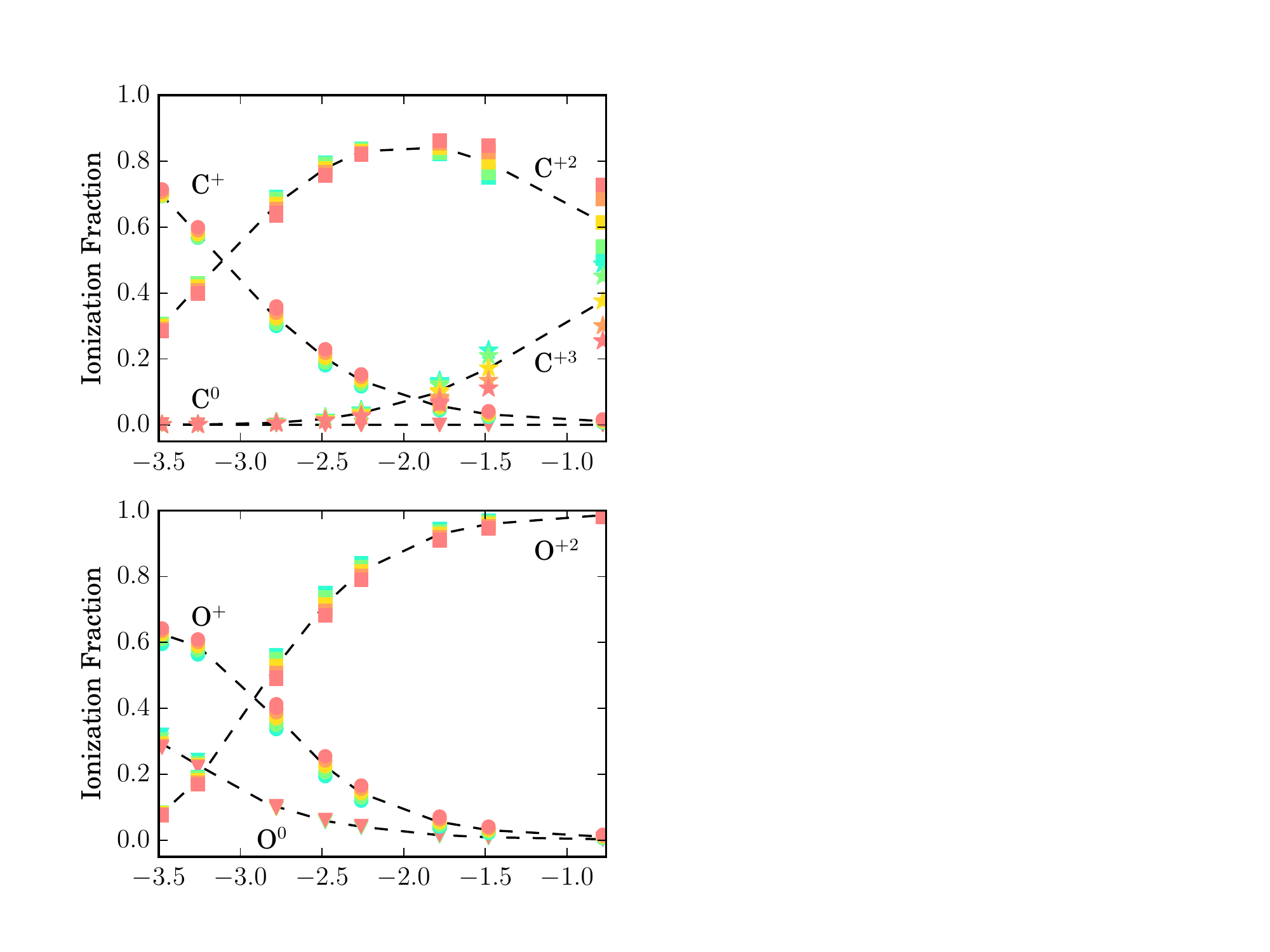}
	\includegraphics[scale=0.9,trim=10mm 65mm 50mm 12mm,clip]{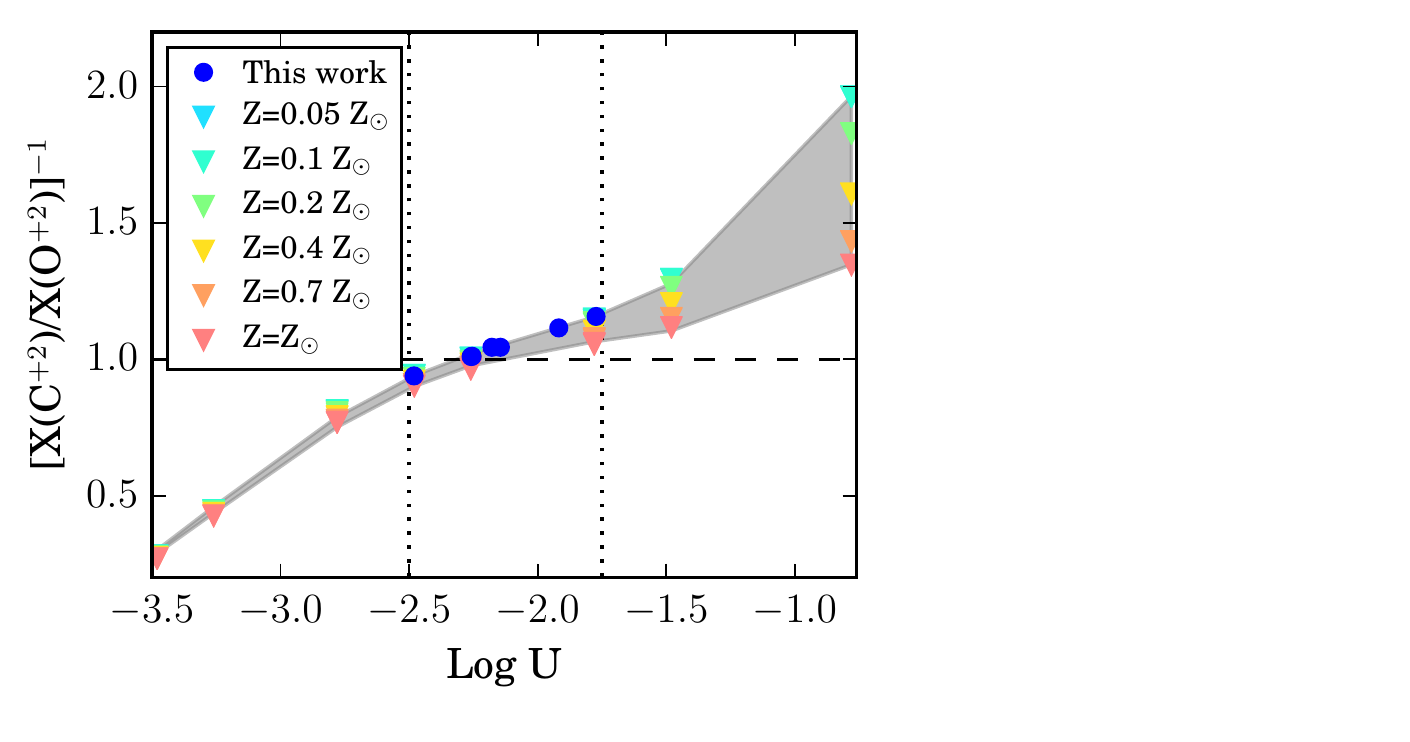} 
\caption{The ionization fraction of C and O species as a function of ionization parameter
for the Z = 0.1 Z$_{\odot}$ (Z = 0.002) stellar models ignoring rotation. 
Symbols are color coded by the gas-phase
oxygen abundance. In the top two panels the different ionic species are designated by
triangles, circles, squares, and stars in order of increasing ionization.
The bottom panel plots the ionization correction factor versus ionization parameter,
 where the seven significant detections in our sample are depicted as blue circles and
 the vertical dotted lines mark the range in log U of our sample. }
\label{fig4}
\end{center}
\end{figure}

\begin{figure} 
\begin{center}
	\includegraphics[scale=0.9,trim=8mm 0mm 50mm 15mm,clip]{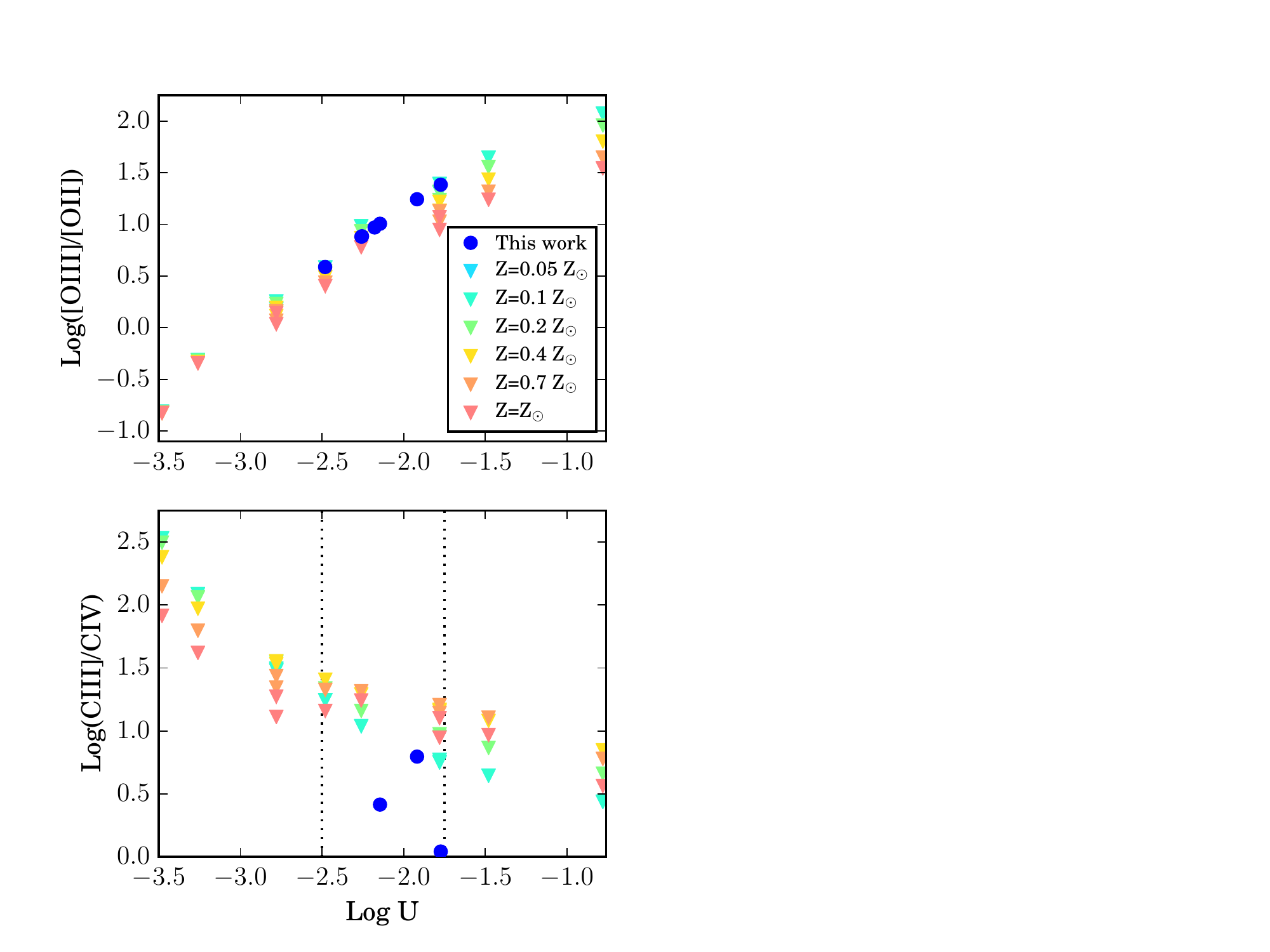}
\caption{Top: Log([\ion{O}{3}] $\lambda5007$/[\ion{O}{2}] $\lambda3727$) versus ionization
parameter from CLOUDY models. 
Symbols are color coded by the gas-phase oxygen abundance.
Following the predicted trend for low metallicities allowed us to estimate the ionization parameter
for our sample based on the [\ion{O}{3}] $\lambda5007$/[\ion{O}{2}] $\lambda3727$ ratio;
the results are plotted as filled blue symbols.
Bottom: Predicted log(\ion{C}{3}] $\lambda\lambda1907,1909$/\ion{C}{4} $\lambda\lambda1548,1550$) 
UV emission line ratios from CLOUDY models versus ionization parameter. 
The three targets in our sample with significant \ion{C}{4} emission
are plotted as filled blue circles, demonstrating their anomalously strong \ion{C}{4} emission.
The range in log U of our sample is depicted by vertical dotted lines.}
\label{fig5}
\end{center}
\end{figure}

We estimate the ICF as a function of the ionization parameter using CLOUDY \citep{ferland13}.
Starburst99 models \citep{leitherer99}, with and without rotation, for two different metallicities 
(Z = 0.1 Z$_{\odot}$, Z = 0.7 Z$_{\odot}$)  
were considered for a region of continuous star formation.
Interestingly, for models assuming a stellar population age of 100 Myr,
stellar metallicity has a much larger effect on the input ionizing 
spectrum than including rotation.
The ionization fraction of C and O species as a function of ionization parameter
are shown in Figure~\ref{fig4} for the 
Z = 0.1 Z$_{\odot}$ (Z = 0.002) stellar models. 
In Figure~\ref{fig5} we plot the predicted [\ion{O}{3}] $\lambda5007$/[\ion{O}{2}] $\lambda3727$ 
emission line ratio versus ionization parameter from our CLOUDY models. 
We calculated the log([\ion{O}{3}]/[\ion{O}{2}]) ratio and direct oxygen abundance  
for our sample from the optical spectra and compared to Figure~\ref{fig5} to determine log U. 

The resulting ionization parameters for objects in our sample are plotted 
against the models in Figures~\ref{fig4} and ~\ref{fig5}.
Vertical dotted lines mark the range of our sample: $-2.48 <$ log U $< -1.77$,
which corresponds with a C$^{+2}$/O$^{+2}$ ICF of 0.94$-$1.16.
We estimate the uncertainty in the ICF as the scatter amongst the different
models considered (stellar metallicity and rotation effects) at a given O$^{+2}$ volume fraction.
Ionic and total C abundances, as well as the corrected C/O ratio, are given in Table~\ref{tbl3}.

Further evidence for the importance of applying an ICF comes from
the visible \ion{C}{4} emission in three of our targets in Figure~\ref{fig2}.
While \ion{C}{4} can be directly measured in these three galaxies,
the observed emission doublet is anomalously strong relative to \ion{C}{3}] 
in comparison to the CLOUDY model predictions shown in Figure~\ref{fig5}; 
additionally, it can be complicated by stellar contributions.
Alternatively, we use the ICF to estimate the C$^{+3}$ contribution,
allowing all seven of our targets to be treated uniformly. 

The previous CEL and RL studies compared to in this work
have all applied ionization corrections from \citet{garnett99}.
However, our updated photoionization models 
produce ICFs which agree with the ICFs 
determined by \citet{garnett99} within less than 0.05 dex, on average.
Therefore any systematic differences between the CEL and RL abundances presented here 
or between literature CEL abundances and this work are not related to choice of ICF.


\begin{deluxetable*}{lccccccc}
\tabletypesize{\scriptsize}
\tablewidth{0pt} 
\setlength{\tabcolsep}{3pt}
\tablecaption{ Ionic and Total Abundance for HST/COS Compact Dwarf Galaxies }
\tablewidth{0pt}
\tablehead{
\CH{Target} 	        			& \CH{J082555}      	& \CH{J104457}      	& \CH{J120122}     	& \CH{J124159}      	& \CH{J122622}      	& \CH{J122436}    	 & \CH{J124827}}
\startdata
\multicolumn{8}{c}{Properties Derived from Optical Spectra} \\
\hline \\
T$_e$ [O~\iii] (K)			& 19,300$\pm$400	& 19,600$\pm$500	& 18,300$\pm$600	& 15,800$\pm$900	& 15,300$\pm$300	& 15,200$\pm$300	& 15,800$\pm$800	\\
T$_e$ [N~\ii] (K)$^a$		& 16,500$\pm$300	& 16,300$\pm$300	& 15,800$\pm$400	& 14,100$\pm$600	& 13,700$\pm$200	& 13,600$\pm$200	& 14,100$\pm$600	\\		
n$_e$ C~\iii] (cm$^{-3})^b$	& 36,600 			& 100			& 100			& 100			& 100			& 12,000			& 100			\\
n$_e$ [S~\ii] (cm$^{-3})$ 		& 260			& 260			& 30				& 150			& 20				& 140			& 10				\\
\\
O$^+$/H$^+$ (10$^{5})$		& 0.18$\pm$0.02	& 0.17$\pm$0.10	& 0.36$\pm$0.11	& 0.87$\pm$0.10	& 2.15$\pm$0.09	& 1.13$\pm$0.06$^c$ & 0.79$\pm$0.08$^c$ \\
O$^{+2}$/H$^+$ (10$^{5})$	& 2.19$\pm$0.07	& 2.62$\pm$0.03	& 2.44$\pm$0.13	& 4.54$\pm$0.48	& 5.84$\pm$0.18	& 5.82$\pm$0.25	& 5.62$\pm$0.53	\\
12 + log(O/H)				& 7.37$\pm$0.01	& 7.45$\pm$0.02	& 7.45$\pm$0.03	& 7.73$\pm$0.04	& 7.90$\pm$0.01	& 7.84$\pm$0.02	& 7.81$\pm$0.03	 \\
\\
N$^{+}$/H$^+$ (10$^{7})$		& 0.94$\pm$0.07	& 0.56$\pm$0.08	& 1.66$\pm$0.29	& 3.46$\pm$0.48	& 3.82$\pm$0.14	& 4.30$\pm$0.38	& 2.80$\pm$1.19	\\
log(N/O)					& -1.28$\pm$0.04	& -1.47$\pm$0.07	& -1.34$\pm$0.08	& -1.41$\pm$0.08	& -1.77$\pm$0.02	& -1.46$\pm$0.04	& -1.45$\pm$0.15	\\
\\
S$^{+}$/H$^+$ (10$^{7})$		& 0.45$\pm$0.02	& 0.37$\pm$0.02	& 0.85$\pm$0.07	& 2.10$\pm$0.19	& 2.23$\pm$0.08	& 1.97$\pm$0.09	& 1.68$\pm$0.24	\\
S$^{+2}$/H$^+$(10$^{7})$	& 2.66$\pm$0.13	& 2.30$\pm$0.12	& 4.03$\pm$0.29	& 6.37$\pm$0.56	& 8.42$\pm$0.05	& \nodata			& \nodata			\\
ICF						& 2.962			& 3.825			& 2.042			& 1.788			& 1.397			& \nodata			& \nodata			\\
log(S/O)					& -1.41$\pm$0.04	& -1.44$\pm$0.04	& -1.45$\pm$0.05	& -1.55$\pm$0.06	& -1.73$\pm$0.04	& \nodata			& \nodata			\\
\hline \\
\multicolumn{8}{c}{Properties Derived from UV Spectra} \\
\hline \\
C$^{+2}$/H$^+$ (10$^{5})$	& 3.12$\pm$0.40	& 1.91$\pm$0.20	& 3.90$\pm$1.17	& 6.25$\pm$2.61	& 0.053$\pm$0.001	& 10.5$\pm$1.8	& 4.20$\pm$1.01	\\
C$^{+3}$/H$^+$ (10$^{5})$	& 5.83$\pm$0.13	& 2.00$\pm$0.19	& \nodata			& \nodata			& \nodata			& \nodata			& 2.34$\pm$0.78	\\
O$^{+2}$/H$^+$ (10$^{5})$	& 6.80$\pm$0.16	& 11.3$\pm$0.07	& 11.5$\pm$2.3		& 41.6$\pm$11.6	& 0.296$\pm$0.001	& 40.5$\pm$4.5	& 18.7$\pm$4.1	\\
C$^{+2}$/O$^{+2}$			& 0.46$\pm$0.06	& 0.17$\pm$0.02	& 0.34$\pm$0.09	& 0.15$\pm$0.07	& 0.18$\pm$0.02	& 0.26$\pm$0.05	& 0.22$\pm$0.10	\\
log(C~\iii]/C~\iv)			& 0.796			& 0.042			& \nodata			& \nodata			& \nodata			& \nodata			& 0.415			\\
log U						& -1.92			& -1.77			& -2.18			& -2.26			& -2.48			& -2.26			& -2.15			\\
ICF						& 1.115$\pm$0.085	& 1.157$\pm$0.070	& 1.044$\pm$0.045	& 1.011$\pm$0.040	& 0.939$\pm$0.040	& 1.010$\pm$0.040 	& 1.044$\pm$0.045	\\
log(C/O)					& -0.29$\pm$0.06	& -0.71$\pm$0.06	& -0.45$\pm$0.11	& -0.82$\pm$0.16	& -0.77$\pm$0.05	& -0.59$\pm$0.07	& -0.63$\pm$0.17	\\
\enddata
\tablecomments{
Ionic and total abundance calculations for our compact dwarf galaxy sample.
The T$_e$~[\ion{O}{3}] electron temperature, n$_e$~[\ion{S}{2}] density, and oxygen abundance are determined using the SDSS optical spectra. 
As shown in Table~\ref{tbl2}, [\ion{O}{2}] $\lambda3727$ was observed in only two of our seven galaxies, 
so O$^{+}$/H$^+$ ionic abundance was determined from the [\ion{O}{2}] $\lambda\lambda7320,7330$ lines unless otherwise notes. 
C/O abundances were calculated using the C$^{+2}$/O$^{+2}$ ratio and corrected for the contribution from other ionization species. 
The C/O ionization parameter is determined using 12+log(O/H), log([\ion{O}{3}]/[\ion{O}{2}]), and the model photoionization 
diagnostic curves shown in Figures~\ref{fig4} and ~\ref{fig5}.\\
$^{a}$ T$_e$[\ion{N}{2}] determined from T$_e$[\ion{O}{3}] and the \citet{garnett92} relationship. \\
$^{b}$ n$_e$\ion{C}{3}] is set equal to 100 when the \ion{C}{3}] $\lambda1907/\lambda1909$ 
ratio is greater than the low density theoretical limit. \\
$^{c}$ O$^{+}$/H$^+$ ionic was determined using [\ion{O}{2}] $\lambda3727$ line.}
\label{tbl3}
\end{deluxetable*}


\section{RESULTS} \label{sec5}


\subsection{The Optimum C/O Sample} \label{sec5.1}

Observations of collisionally excited C and O emission exist for only a small number of galaxies.
In order to assemble the most comprehensive picture to date of C/O determinations,
we combine our new observations with previously published values.
The first detailed abundance study of the UV spectrum of an \ion{H}{2} region was that
of the Orion Nebula by \citet{torres-peimbert80} utilizing International Ultraviolet Explorer (IUE) 
observations. 
As reported by \citet{dufour84}, C/O abundances were measured for perhaps two dozen 
\ion{H}{2} regions with the IUE, yet most were of low signal-to-noise detections 
and often combined the UV and optical CELs, the exception being the study of N81
by \citet{dufour82}.
Later, the first FOS/HST observations of collisionally excited carbon and 
oxygen emission lines in the UV were taken by \citet{garnett95,garnett97} 
for 7 nearby metal-poor dwarf galaxies.
However, \ion{O}{3}] $\lambda1666$ was detected at a strength above 3$\sigma$ in only 4 of these targets.
Kobulnicky \& Skillman (1997,1998) furthered the study of C/O, adding measurements in
3 \ion{H}{2} regions in NGC 5253 and 3 metal-poor dwarf galaxies respectively.
Unfortunately, the UV \ion{O}{3}] lines were measured as upper limits, and,
consequently, C/O abundances were determined using \ion{C}{3}] $\lambda1909$ with [\ion{O}{3}] $\lambda5007$.
While Kobulnicky \& Skillman (1997,1998) were careful to use multiple FOS gratings to observe
the necessary optical and UV lines with the same instrument and thus avoid aperture effects,
the wavelength difference of lines used makes the resulting C/O value very sensitive to
reddening uncertainties and choice of extinction curve.
\citet{izotov99} recalculated C/O abundances in a uniform manner for the above studies,
but, again, they used the combined UV/optical approach subject to large extinction uncertainties.

The optimal way to constrain the relationship of C with O 
is to compare all targets in a uniform way.
Because both \ion{C}{3}] and \ion{O}{3}] are doublet 
emission lines in the UV, there are a number of ways the 
observed lines can be combined to determine C$^{+2}$/O$^{+2}$.
Past observations with the IUE and HST/FOS have not resolved the \ion{C}{3}] 
$\lambda1907,1909$ lines, and so these studies used the total \ion{C}{3}] 
$\lambda1907 + \lambda1909$ intensity in their calculations.
While the \ion{O}{3}] $\lambda\lambda1661,1666$ lines can be resolved, 
the $\lambda1661$ line is often too weak to detect significantly.
Since the ratio of the two \ion{O}{3}] lines is physically fixed, nothing is lost 
in only using the stronger $\lambda1666$ line in determining the C/O ratio.
Thus, in order to build upon past studies, we choose to use the UV
\ion{C}{3}] $\lambda\lambda1907,1909$ / \ion{O}{3}] $\lambda1666$ 
line ratio in our C/O abundance determinations.
Further, we set a minimum criterion of 3$\sigma$ for a line to be 
considered significantly detected, although we also consider lesser 
detections in the literature for completeness.
Of studies to date, only 5 nearby galaxies meet our best sample criteria of having:
(1) UV CEL \ion{O}{3}] and \ion{C}{3}] detections and (2) an optical direct oxygen 
abundance, with all relevant lines detected at a strength of 3$\sigma$ or greater.
Four of the five targets come from G95: C1543+091, 
NGC~2363, SBS~0335-052, and SMC~N88A \citep[also reported by][]{kurt99}, 
with the addition of N81 from \citet{dufour82}.
Adding these to our sample creates a combined optimal sample of 12 objects.

If we relax the criteria to include less significant \ion{O}{3}] $\lambda1666$ 
observations, six additional targets can be considered: 30~Dor and T1214-277 
from G95, I~Zw~18 NW and I~Zw~18 SE from \citet{garnett97}, 
VS~44 from \citet{garnett99}, and Mrk~996 from \citet{thuan96}. 
For these 11 additional literature targets we compute the C and O abundances in a 
uniform manner following the method outlined in Sections~\ref{sec3} and ~\ref{sec4}.
References for the UV and optical emission line measurements are given in Table~\ref{tblA1}.
Emission line intensities and derived ionic and total abundances for the 
literature targets are provided in Tables~\ref{tblA2} $-$~\ref{tblA5}.

The optimal sample presented here is the largest sample to date 
of UV C/O abundances calculated in a uniform manner.
Note that C/O abundances determined from 
\ion{C}{3}] $\lambda\lambda1907,1909$/\ion{O}{3}]$\lambda1666$ and
\ion{C}{3}] $\lambda\lambda1907,1909$/[\ion{O}{3}]$\lambda5007$ 
can differ by up to 0.5 dex in the same source.
Since these differences are not systematic and a specific cause has not been identified, 
we have avoided using C/O determinations that use an optical and UV CEL line ratio.

\subsection{Relative C/O Abundances} \label{sec5.2}

The C/O abundances for our sample are plotted versus 
oxygen abundance in the left panel of Figure~\ref{fig6a}.
The 12 nearby dwarf galaxies that comprise our optimal sample are 
plotted as filled blue (this work) and purple (literature sources) circles.
For comparison, we have included RL C/O abundances as filled 
green squares (see Section~\ref{sec5.4}).

The dotted line in Figure~\ref{fig6a} is the least-squares fit 
to six dwarf galaxies using HST/FOS from G95 (their equation 5),
valid for $7.3 <$ 12+log(O/H) $< 8.4$.
Using this line as a guide, the 12 significant C/O detections generally 
agree with the trend of increasing C/O with O/H, with a few exceptions.
In fact, nine of the 12 galaxies are within 2$\sigma$ of the G95 relationship;
the exceptions are J082555, J104457, and J120122 from this work.
However, without assuming a correlation a priori, the 12 points in 
our sample in the metallicity range of 7.2 $\le$ 12+log(O/H) $\le$ 
8.2 in Figure~\ref{fig6a} demonstrate no trend in C/O with O/H.
The dashed line in Figure~\ref{fig6a} is the weighted mean of the significant
UV CEL C/O detections (log(C/O) $=-0.62$), which has a dispersion of $\sigma=0.25$ dex.
Visually, the dotted and dashed lines are comparable in their ability to fit the data in that
the large dispersion vitiates the presence of any potential real trend.
Thus, our uniform sample of C/O observations may represent one of two scenarios:
either our observations are consistent with the trend found by G95, albeit with outliers,
or there is significantly more variability in the C/O relationship than previously thought.

The most extreme outlier from the G95 relationship in Figure~\ref{fig6a} is J082555.
We further investigated this target with respect to the C/O relationship 
by remeasuring the dependent variable: 12+log(O/H).
For this purpose, we obtained follow up MMT optical spectra, as described in 
Appendix~\ref{A3}.  
At the low redshift of J082555, the [ion{O}{2}] $\lambda\lambda3727,3729$ lines 
are not included in the SDSS spectrum, which prompted further investigation.
We find that the SDSS spectrophotometry is consistent with our MMT observations,
confirming the low oxygen abundance of J082555.
At this oxygen abundance, J082555 appears to be significantly carbon enhanced 
relative to other metal-poor dwarf galaxies. 

In the proceeding sections we discuss the potential sources of this C/O distribution and 
compare to other measures of C/O, namely RL studies and stellar abundances. 

\subsection{Relative C/N Abundances} \label{sec5.3}
In Figure~\ref{fig6b} we plot the C/N ratio versus oxygen abundance
using the same symbol designation as Figure~\ref{fig6a}.
In this plot, the dashed line represents the weighted mean C/N of the 
12 targets with significant C/O detections.
Because there are no significant outliers from this flat relationship to the extent 
seen in Figure~\ref{fig6a}, our data are consistent with no trend in C/N with O/H.
This result contrasts the increasing relationship found by G95.

To reiterate, past studies of nebular C/O abundances have found an
increasing trend relative to O/H with significant dispersion. 
The updated observations presented here are generally consistent with 
an increasing relationship, with significant outliers, or could be revealing a plateau 
with large scatter \citep[similar to the N/O trend seen for dwarf galaxies,
e.g.,][]{garnett90,thuan95,vanzee98a,berg12}.
In early studies of nebular N/O, an increasing trend with O/H 
was found from observations of a handful of dwarf galaxies. 
As larger sets of N/O data were collected, a different story emerged
in which the N/O relationship was found to plateau for 12+log(O/H) $\lesssim 8$,
but with significant scatter.
The flat trend seen in Figure~\ref{fig6b} may indicate that carbon 
follows a similar pattern to nitrogen. 

\subsection{Comparison with Recombination Line Abundances} \label{sec5.4}

The brightest C and O recombination lines are \ion{C}{2} 
$\lambda4267$ and \ion{O}{2} $\lambda4650$ respectively.
While this combination of RLs is relatively insensitive to reddening 
corrections and electron temperature, optical RLs are inherently 
faint and therefore observationally challenging, especially at low metallicities.
In this regard, UV CELs in low-metallicity galaxies (12+log(O/H) $<$ 8.0) 
are complimentary to the high metallicity RL work.

\begin{figure*}
	\subfigure[]{\includegraphics[scale=0.45,angle=0,trim=0mm 0mm 0mm 0mm]{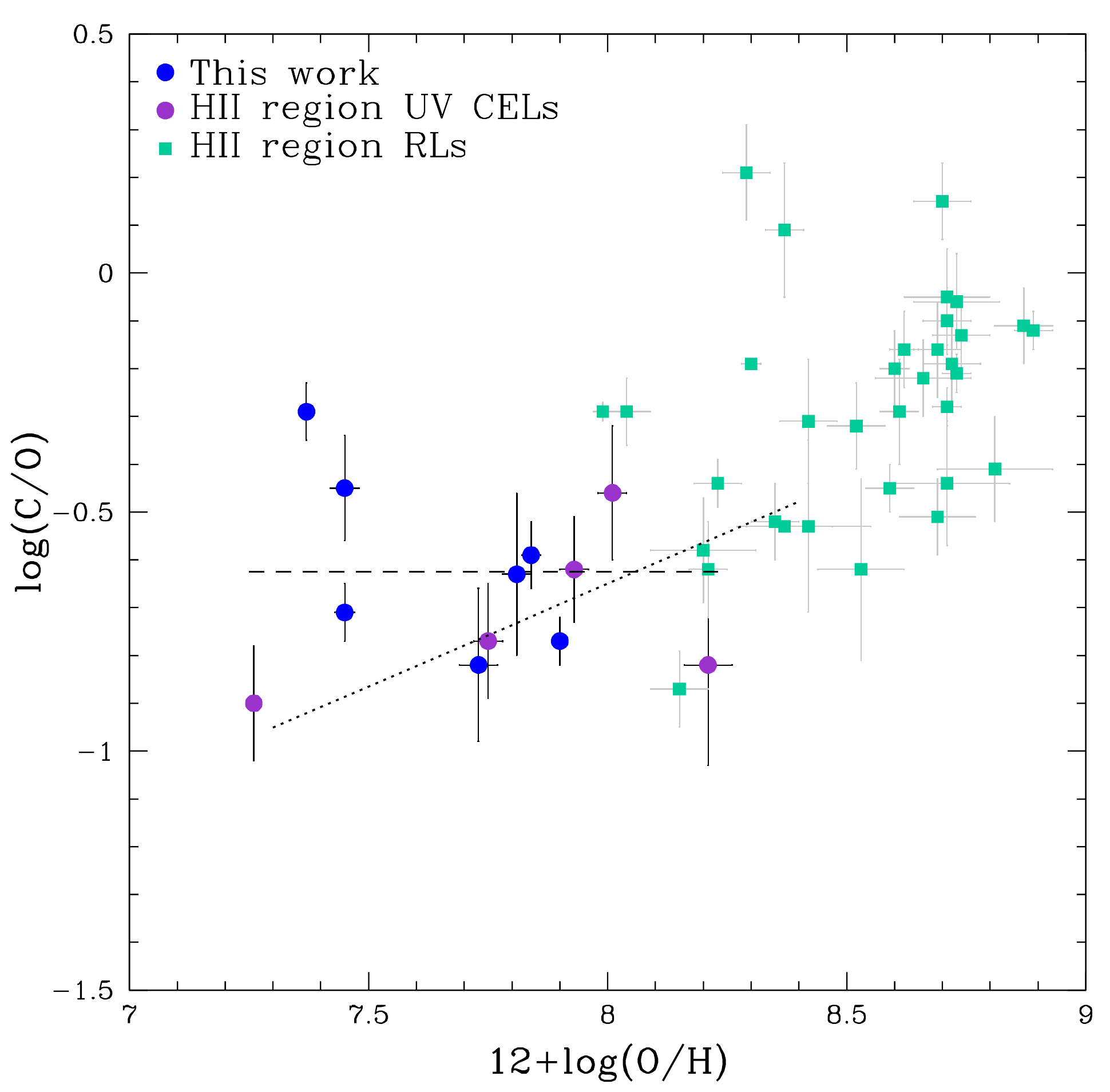}\label{fig6a}} 
        	\subfigure[]{\includegraphics[scale=0.45,angle=0,trim=0mm 0mm 0mm 0mm]{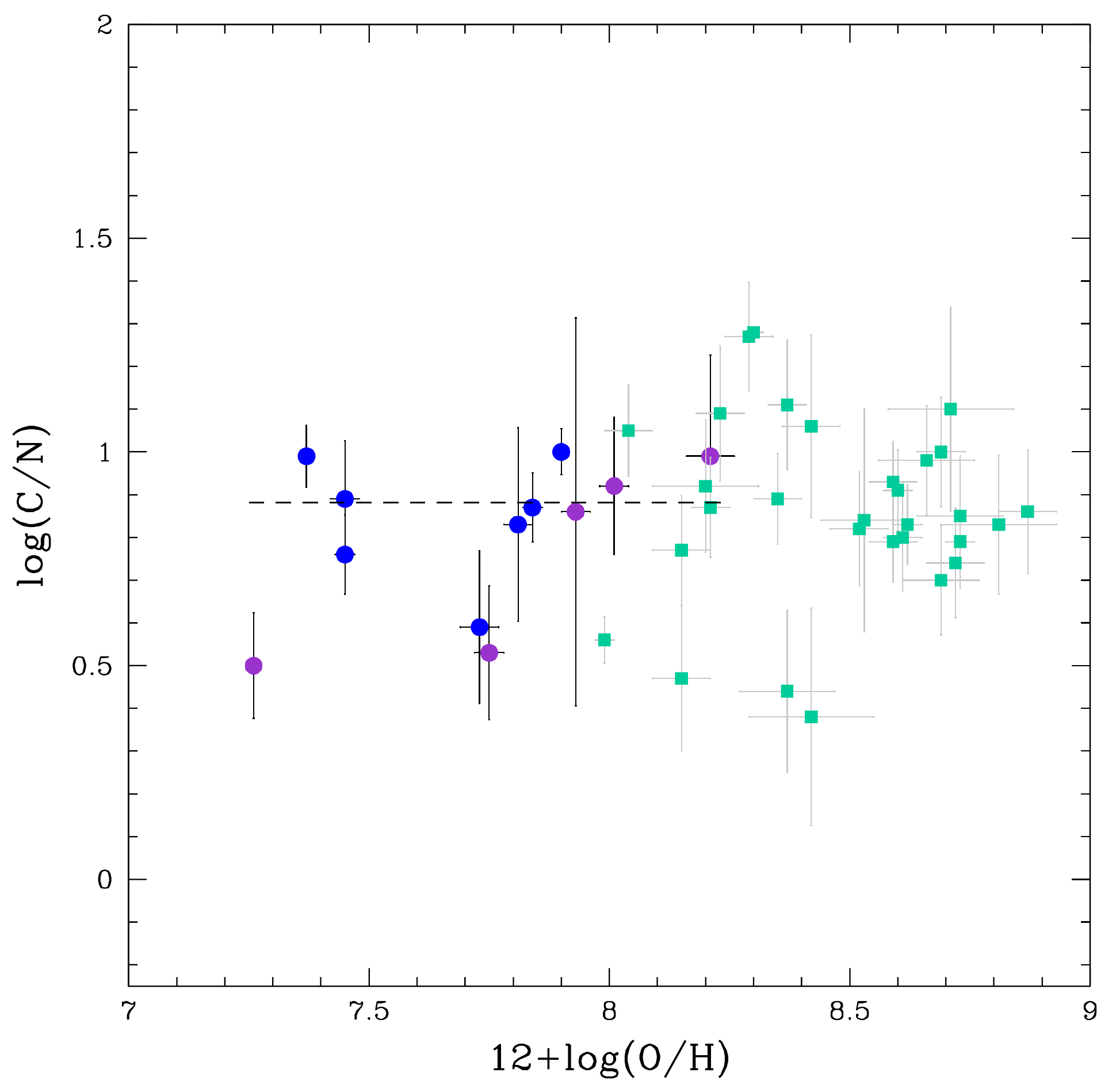}\label{fig6b}}
\caption{
(a) Carbon to oxygen ratio vs. oxygen abundance for star forming galaxies.
Our HST/COS observations are plotted as filled blue circles.
There are five additional targets in the literature that meet the criteria of 
having measured direct oxygen abundances and C/O abundances determined 
from UV CELs with strengths of 3 $\sigma$ or greater, which are plotted as purple circles.
Located at larger oxygen abundances, green filled squares represent star forming 
galaxies with recombination line abundance determinations 
\citep{esteban02,pilyugin05,garcia-rojas07,lopez-sanchez07,esteban09,esteban14}.
The dotted line is the least-squares fit from G95 and the dashed line is the 
weighted mean of the significant CEL C/O detections (filled purple and blue circles). 
(b) Carbon to nitrogen abundance vs. oxygen abundance.
While a significant amount of scatter is present ($\sim0.5$ dex), C/N appears 
to be relatively constant across oxygen abundance, suggesting carbon may
follow nitrogen in originating from primary (secondary) production 
at low (high) values of O/H. The dashed line marks the weighted mean of the 
significant CEL C/O detections.}
\end{figure*}

The C/O relationship has been studied for several nearby galaxies using RLs
\citep{esteban02,peimbert05,garcia-rojas07,lopez-sanchez07,esteban09,esteban14}.
In Figures~\ref{fig6a} and \ref{fig6b} we plot the results from these RL studies 
as filled green squares in comparison to our CEL sample, where the RL 
observations fill in the C/O relationship for 12+log(O/H) $> 8.0$.
\citet{esteban14} found C/O for six extragalactic \ion{H}{2} regions determined from
RLs and the $\lambda1909$/$\lambda5007$ UV-to-optical CEL ratio to be 
consistent within their uncertainties.
Taken together, CELs and RLs seem to follow a general increasing trend in Figure~\ref{fig6a}.
However, we caution against forming any conclusions about the trend of C/O 
at low oxygen abundance, as there are too few data points to do so. 

Note that for \ion{H}{2} regions, RL absolute abundances are found to be higher 
than those measured from CELs by up to 0.35 dex \citep[e.g.,][]{garcia-rojas07}; 
this difference is known as the ``Abundance Discrepancy Problem".
Because RLs and CELs are each useful over a different range in 
oxygen abundance, we find the benefits of using both RLs and CELs to interpret 
the C/O relationship outweighs this problem.  
Inspection of Figure~\ref{fig6a} shows that accounting for the discrepancy
by shifting the CEL abundances a few dex toward greater O/H values would not 
alter the appearance of an increasing C/O trend.  

Based on the data in Figure~\ref{fig6a}, we conclude that RL C/O abundances 
are more easily observed at higher metallicities (due to their linear dependence 
on abundance), but show a similar degree of dispersion to CEL abundances 
in the low metallicity regime.
Ideally, we would like to measure the C/O relationship over a broad range of 
oxygen abundance for both CELs and RLs independently. 
However, given the lack of RL detections in nearby spiral galaxies from 
$2-6$ hour observations on the Large Binocular Telescope as part of the 
Chemical Abundances of Spirals project \citep[CHAOS,][]{berg15},
this remains a challenging task.
Significantly expanding the sample of RL detections will require large integrations times 
on large telescopes to observe high-surface brightness \ion{H}{2} regions in nearby galaxies.


\section{SOURCES OF CARBON} \label{sec6}


Carbon is primarily produced by He burning through the triple-$\alpha$ process. 
However, this reaction can occur in both massive ($M > 8 M_{\odot}$) 
and low- to intermediate-mass ($1M_{\odot} < M < 8M_{\odot}$) stars.
Previously, the C/O ratio has been observed to increase with increasing 
O/H in both individual stars above 12+log(O/H) = 8.0 and in galaxies and \ion{H}{2} regions.
This trend has been studied by many authors, using a variety 
of chemical evolution models and stellar yields. 
Some find that the data are best explained by C arising almost 
exclusively from quasi-secondary production in 
massive stars, in which case the trend in C/O versus O/H is due to 
metallicity-dependent stellar winds, as mass loss and ISM enrichment 
are greater at higher metallicities \citep{maeder92,henry00}. 
Rotation may also play a significant role in carbon production. 
\citet{meynet03} showed that rotating stellar models predict enhanced 
yields of $^{12}$C in general, and predict the production of primary 
$^{13}$C at very low metallicity.

Some authors find that carbon production is dominated by primary production from 
low- to intermediate-mass stars ($1 M_{\odot} < M < 3 M_{\odot}$) and that the 
C/O versus O/H trend is therefore an evolutionary effect due to the delayed release 
of carbon relative to oxygen (which is produced almost exclusively by massive stars) 
in younger and less metal-rich systems \citep[e.g.,][]{chiappini03}. 
\citet{carigi05} find that massive and low- to intermediate-mass stars contribute 
roughly equal amounts of carbon in the solar vicinity. 
Conversely, \citet{akerman04} find that the metallicity-dependent mass loss of high 
mass stars is the main factor in producing the C/O versus O/H trend, with a small 
contribution from delayed C production in lower mass stars.

To examine whether carbon is produced in a similar manner to nitrogen,
we refer once again to the plot of C/N versus oxygen abundance in Figure~\ref{fig6b}.
The relatively constant values of C/N across the large range in 
oxygen abundance suggest that carbon and nitrogen production
are dominated by similar mechanisms.
For N/O, a plateau is observed at low oxygen abundance, presumably 
as a result of primary nitrogen production by intermediate mass stars,
whereas secondary nitrogen becomes prominent at higher metallicities 
(12+log(O/H) $>$ 8.3) resulting in an increasing N/O ratio with increasing 
oxygen abundance \citep{henry00}. 
If correct, the flat C/N correlation argues for metallicity-independent primary carbon 
enrichment of the interstellar medium for the low metallicity galaxies investigated here.
Quasi-secondary carbon production would then become prominent at higher metallicities,
but may still play a role at low metallicity, inducing the observed scatter.  

\subsection{Comparison with Stellar Abundances} \label{sec6.1} 
Another approach to understanding the gas phase carbon abundance is to
compare to the abundance pattern of stars, as stellar and nebular abundances 
are expected to share the same chemical composition once evolutionary 
effects in stars are accounted for \citep[e.g.,][]{bresolin09a,simon-diaz11}.
In Figure~\ref{fig7} we reproduce the same \ion{H}{2} region RL and CEL C/O 
abundances from Figure~\ref{fig6a}, but plotted relative to a larger range in O/H
to allow comparison to C/O abundances of Galactic disk stars \citep[4-pointed stars;][]{gustafsson99}
 and halo stars \citep[open triangles;][]{akerman04,fabbian09}, 
as well as metal-poor (12+log(O/H) $< 7.0$) damped Ly$\alpha$ absorbers 
\citep[DLAs; e.g.,][]{cooke11}. 
The sample of Galactic halo stars appear to lie along a plateau just below our sample
at low oxygen abundance and with smaller scatter than our extragalactic nebular abundances.
However, the elevated C relative to O seen in some metal-poor dwarf galaxies 
could be due to the effects of a significant population of stars with C-rich winds, 
such as Wolf-Rayet stars.

Large deviations from the C/O relationship have been measured
for carbon-enhanced metal-poor (CEMP) stars, which are iron poor stars that 
exhibit elevated carbon \citep[{[C/Fe] $\ge +1.0$ and [Fe/H] $\le -2.0$;}][]{beers05}. 
Interestingly, the frequency of CEMP-stars has been observed to increase
with declining [Fe/H] \citep[e.g.,][]{spite13}.
Of particular interest to our low-metallicity sample is the subclass of 
CEMP stars, CEMP-no stars, which are classified by their lack of strong 
neutron-capture-element enhancements and are typically found at lower 
metallicity than other subclasses \citep{aoki07}.
The chemical compositions of CEMP-no stars are well matched to the yields
of faint, primordial SNe \citep[e.g.,][]{nomoto13,marassi14,tominaga14}.
Under this assumption, first generation / Population III low-energy SNe experienced 
minimal mixing and large fallback such that the ejection of their outer layers enriched
the ISM with C-rich, Fe-poor material.
CEMP stars then formed as second generation, Fe-deficient stars from the enriched gas.
Alternatively, CEMP abundances have also been explained by the chemical contributions 
of massive metal-free ``spinstars", where stellar winds of rapidly-rotating mega metal-poor 
stars (MMP; [Fe/H] $< -6.0$) with partial mixing eject significant amounts of CNO 
\citep{hirschi06,meynet10,maeder15}. 

Under the assumption that primordial faint SNe dominated early metal enrichment,
\citet{salvadori15} investigated the frequency of CEMP stars in 10 Local Group dwarf galaxies
with similar masses, luminosities, and gas phase oxygen abundances as 
the targets in this study \citep[see properties of Local Group dwarf galaxies in][]{mateo98}.
They found that CEMP-no stars should be present in all dwarf galaxies,
but with a relative fraction that is dependent on both luminosity and metallicity.
According to this study, the presence of CEMP-no stars relative to the total number of stars
can be as significant as 50\% in the most metal-poor ([Fe/H] $< -4.75$), low-luminosity
($L < 10^4 L_{\odot}$) dwarfs down to 0.02\% in more metal-rich ([Fe/H] $> -3$),
luminous ( $L > 10^7 L_{\odot}$) dwarf galaxies. 
While the chemical evolution history of each galaxy is different, 
the processes which create CEMP-no stars likely play a varying, 
but important role in enriching a galaxy's ISM with carbon at early times.

\subsection{Dispersion in C/O} \label{sec6.2}

Compilations of previous nebular abundance studies have found a 
linearly increasing C/O vs O/H relationship with relatively small dispersion, 
but the simplicity of this relationship may be serendipitous.
By recalculating the literature sample in the uniform manner presented in 
Sections~\ref{sec3} and \ref{sec4}, we see the trend in C/O vs O/H 
(Figure~\ref{fig6a}) shows a weaker correlation than in the original studies.
Considering the entire optimal sample presented in Section~\ref{sec5.1}, 
an increasing C/O trend emerges that is consistent with G95, 
but significant scatter (or outliers) is (are) apparent.
\citet{garnett90} discuss the likelihood of large fluctuations in C/O at a fixed O/H
as a result of delayed C ejection from intermediate-mass stars of various 
starburst episodes. 
Based on the available data of the day, \citet{izotov99} found remarkably small dispersion
in the C/O ratio, claiming it as evidence against time-delayed production of C in 
the lowest-metallicity blue compact galaxies. 

We have already discussed two potential sources of variations in the C/O abundance
of dwarf galaxies: the relative fraction of first generation faint SNe dominating early 
metal-enrichment and the delayed ejection from intermediate-mass stars.
Since the star formation histories of dwarf galaxies tend to be bursty \citep[e.g.,][]{jlee09}, 
delayed C ejection from intermediate-mass stars of various starburst episodes could
result in large fluctuations in C/O at a fixed O/H.

Variations in the initial mass function (IMF) may contribute to 
the dispersion in carbon abundance at a given oxygen abundance.
If both massive and intermediate-mass stars are important for carbon 
production, then variations in the relative number of intermediate-mass to 
massive stars would also produce significant dispersion.
If star formation is triggered in a low-mass cloud, 
the high mass end of the IMF may be poorly populated \citep[e.g.,][]{fumagalli11}.
Therefore, the ISM enrichment of oxygen by massive stars in the low-mass, 
metal-poor galaxies studied here could be subject to statistical fluctuations in their IMFs.

\begin{figure*} 
\begin{center}
	\includegraphics[scale=0.55,angle=0,trim=0mm 0mm 0mm 0mm]{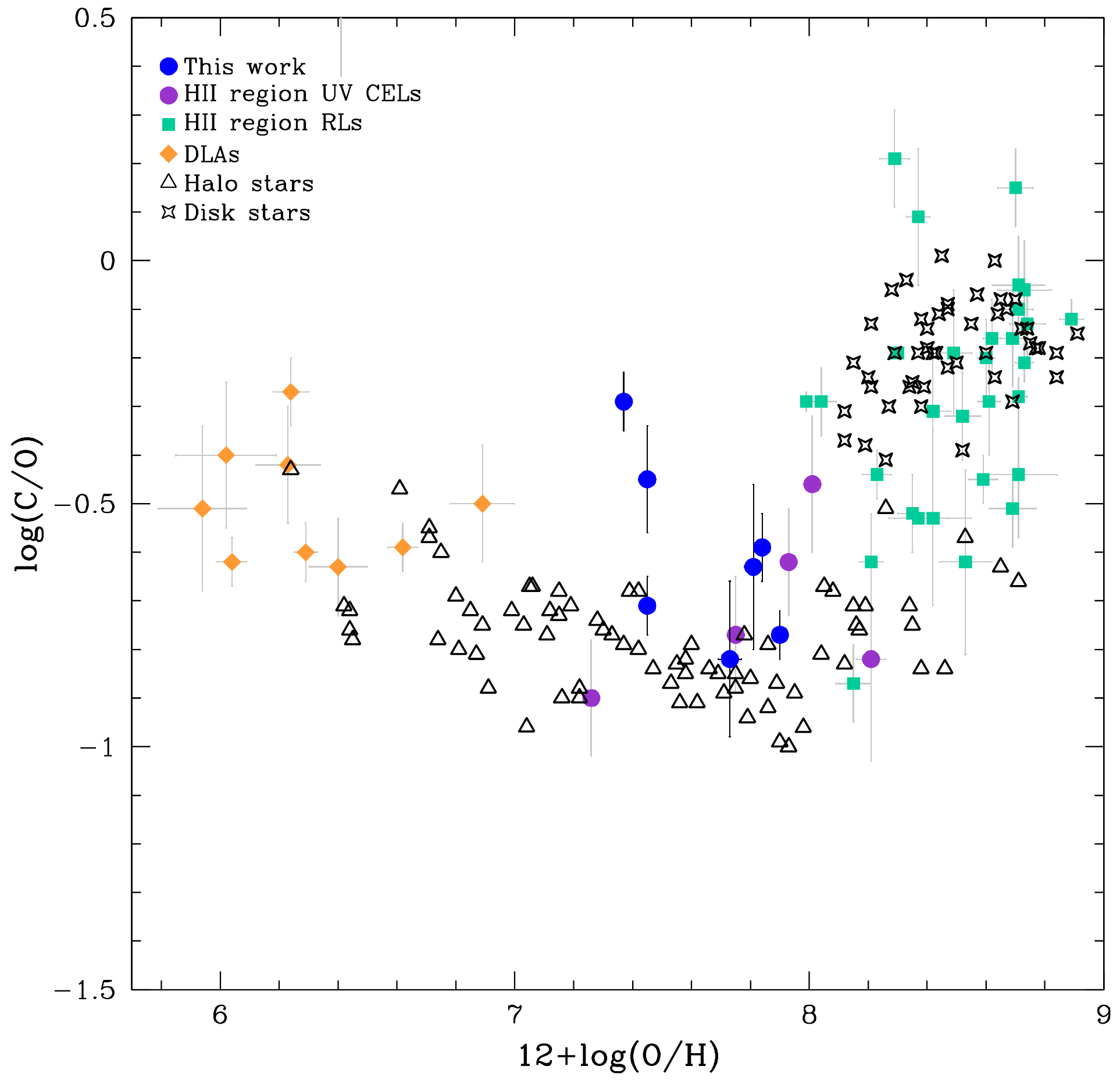}
\caption{Carbon vs. oxygen abundances for a variety of object types.
\ion{H}{2} region designations are the same as in Figure~\ref{fig6a}.
Milky Way stars from both galactic halo star \citep[triangles;][]{akerman04,spite05,fabbian09}
and disk star measurements \citep[4-pointed stars;][]{gustafsson99}. 
Damped Lyman alpha systems from \citet{cooke11} to extend coverage to lower oxygen
abundances \citep[orange diamonds;][]{cooke11}.}
\label{fig7}
\end{center}
\end{figure*}

A truncation of the upper IMF may also be responsible for the relatively high C/O 
ratios seen in the most metal-poor damped Ly$\alpha$ absorbers with oxygen 
abundances of 12+log(O/H) $< 7.0$ \citep[e.g.,][]{pettini08,cooke11,tsujimoto11}.
DLAs from \citet[][orange diamonds]{cooke11} are added to Figure~\ref{fig7}, 
where they follow the trend of Galactic halo stars of increasing C/O as O/H 
decreases \citep[e.g.,][]{fabbian09}.
\citet{mattsson10} suggest an evolving, top-heavy IMF during the early stages 
of Galactic evolution to explain the observed declining trend in C/O with 
increasing O/H in the solar neighborhood. 
Other authors argue for a trend in IMF slope with galaxy mass,
in the sense that low-mass galaxies form stars with flatter IMFs \citep{brown12,kalirai13},
while galaxies more massive than the Milky Way have steeper IMFs \citep{conroy12}.
Given the interest in possibilities of a non-universal IMF in low-mass metal-poor
galaxies, a better characterization of the behavior of C/O is vital.


\subsection{Chemical Evolution Models} \label{sec6.3}

We now present an analysis of our C/O results for the seven low 
metallicity dwarf galaxies over a broad metallicity range, where 
metallicity is assumed to be tracked by oxygen abundance. 
Figure~\ref{fig7} shows the behavior of log(C/O) versus 12+log(O/H) 
between roughly 6-9 for the latter parameter. 
Abundances were taken from studies of several object types, 
the symbols for which are indicated in the legend.
As the value of 12+log(O/H) increases from 6 to 9, the data indicate 
that log(C/O) starts at around $-$0.25 and becomes smaller, declining 
to a minimum value of roughly $-$0.9 at 12+log(O/H) = 7.8. 
This is followed by a steady increase up to a maximum of 0.0 near 12+log(O/H) = 8.9. 
Note that the positions of our seven low metallicity dwarf galaxies fall in 
the region in which the observed C/O profile passes through its lowest point. 

We point out that the clear trends in C/O over a broad metallicity range that are 
exhibited by the data in Figure~\ref{fig7} are produced by an assemblage of 
abundance measurements of distinctly different object types.
The chemical data from these different objects represent the abundance ratios of 
the interstellar medium at different evolutionary times, where high redshift DLAs are 
considered to be the progenitors of present day spiral galaxies.
Therefore, the presence of an aggregate pattern suggests the existence of a universal thread
governing chemical evolution, in spite of the fact that each data point 
in Figure~\ref{fig7} represents an object whose provenance is unique.
For example, processes related to star formation, as well as mass flows and
stellar feedback, are strongly influenced by local conditions within a host galaxy.
In turn, these processes can all vary with a galaxy's total mass and rotational
properties, as well as local surface densities.
Thus, a universal pattern in abundance trends must result from the details of stellar
nucleosynthesis which, for the most part, remain robust despite environmental conditions\footnotemark[12].
We have taken this as our working hypothesis.

\footnotetext[12]{One important exception to this conclusion may be differences in stellar 
rotation rates which some stellar evolution models suggest may have a noticeable effect on stellar
yields, in particular at low metallicity \citep{meynet06,hirschi07}.}

\begin{figure}
\begin{center}
	\includegraphics[scale=0.43,trim=0mm 0mm 0mm 0mm,clip]{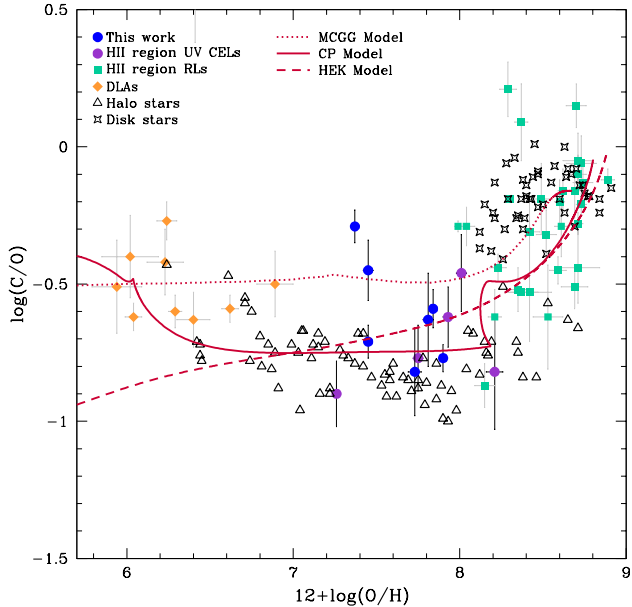}
\caption{Carbon vs. oxygen abundances from Figure~\ref{fig7} in comparison
to chemical evolution models from \citet[][dashed line]{henry00}, \citet[][dotted line]{molla15},
and \citet[][solid line]{carigi11}. 
The Carigi \& Peimbert model is the most successful at reproducing the general trend
seen for the collective log(C/O) data over the observed range in oxygen abundance. }
\label{fig8}
\end{center}
\end{figure}

In an effort to understand the observed profile shape of C/O with O/H
in the context of stellar nucleosynthesis, we compare three model chemical evolution tracks
in Figure~\ref{fig8} with the data presented in Figure~\ref{fig7}.
Each of the tracks represents a one-zone chemical evolution model for the solar vicinity. 
The three models are taken from papers by \citet[dashed line; HEK]{henry00}, 
\citet[solid line; CP]{carigi11}, and \citet[dotted line; MCGG]{molla15}. 
The instantaneous recycling approximation was relaxed in all three models, i.e., 
differences in stellar lifetimes were taken into account when computing the mass of 
each element released by stars as a function of time. 
Details of each model can be found in the respective references. 
Below we provide the salient traits of each for the purpose of this study. 

The HEK model included effects of infall (with a timescale of 4~Gyr) 
but not outflow and employed a metallicity-sensitive star formation 
efficiency along with the initial mass function of \citet{salpeter55}. 
The lower and upper limits on stellar mass were 1 and 120~M$_\odot$, respectively. 
In addition, the massive star ($9-120$~M$_\odot$) yields of \citet{maeder92} for carbon, 
nitrogen, and oxygen were used, while the yields of these same elements for the 
low and intermediate mass stars (LIMS) of $1-8$~M$_\odot$ were taken from \citet{vg97}. 
We note that HEK adjusted their massive star carbon yields slightly upward in 
order to force agreement with measured C/O levels when 12+log(O/H) $>$ 8.0.

The MCGG model employed the initial mass function by \citet{kroupa01}. 
Their massive star yields for CNO were taken from \citet{limongi03} and \citet{chieffi04}, 
while their LIMS yields for those elements were those published by \citet{gavilan05,gavilan06}. 
The lower and upper stellar mass limits were 0.8 and 100~M$_\odot$, respectively. 
Gas infall from the halo to the disk is assumed, with a collapse timescale of 
$\tau$=7.5~Gyr for a galactocentric distance of 8~kpc. 
The star formation process takes place in two steps, with molecular clouds forming first 
from the diffuse gas, followed by the creation of stars from cloud-cloud collisions. 
These processes are assumed to occur with efficiencies of 0.40 and 0.20, respectively. 
No outflows are assumed in this model.

The CP model considered that the solar vicinity ($r=8$ kpc) formed 
from primordial accretion in a double-infall scenario.
The halo was built with a timescale of $\tau = 0.5$ Gyr during the first Gyr and 
then the disk formed with  $\tau=8$~Gyr.  
The star formation rate was similar to the Kennicutt law \citep{kennicutt98}, and its efficiency was five 
times higher during the halo-forming phase than it is during the disk-forming phase.
The initial mass function of \citet{kroupa93} was applied in the $8 - 80~M_\odot$  
range for Pop~III, and in the $0.08-80~M_\odot$  range for all other populations. 
The metal-dependent yields for $M <  8 M_\odot$  were taken from \citet{marigo96,marigo98} 
and \citet{portinari98} and from \citet{hirschi07}, \citet{meynet02}, 
and \citet[high mass loss rate]{maeder92} for M $>$ 8~M$_\odot$. No outflow was included.

All three models are reasonably consistent with steady increases in log(C/O) for 
12+log(O/H) $\geq 8.1$. For $6.8 \leq$ 12+log(O/H) $\leq 8.1$ the tracks from MCGG 
and CP show little change in log(C/O) while HEK's curve rises slightly, all with increasing metallicity. 
As metallicity declines below 12+log(O/H) = 6.8 the curve of HEK falls slightly 
and that of MCGG remains flat. 
However, only CP's model track predicts the rise in log(C/O) seen for 
DLAs and halo stars for 12+log(O/H) $<$ 6.8. 

Clearly, the model by CP is the most successful at reproducing the details of the 
observed log(C/O) versus 12+log(O/H) behavior.
By considering yields of massive stars that depend on the metallicity {\it as well as rotation}, 
the CP model is able to predict the inverse relation between these two parameters below 
12+log(O/H) $\leq 6.8$, in addition to matching the trends for metallicities above this level.
In their model, metal-poor massive stars produce low C/O yields, 
but massive stars of high metallicity (and rotating stars with null metallicity) produce high C/O yields. 
In particular, the behavior below 6.8 is due to the early, rapid and high contribution 
to C/O from massive Population~III stars whose yield predictions by 
\citet{hirschi07} include the effects of stellar rotation. 
The plateau at higher metallicities up to 12+log(O/H) = 8.1 is due to a nearly constant 
ratio of carbon and oxygen yields from massive Population~II stars. 
At higher metallicities, the semi-loop is due to the primordial infall at the beginning of the 
disk formation and is followed by the delayed contribution of carbon by Population~II 
LIMS along with a high contribution to C/O from Population~I massive stars. 

While the CP model does an adequate job of matching what we might consider 
the observed average value of log(C/O) in the region between 7.4 and 8.0 in 12+log(O/H), 
where our seven low metallicity dwarf galaxies are located, it fails to reproduce 
the {\it contour} of the broad dip in log(C/O). 
As this region is outlined by Galactic halo stars and confirmed to be universal 
by the presence of both extragalactic \ion{H}{2} regions as well as our dwarf galaxies, 
the dip is most likely due to the behavior of stellar yields of carbon and oxygen in this 
metallicity range, a behavior that is thus far not predicted by 
the three yield sets that are tested by the models described here.

One should use caution in using such chemical evolution models to 
interpret individual objects with contrasting physical properties or directly 
compare the vastly different samples plotted in Figure~\ref{fig7}. 
It is likely that extant scatter around the trends are related to important differences 
in events or conditions connected with individual host galaxies.
Examples could include feedback, element-selective outflows, variations in the
initial mass function, and the relative fraction of first generation faint SNe
dominating early metal-enrichment.
For instance, many metal poor stars in dwarf galaxies show abundance ratios 
similar to the metal poor halo stars \citep{tolstoy09}, however, some dwarf galaxies
with inferred bursty star formation histories and truncated IMFs may require
outflows to reproduce their observed abundance ratios.
For this reason, we feel our model analysis based throughout upon the arbitrary
choice of conditions in the solar neighborhood is valid for testing the hypothesis 
that universal patterns of stellar nucleosynthesis serve as the basis for the existence
of the trends in C/O over a large metallicity range. 

In summary, above 12+log(O/H) = 8.1, C production is apparently governed by a 
secondary effect in which higher metallicity is responsible for increased C yields
from massive stars.
This effect is possibly mediated by an increase in the mass of dust in stellar
atmospheres as metallicity increases, resulting in greater wind-driven mass loss.


\section{SUMMARY AND CONCLUSIONS} \label{sec7}

To improve our understanding of carbon production,
we have obtained carbon and oxygen abundances of 7 nearby, 
low-metallicity, high-ionization dwarf galaxies using COS on HST.
In order to minimize sources of discrepancies and uncertainty,
we define our sample with the following criteria:
\begin{enumerate}
\item C/O abundance is calculated from the UV \ion{C}{3}] 
\W\W1907,1909 and \ion{O}{3}] \W1666 lines such that a single 
observational setup is used and the reddening law is relatively flat.
\item The nebular physical conditions - $T_e$, $n_e$, O/H - are 
determined using the direct method from complimentary optical observations.
\item All lines are detected at a strength of 3$\sigma$ or greater 
to reduce spurious detections.
\end{enumerate}
Our C/O sample was expanded to include sources from the literature 
that met the outlined criteria, for which we recalculated the C/O and 
O/H abundances to create a uniform sample of 12 objects.

Prior to the present study, the cumulative database of nebular C/O 
abundances of nearby galaxies points to an increasing relationship 
between C/O versus O/H abundance \citep[e.g.,][]{garnett95}.
This relationship is based upon samples of few observations and a variety 
of abundance determination methods, yet is what competing models of low- 
to intermediate-mass and massive stellar carbon production are based on.
With the addition of our new observations at low oxygen abundance,
we find that the simple trend of C/O increasing with O/H is no longer clear due 
to few measurements with significant scatter.
In fact, the current sample, spanning 7.2 $\le$ 12+log(O/H) $\le$ 8.2,
demonstrates no trend in C/O at low metallicity. 
Further observations are needed to determine the trend and dispersion
of C/O in this metallicity regime.

We examined the C/N ratio for our sample and found it to be constant (but with 
significant scatter) over a large range in oxygen abundance, indicating carbon is 
predominantly produced by the same nucleosynthetic mechanisms as nitrogen.
In this scenario, primary production of carbon results in a flat trend in C/O 
at low metallicity and transitions to an increasing relationship with
quasi-secondary production at higher metallicities. 
If correct, this result, in addition to the flattening predicted by chemical evolution models,
would support a constant C/O relationship at low nebular metallicities (12+log(O/H) 
$\le$ 8.0), in contrast to previous studies. 

Potential scenarios to explain the scatter in nebular C/O at low metallicity include
contributions from both low- to intermediate-mass and high-mass stars,
various levels of carbon in a galaxy's primordial chemistry due to low-energy
SNe that produced CEMP-no stars, and under populating the high mass end of the IMF. 
These theories are inherently speculative because they are based on a small sample, 
and thus larger samples of UV C/O observations are needed.
We have established the first UV C/O analysis of local galaxies with COS on HST,
and argue that, given the currently available instruments/telescopes, 
it is the best method to adopt.


\acknowledgements

DAB and DKE are supported by the US National Science Foundation 
through the Faculty Early Career Development (CAREER) Program, grant AST-1255591.
This work was supported by NASA through grant GO-13312 
from the Space Telescope Institute, which is operated by
Aura, Inc., under NASA contract NAS5-26555. 
This paper also used observations obtained at the MMT Observatory, 
a joint facility of the Smithsonian Institution and the University of Arizona. 
MMT observations were obtained as part of the University of Minnesota's 
guaranteed time on Steward Observatory facilities through membership in the 
Research Corporation and its support for the Large Binocular Telescope. 

Funding for the SDSS and SDSS-II has been provided by the Alfred P. Sloan Foundation, 
the Participating Institutions, the National Science Foundation, the U.S. Department of Energy, 
the National Aeronautics and Space Administration, the Japanese Monbukagakusho, 
the Max Planck Society, and the Higher Education Funding Council for England. 
The SDSS Web Site is http://www.sdss.org/.

The SDSS is managed by the Astrophysical Research Consortium for the Participating Institutions. 
The Participating Institutions are the American Museum of Natural History, Astrophysical Institute Potsdam, 
University of Basel, University of Cambridge, Case Western Reserve University, University of Chicago, 
Drexel University, Fermilab, the Institute for Advanced Study, the Japan Participation Group, 
Johns Hopkins University, the Joint Institute for Nuclear Astrophysics, 
the Kavli Institute for Particle Astrophysics and Cosmology, the Korean Scientist Group, 
the Chinese Academy of Sciences (LAMOST), Los Alamos National Laboratory, 
the Max-Planck-Institute for Astronomy (MPIA), the Max-Planck-Institute for Astrophysics (MPA), 
New Mexico State University, Ohio State University, University of Pittsburgh, University of Portsmouth, 
Princeton University, the United States Naval Observatory, and the University of Washington.

This research has made use of NASA's Astrophysics Data System Bibliographic Services and the 
NASA/IPAC Extragalactic Database (NED), which is operated by the Jet Propulsion Laboratory, 
California Institute of Technology, under contract with the National Aeronautics and Space Administration.


\appendix


\section{SUPPLEMENTAL OBSERVATIONS}

In Tables~\ref{tblA1}$-$\ref{tblA5} we present references for the supplementary UV 
and optical data used in this work, the emission line intensities adopted from these 
sources, and the subsequently recalculated ionic and total abundances.

\begin{deluxetable}{lccc}[H]
\tabletypesize{\scriptsize}
\tablewidth{0pt}
\setlength{\tabcolsep}{3pt}
\tablecaption{Literature UV and Optical Emission Line Flux Sources}
\tablewidth{0pt}
\tablehead{}
\startdata
{} 				& {}					& \multicolumn{2}{c}{Literature Source}		\\
\cline{3-4}								
{Galaxy} 			& {3$\sigma$ Sample?}	& {UV}      				& {Optical}            	\\
\hline 
{N81}			& \checkmark			& \citet{dufour82}		& \citet{dufour82}	\\
{C1543+091}		& \checkmark			& \citet{garnett95}		& \citet{campbell86}	\\
{NGC~2363}		& \checkmark			& \citet{garnett95}		& \citet{peimbert86}	\\
{SBS~0335-052E}	& \checkmark			& \citet{garnett95}		& \citet{izotov09} 	\\
{30~Dor}			& 					& \citet{garnett95}		& \citet{peimbert03}	\\
{T1214-277}		&					& \citet{garnett95}		& \citet{guseva11}	\\
{Mrk~996}			&					& \citet{thuan96}		& \citet{thuan96}	\\
{I~Zw~18 NW}		&					& \citet{garnett97}		& \citet{skillman93}	\\
{I~Zw~18 SE}		&					& \citet{garnett97}		& \citet{skillman93}	\\
{VS~44}			&					& \citet{garnett99}		& \citet{berg13}		\\
{SMC~N88A\_barr}	& \checkmark			& \citet{kurt99}			& \citet{kurt99}		\\
\enddata
\label{tblA1}
\end{deluxetable}


\begin{deluxetable}{lccccc}[h]
\tabletypesize{\scriptsize}
\tablewidth{0pt}
\setlength{\tabcolsep}{3pt}
\tablecaption{ Emission-Line Intensities from Literature Sources with C/O Detections}
\tablewidth{0pt}
\tablehead{
\CH{} & \multicolumn{5}{c}{$I(\lambda)/I(\mbox{H}\beta)$} }
\startdata
{}                       			
{Ion}                       			& {N81}      		& {C1543+091}      	& {NGC~2363}      	& {SBS0335-052E}   	& {SMC N88A\_bar} 	\\
\hline
{O~\iii] $\lambda$1666.15}  & 0.114$\pm$0.036$^a$ & 0.060$\pm$0.013	& 0.010$\pm$0.002	& 0.145$\pm$0.044	& 0.152$\pm$0.037	\\
{C~\iii] $\lambda$1908.73}$^a$	& 0.302$\pm$0.045 	& 0.175$\pm$0.008	& 0.045$\pm$0.001	& 0.261$\pm$0.022	& 0.904$\pm$0.075	\\
\hline
{[O~\ii]~$\lambda$3727}   		& 1.294$\pm$0.183	& 0.580$\pm$0.021	& 0.581$\pm$0.023	& 0.235$\pm$0.006	& 0.355$\pm$0.027	\\
{[Ne~\iii]~$\lambda$3868}		& 0.384$\pm$0.054	& 0.403$\pm$0.028	& 0.555$\pm$0.023	& 0.412$\pm$0.008	& 0.679$\pm$0.052	\\
{H$\delta$ $\lambda$4101} 	& 0.282$\pm$0.040	& 0.265$\pm$0.016	& 0.247$\pm$0.021	& 0.260$\pm$0.006	& 0.294$\pm$0.023	\\ 	   		
{H$\gamma$ $\lambda$4340}	& 0.462$\pm$0.065	& 0.496$\pm$0.027	& 0.470$\pm$0.022	& 0.474$\pm$0.010	& 0.479$\pm$0.037	\\
{[O~\iii]~$\lambda$4363}  		& 0.064$\pm$0.009	& 0.131$\pm$0.009	& 0.136$\pm$0.010	& 0.112$\pm$0.003  & 0.139$\pm$0.011	\\	 	
{H$\beta$ $\lambda$4861}  	& 1.000$\pm$0.141	& 1.000$\pm$0.040	& 1.000$\pm$0.028	& 1.000$\pm$0.021	& 1.000$\pm$0.078	\\	
{[O~\iii]~$\lambda$4958}  		& 1.768$\pm$0.250	& 1.932$\pm$0.078	& 2.342$\pm$0.062	& 1.095$\pm$0.022	& 2.623$\pm$0.201	\\
{[O~\iii]~$\lambda$5006}  		& 5.243$\pm$0.742	& 5.635$\pm$0.227	& 7.070$\pm$0.199	& 3.152$\pm$0.065	& 7.633$\pm$0.588	\\		
{[N~\ii]~$\lambda$6548}   		& \nodata			& 0.013$\pm$0.008	& 0.009$\pm$0.040	& \nodata			& 0.015$\pm$0.005	\\
{H$\alpha$ $\lambda$6562} 	& 2.832$\pm$0.400	& 2.779$\pm$0.093	& 2.790$\pm$0.075	& 2.738$\pm$0.059	& 2.805$\pm$0.217	\\
{[N~\ii]~$\lambda$6583}   		& 0.062$\pm$0.009	& 0.032$\pm$0.008	& 0.023$\pm$0.040	& 0.009$\pm$0.001	& 0.036$\pm$0.006	\\
{[S~\ii]~$\lambda$6716}   		& 0.031$\pm$0.004	& 0.049$\pm$0.006	& 0.044$\pm$0.040	& 0.017$\pm$0.001	& 0.012$\pm$0.002	\\
{[S~\ii]~$\lambda$6730}   		& 0.059$\pm$0.008	& 0.036$\pm$0.005	& 0.032$\pm$0.049	& 0.017$\pm$0.001	& 0.023$\pm$0.003	\\
\enddata
\tablecomments{Emission-line intensities for galaxies supplemented from the 
literature with $3\sigma$ \ion{O}{3}] and \ion{C}{3}] detections.
The first column lists the vacuum wavelengths of the observed ions for wavelengths 
of $\lambda < 2000$ \AA, and approximate air wavelengths for optical emission lines.
The flux values for each object listed are reddening corrected intensity ratios relative to H$\beta$. \\
$^a$ Blended line measurement. }
\label{tblA2}
\end{deluxetable}

\begin{deluxetable}{lcccccc}[h]
\tabletypesize{\scriptsize}
\tablewidth{0pt}
\setlength{\tabcolsep}{3pt}
\tablecaption{Emission-Line Intensities from Literature Sources with C/O Non-Detections}
\tablewidth{0pt}
\tablehead{
\CH{} & \multicolumn{6}{c}{$I(\lambda)/I(\mbox{H}\beta)$} }
\startdata
{}                       			
{Ion}                       			& {30~Dor}      		& {T1214-277}      	& {Mrk~996}   		& {I~Zw~18~NW} 	& {I~Zw~18~SE}      	& {VS~44}			\\
\hline
{O~\iii] $\lambda$1666.15}	& 0.111$\pm$0.087	& 0.007$\pm$0.005	& 0.401$\pm$0.087	& 0.270$\pm$0.135	& 0.290$\pm$0.145 	& 0.017$\pm$0.014 	\\
{C~\iii] $\lambda$1908.73}$^a$	& 1.309$\pm$0.045	& 0.017$\pm$0.002	& 4.734$\pm$1.980	& 0.550$\pm$0.091	& 0.470$\pm$0.071 	& 0.057$\pm$0.007 	\\
\hline
{[O~\ii]~$\lambda$3727}   		& 1.526$\pm$0.030	& 0.308$\pm$0.007	& 1.029$\pm$0.058	& 0.264$\pm$0.013	& 0.466$\pm$0.026	& 2.470$\pm$0.112	\\
{[Ne~\iii]~$\lambda$3868}		& 0.415$\pm$0.011	& 0.413$\pm$0.009	& 0.751$\pm$0.048	& 0.153$\pm$0.008	& 0.154$\pm$0.011	& 0.090$\pm$0.004	\\
{H$\delta$ $\lambda$4101} 	& 0.307$\pm$0.007	& 0.294$\pm$0.006	& 0.319$\pm$0.029	& 0.269$\pm$0.013	& 0.257$\pm$0.015	& 0.260$\pm$0.011	\\ 	   		
{H$\gamma$ $\lambda$4340}	& 0.529$\pm$0.012	& 0.509$\pm$0.010	& 0.530$\pm$0.036	& 0.469$\pm$0.020	& 0.435$\pm$0.020	& 0.480$\pm$0.014	\\
{[O~\iii]~$\lambda$4363}  		& 0.035$\pm$0.001	& 0.173$\pm$0.004	& 0.220$\pm$0.025  & 0.062$\pm$0.004	& 0.044$\pm$0.007	& 0.008$\pm$0.001	\\	 	
{H$\beta$ $\lambda$4861}  	& 1.000$\pm$0.028	& 1.000$\pm$0.020	& 1.000$\pm$0.044	& 1.000$\pm$0.040	& 1.000$\pm$0.040	& 1.000$\pm$0.028	\\	
{[O~\iii]~$\lambda$4958}  		& 1.678$\pm$0.039	& 1.719$\pm$0.035	& 0.925$\pm$0.040	& 0.648$\pm$0.026	& 0.604$\pm$0.025	& 0.650$\pm$0.016	\\
{[O~\iii]~$\lambda$5006}  		& 5.003$\pm$0.131	& 5.118$\pm$0.103	& 2.706$\pm$0.107	& 1.950$\pm$0.077	& 1.717$\pm$0.070	& 1.960$\pm$0.064	\\		
{[N~\ii]~$\lambda$6548}   		& 0.038$\pm$0.001	& 0.003$\pm$0.001	& \nodata			& 0.045$\pm$0.001	& 0.055$\pm$0.002	& 0.124$\pm$0.006	\\
{H$\alpha$ $\lambda$6562} 	& 2.859$\pm$0.058	& 2.750$\pm$0.058	& 2.701$\pm$0.106	& 2.760$\pm$0.407	& 2.760$\pm$0.089	& 2.910$\pm$0.133	\\
{[N~\ii]~$\lambda$6583}   		& 0.115$\pm$0.002	& 0.008$\pm$0.001	& 0.153$\pm$0.011	& 0.008$\pm$0.001	& 0.013$\pm$0.001	& 0.300$\pm$0.021	\\
{[S~\ii]~$\lambda$6716}   		& 0.071$\pm$0.001	& 0.022$\pm$0.001	& 0.052$\pm$0.005	& 0.024$\pm$0.001	& 0.041$\pm$0.001	& 0.210$\pm$0.011	\\
{[S~\ii]~$\lambda$6730}   		& 0.066$\pm$0.001	& 0.014$\pm$0.001	& 0.051$\pm$0.005	& 0.021$\pm$0.001	& 0.029$\pm$0.001	& 0.160$\pm$0.010	\\
\enddata
\tablecomments{ Same as Table~\ref{tblA2} but for galaxies supplemented from the 
literature with $< 3\sigma$ \ion{O}{3}] and \ion{C}{3}] detections. \\
$^a$ Blended line measurement. }
\label{tblA3}
\end{deluxetable}


\begin{deluxetable}{lccccc}[h]
\tabletypesize{\scriptsize}
\tablewidth{0pt} 
\setlength{\tabcolsep}{3pt}
\tablecaption{ Ionic and Total Abundance for Literature Sources with C/O Detections}
\tablewidth{0pt}
\tablehead{
\CH{Target} 	        			& \CH{N81}      		& \CH{C1543+091}  	& \CH{NGC~2363}   	& \CH{SBS0335-052E} & \CH{SMC N88A\_bar}}
\startdata
\multicolumn{6}{c}{Properties Derived from Optical Spectra} \\
\hline \\
T$_e$ [O~\iii] (K)			& 12,200$\pm$800	& 16,300$\pm$600	& 14,900$\pm$500	& 20,600$\pm$400	& 14,300$\pm$600	\\
T$_e$ [N~\ii] (K)$^a$		& 11,500$\pm$600	& 14,400$\pm$400	& 13,400$\pm$400	& 17,400$\pm$300	& 13,000$\pm$400	\\		
n$_e$ [S~\ii] (cm$^{-3})$ 		& 9290			& 80				& 70				& 530			& 8030			\\
\\
O$^+$/H$^+$ (10$^{5})$		& 6.25$\pm$1.29	& 0.55$\pm$0.04	& 0.69$\pm$0.05	& 0.14$\pm$0.01	& 1.02$\pm$0.12	\\
O$^{+2}$/H$^+$ (10$^{5})$	& 9.92$\pm$1.57	& 5.03$\pm$0.34	& 7.74$\pm$0.56	& 1.68$\pm$0.05	& 9.30$\pm$0.78	\\
12 + log(O/H)				& 8.21$\pm$0.05	& 7.75$\pm$0.03	& 7.93$\pm$0.03	& 7.26$\pm$0.01	& 8.01$\pm$0.03	\\
\\
N$^{+}$/H$^+$ (10$^{7})$		& 9.69$\pm$1.80	& 2.80$\pm$0.72	& 2.29$\pm$3.99	& 0.55$\pm$0.03	& 4.25$\pm$0.75	\\
log(N/O)					& -1.81$\pm$0.11	& -1.30$\pm$0.10	& -1.48$\pm$0.44	& -1.40$\pm$0.03	& -1.38$\pm$0.08	\\
\hline \\
\multicolumn{6}{c}{Properties Derived from UV Spectra} \\
\hline \\
C$^{+2}$/H$^+$ (10$^{5})$	& 1.83$\pm$0.08	& 0.20$\pm$0.01	& 0.08$\pm$0.01	& 0.10$\pm$0.01	& 2.07$\pm$0.15	\\
O$^{+2}$/H$^+$ (10$^{5})$	& 15.8$\pm$0.6	& 1.25$\pm$0.02	& 0.36$\pm$0.01	& 0.88$\pm$0.04	& 4.86$\pm$0.18	\\
C$^{+2}$/O$^{+2}$			& 0.12$\pm$0.07	& 0.16$\pm$0.05	& 0.22$\pm$0.06	& 0.12$\pm$0.04	& 0.43$\pm$0.14	\\
log U						& -2.50			& -2.16			& -2.07			& -2.03			& -1.83			\\
ICF						& 0.933			& 1.049			& 1.074			& 1.085			& 1.141			\\
log(C/O)					& -0.82$\pm$0.21	& -0.77$\pm$0.12	& -0.62$\pm$0.11	& -0.90$\pm$0.12	& -0.46$\pm$0.14	\\
\enddata
\tablecomments{Ionic and total abundance calculations for targets listed in Table~\ref{tblA2}.
All conventions are the same as in Table~\ref{tbl3}, except O$^{+}$/H$^+$ ionic abundances were 
determined from the [\ion{O}{2}] $\lambda3727$ line.}
\label{tblA4}
\end{deluxetable}

\begin{deluxetable}{lcccccc}[h]
\tabletypesize{\scriptsize}
\tablewidth{0pt} 
\setlength{\tabcolsep}{3pt}
\tablecaption{ Ionic and Total Abundance for Literature Sources with C/O Non-Detections}
\tablewidth{0pt}
\tablehead{
\CH{Target} 	        			& \CH{30~Dor}      	& \CH{T1214-277} 	& \CH{Mrk~996} 	& \CH{I~Zw~18~NW} & \CH{I~Zw~18~SE} 	& \CH{VS~44} }
\startdata
\multicolumn{7}{c}{Properties Derived from Optical Spectra} \\
\hline \\
T$_e$ [O~\iii] (K)			& 10,200$\pm$100	& 20,100$\pm$300	& 25,000$\pm1000^a$ &19,400$\pm$900 & 17,000$\pm$1500		& 8,700$\pm$200	\\
T$_e$ [N~\ii] (K)$^a$		& 10,200$\pm$200	& 17,000$\pm$200	& 20,500$\pm$200	& 16,600$\pm$600	& 14,900$\pm$1000		& 9,100$\pm$200	\\		
n$_e$ [S~\ii] (cm$^{-3})$ 		& 500			& 100			& 620			& 330			& 50					& 130			\\
\\
O$^+$/H$^+$ (10$^{5})$		& 5.72$\pm$0.36	& 0.18$\pm$0.01	& 0.41$\pm$0.02	& 0.17$\pm$0.02	& 0.39$\pm$0.06		& 14.3$\pm$1.2	\\
O$^{+2}$/H$^+$ (10$^{5})$	& 16.6$\pm$0.4	& 2.87$\pm$0.07	& 1.01$\pm$0.05	& 1.17$\pm$0.08	& 1.39$\pm$0.21		& 11.9$\pm$0.7		\\
12 + log(O/H)				& 8.35$\pm$0.01	& 7.48$\pm$0.01	& 7.15$\pm$0.02	& 7.13$\pm$0.03	& 7.25$\pm$0.05		& 8.42$\pm$0.02	\\
\\
N$^{+}$/H$^+$ (10$^{7})$		& 22.3$\pm$1.2	& 0.54$\pm$0.02	& 7.16$\pm$0.63	& 0.55$\pm$0.05	& 1.08$\pm$0.15		& 79.0$\pm$7.4	\\
log(N/O)					& -1.41$\pm$0.03	& -1.52$\pm$0.02	& -0.76$\pm$0.04	& -1.50$\pm$0.05	& -1.56$\pm$0.08		& -1.26$\pm$0.05	\\
\hline \\
\multicolumn{7}{c}{Properties Derived from UV Spectra} \\
\hline \\
C$^{+2}$/H$^+$ (10$^{5})$	& 28.0$\pm$1.3	& 0.007$\pm$0.001	& 0.90$\pm$1.79	& 0.28$\pm$0.03	& 0.43$\pm$0.03		& 4.82$\pm$0.03	\\
O$^{+2}$/H$^+$ (10$^{5})$	& 68.0$\pm$5.9	& 0.045$\pm$0.001	& 1.07$\pm$0.09	& 2.21$\pm$0.30	& 4.65$\pm$0.67		& 51.6$\pm$0.7	\\
C$^{+2}$/O$^{+2}$			& 0.41$\pm$0.32	& 0.16$\pm$0.12	& 0.85$\pm$0.42	& 0.13$\pm$0.07	& 0.93$\pm$0.66		& 0.09$\pm$0.08	\\
log U						& -2.57			& -1.94			& -2.65			& -2.27			& -2.53				& -3.04			\\
ICF						& 0.899			& 1.109			& 0.861			& 1.017			& 0.918				& 0.620			\\
log(C/O)					& -0.43$\pm$0.25	& -0.74$\pm$0.25	& -0.14$\pm$0.18	& -0.89$\pm$0.20	& -1.07$\pm$0.23		& -1.24$\pm$0.27	\\
\enddata
\tablecomments{Ionic and total abundance calculations for targets listed in Table~\ref{tblA3}.
All conventions are the same as in Table~\ref{tbl3}, except O$^{+}$/H$^+$ ionic abundances were 
determined from the [\ion{O}{2}] $\lambda3727$ line. \\
$^a$ 25,000 K is the upper limit in the IDL routine IM\_TEMDEN.PRO, the program used to calculate $T_e$.}
\label{tblA5}
\end{deluxetable}


\section{FOLLOW-UP OPTICAL MMT SPECTRUM FOR J082555}
Our C/O sample was chosen from the SDSS based on properties of their optical spectra.
One of the highest ionization targets from our sample, J082555 (log([\ion{O}{3}]/[\ion{O}{2}]) = 1.11),
is also the most significant outlier from the trend of C/O with O/H in Figures~\ref{fig6a}-\ref{fig8}.
This makes the physical properties of J082555 particularly interesting, and so we want to be
confident in the spectrophotometry of the optical spectrum.
Additionally, due to the limited blue wavelength coverage of the SDSS spectrograph, many
targets in our sample, including J08222, did not have [\ion{O}{2}] \W3727 line measurements.
Instead, the red [\ion{O}{2}] \W\W7320,7330 doublet was used to determine the 
O$^{+}$/H$^{+}$ abundance, despite the fact that it is complicated by lying on top of stellar absorption 
features as well as lying just inside the blue-end of strong OH Meinel band emission.

For these reasons we observed a follow-up optical spectrum of J082555 using the Blue Channel 
spectrograph on the MMT 6.5-m telescope on 11 and 12 November 2015 in order to measure 
the blue [\ion{O}{2}] $\lambda3727$ emission line and independently confirm the SDSS flux calibration. 
Three 1200-second exposures were taken for each target with a 1.\arcsec5 wide long slit 
oriented at a position angle equal to the parallactic angle at the midpoint of the observations
to minimize differential light loss. 
The spectra were reduced with the standard techniques outlined in \citet{berg12}. 
Emission line flux measurements and chemical abundance calculations were performed 
in the manner described in Sections~\ref{sec3.3} and \ref{sec4} respectively. 

The resulting de-reddened emission line intensities are given in Table~\ref{tblA6},
followed by the ionic and total abundances in Table~\ref{tblA7}.
The corresponding SDSS values are tabulated alongside the MMT data for comparison.
From Table~\ref{tblA6} we conclude that the spectrophotometry of the SDSS and MMT
spectra of J082555 are generally well matched. 
However, there is some difference in the $T_e$ in Table~\ref{tblA7}, 
resulting in a 0.06 dex in log(O/H).
Because the C/O abundance is determined from the UV spectrum alone (i.e., 
it is independent of the oxygen abundance determined from the optical spectrum),
a much larger offset in log(O/H) would be needed for J082555 to no longer be an outlier.

\clearpage

\begin{deluxetable}{lcc}[h]
\tabletypesize{\scriptsize}
\tablewidth{0pt}
\setlength{\tabcolsep}{3pt}
\tablecaption{ Comparison of Emission-Line Intensities from SDSS and MMT Observations of J082555}
\tablewidth{0pt}
\tablehead{
\CH{} & \multicolumn{2}{c}{$I(\lambda)/I(\mbox{H}\beta)$} }
\startdata
{Ion}						& {SDSS}			& {MMT}			\\ 
\hline
{[O~\ii]~$\lambda$3727}   		& \nodata			& 0.286$\pm$0.006	\\ 
{H12 $\lambda$3750}		& \nodata			& 0.020$\pm$0.001	\\ 
{H11 $\lambda$3770}       		& \nodata 			& 0.028$\pm$0.001  	\\
{H10 $\lambda$3797}       		& \nodata  		& 0.042$\pm$0.004  	\\ 
{He~I~$\lambda$3819}    		& \nodata	  		& 0.009$\pm$0.001  	\\ 
{H9 $\lambda$3835}        		& 0.076$\pm$0.002  	& 0.062$\pm$0.002  	\\ 
{[Ne~\iii]~$\lambda$3868}		& 0.317$\pm$0.009	& 0.288$\pm$0.06  	\\ 
{He~I+H8$\lambda$3889} 	& 0.230$\pm$0.006  & 0.206$\pm$0.004  	\\ 
{[Ne~\iii]~$\lambda$3967+H7}  & 0.203$\pm$0.006	& 0.245$\pm$0.006  	\\ 
{[Ne~\iii]~$\lambda$4011}		& \nodata			& 0.003$\pm$0.002  	\\ 
{He~I~$\lambda$4026}    		& 0.020$\pm$0.001  	& 0.018$\pm$0.001  	\\ 
{[S~\ii]~$\lambda$4068}   		& \nodata 		 	& \nodata  		\\ 
{H$\delta$ $\lambda$4101} 	& 0.274$\pm$0.008	& 0.251$\pm$0.005  \\ 
{He~I~$\lambda$4120}      	& \nodata 			& 0.002$\pm$0.001  \\ 
{He~I~$\lambda$4143}      	& \nodata  		& 0.003$\pm$0.002  \\ 
{H$\gamma$ $\lambda$4340} 	& 0.486$\pm$0.014	& 0.455$\pm$0.009  \\ 
{[O~\iii]~$\lambda$4363}  		& 0.115$\pm$0.003	& 0.105$\pm$0.003	\\ 
{He~I~$\lambda$4387}      	& \nodata  		& 0.004$\pm$0.001  	\\ 
{He~I~$\lambda$4471}      	& 0.038$\pm$0.001  	& 0.036$\pm$0.001  	\\ 
{[Ar~\iv] $\lambda4711$}		& 0.011$\pm$0.001	& \nodata			\\ 
{[Ar~\iv] $\lambda4740$}		& 0.009$\pm$0.001	& \nodata			\\ 
{H$\beta$ $\lambda$4861}  	& 1.000$\pm$0.028  	& 1.000$\pm$0.029  	\\ 
{He~I~$\lambda$4921}      	& 0.010$\pm$0.001  	& 0.008$\pm$0.001  	\\ 
{[O~\iii]~$\lambda$4958}  		& 1.236$\pm$0.025	& 1.201$\pm$0.034  	\\ 
{[O~\iii]~$\lambda$5006}  		& 3.624$\pm$0.103	& 3.708$\pm$0.105  	\\ 
{He~I~$\lambda$5015}      	& 0.018$\pm$0.001  	& 0.024$\pm$0.001  \\ 
{NI $\lambda$5197}        		& \nodata			& 0.002$\pm$0.001  	\\ 
{Cl~\iii $\lambda5517$}		& \nodata			& 0.003$\pm$0.001	\\ 
{Cl~\iii $\lambda5537$}		& \nodata			& 0.002$\pm$0.001	\\ 
{He~I~$\lambda$5875}      	& \nodata			& \nodata			\\ 
{[O~\i] $\lambda$6300}    		& 0.006$\pm$0.001  & 0.005$\pm$0.001  \\ 
{[S~\iii]~$\lambda$6312}  		& 0.010$\pm$0.001  & 0.009$\pm$0.001  	\\
{[O~\i] $\lambda$6363}    		& \nodata			& \nodata			\\ 
{[N~\ii]~$\lambda$6548}   		& 0.004$\pm$0.001	& 0.006$\pm$0.001  	\\ 
{H$\alpha$ $\lambda$6562} 	& 2.754$\pm$0.078	& 2.656$\pm$0.083  	\\ 
{[N~\ii]~$\lambda$6583}   		& 0.013$\pm$0.001	& 0.013$\pm$0.001  	\\ 
{He~I~$\lambda$6678}      	& 0.024$\pm$0.001  & 0.027$\pm$0.002  	\\ 
{[S~\ii]~$\lambda$6716}   		& 0.026$\pm$0.001	& 0.028$\pm$0.002  	\\ 
{[S~\ii]~$\lambda$6730}   		& 0.022$\pm$0.001	& 0.019$\pm$0.002  	\\ 	
{[O~\ii]~$\lambda$7320}   		& 0.006$\pm$0.001	& \nodata			\\ 
{[O~\ii]~$\lambda$7330}   		& 0.006$\pm$0.001	& \nodata			\\ 
\hline
{E(B$-$V)}				& 0.160$\pm$0.009	& 0.000$\pm$0.012	\\ 
{F$_{H\beta}$}				& 230.8			& 108.9			\\ 
\enddata
\tablecomments{
The first column lists the vacuum wavelengths of the observed ions for wavelengths 
of $\lambda < 2000$ \AA, and approximate air wavelengths for optical emission lines.
The flux values are the reddening corrected intensity ratios relative to H$\beta$.
The last two rows are the extinction and the H$\beta$ raw flux, in units of 
$10^{-16}$ erg s$^{-1}$ cm$^{-2}$. }
\label{tblA6}
\end{deluxetable}


\begin{deluxetable}{lcc}[h]
\tabletypesize{\scriptsize}
\tablewidth{0pt}
\setlength{\tabcolsep}{3pt}
\tablecaption{ Comparison of Ionic and Total Abundance from SDSS and MMT Observations of J082555}
\tablewidth{0pt}
\tablehead{
\CH{}					& {SDSS}			& {MMT}			
}
\startdata
T$_e$ [O~\iii] (K)			& 19,300$\pm$400	& 18,200$\pm$300	\\ 
T$_e$ [N~\ii] (K)			& 16,500$\pm$300	& 15,800$\pm$200	\\ 		
n$_e$ C~\iii] (cm$^{-3}$)		& 36,500 			& 36,500			\\
n$_e$ [S~\ii] (cm$^{-3})$ 		& 299			& 9				\\ 
\\
O$^+$/H$^+$ (10$^{-5})$		& 0.19$\pm$0.01$^a$ & 0.20$\pm$0.01	\\ 
O$^{+2}$/H$^+$ (10$^{-5})$	& 2.21$\pm$0.07	& 2.53$\pm$0.08	\\ 
12 + log(O/H)				& 7.38$\pm$0.01	& 7.44$\pm$0.01	\\ 
\\
C$^{+2}$/H$^+$ (10$^{-5})$	& 3.18$\pm$0.40	& 4.42$\pm$0.60	\\ 
C$^{+3}$/H$^+$ (10$^{-5})$	& 0.60$\pm$0.01	& 0.88$\pm$0.02	\\ 
O$^{+2}$/H$^+$ (10$^{-5})$	& 4.95$\pm$0.12	& 7.53$\pm$0.20	\\ 
C$^{+2}$/O$^{+2}$			& 0.64$\pm$0.05	& 0.59$\pm$0.05	\\ 
log(C3C4)					& 0.727			& 0.701			\\ 
log U						& -2.04			& -2.05			\\
ICF						& 1.081$\pm$0.080	& 1.081$\pm$0.080	\\ 
log(C/O)					& -0.16$\pm$0.04	& -0.20$\pm$0.04	\\ 
\\
N$^{+}$/H$^+$ (10$^{-7})$	& 0.89$\pm$0.03	& 1.01$\pm$0.10	\\ 
log(N/O)					& -1.32$\pm$0.02	& -1.30$\pm$0.05	\\ 
\\
S$^{+}$/H$^+$ (10$^{-7})$	& 0.46$\pm$0.02	& 0.45$\pm$0.03	\\ 
S$^{+2}$/H$^+$(10$^{-7})$	& 2.69$\pm$0.10	& \nodata			\\ 
log(S/O)					& -1.41$\pm$0.04	& \nodata			\\ 
\enddata
\tablecomments{
Ionic and total abundance calculations for SDSS and MMT observations of J082555.
The ionization parameter is determined using log([\ion{O}{3}]/[\ion{O}{2}]) and the model photoionization 
diagnostic curves shown if Figures~\ref{fig4} and ~\ref{fig5}.\\
$^{a}$ O$^{+}$/H$^+$ calculated from [\ion{O}{2}] $\lambda\lambda7320,7330$.}
\label{tblA7}
\end{deluxetable}


\section{IMPROVED SAMPLE SELECTED FOR FUTURE STUDIES} \label{A3}
It is difficult to predict the S/N of emission lines in UV spectra without preliminary UV spectra. 
Within our COS sample we have examples of successful C/O detections and non-detections alike. 
Contrasting this set of objects allows us to determine what parameters predict successful 
observations of significant C and O.
We examined many properties, including GALEX magnitude, redshift, reddening, 
and optical emission line fluxes, but found no correlation with detection of the UV CELs.
However, excitation (indicated by log [\ion{O}{3}] $\lambda5007$/[\ion{O}{2}] $\lambda3727$ ) does play a significant role.
This is not surprising as both \ion{O}{3}] and \ion{C}{3}] require high excitation (and ionization) energies. 
However, the SDSS spectrograph has a wavelength coverage of 3800$-$9200 \AA\,and so 
[\ion{O}{2}] $\lambda3727$ was only observed in 4 of our 12 targets (when $z > 0.02$).

\begin{figure} 
\begin{center}
	\subfigure[]{\includegraphics[scale=0.425]{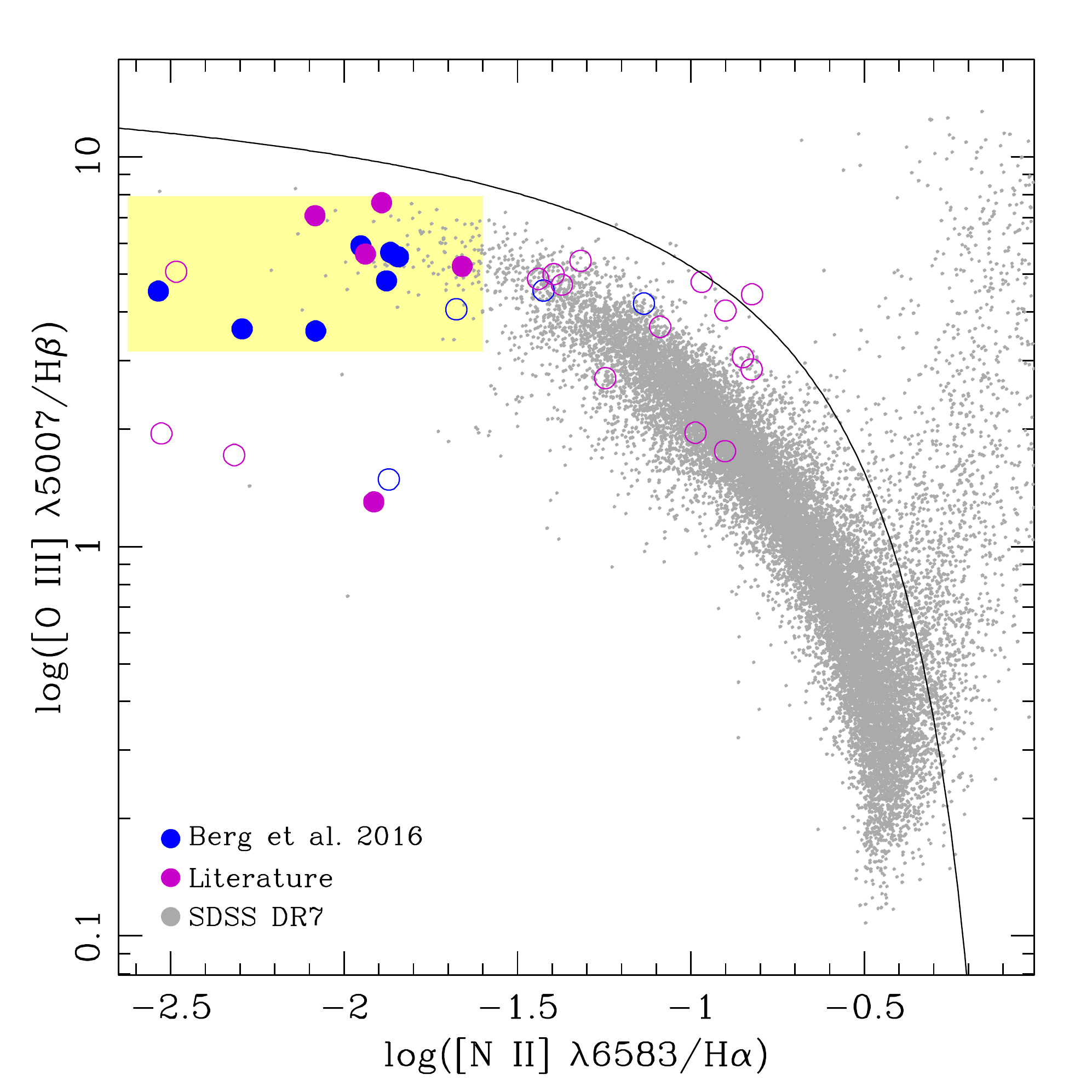}\label{fig9a}}
	\subfigure[]{\includegraphics[scale=0.425]{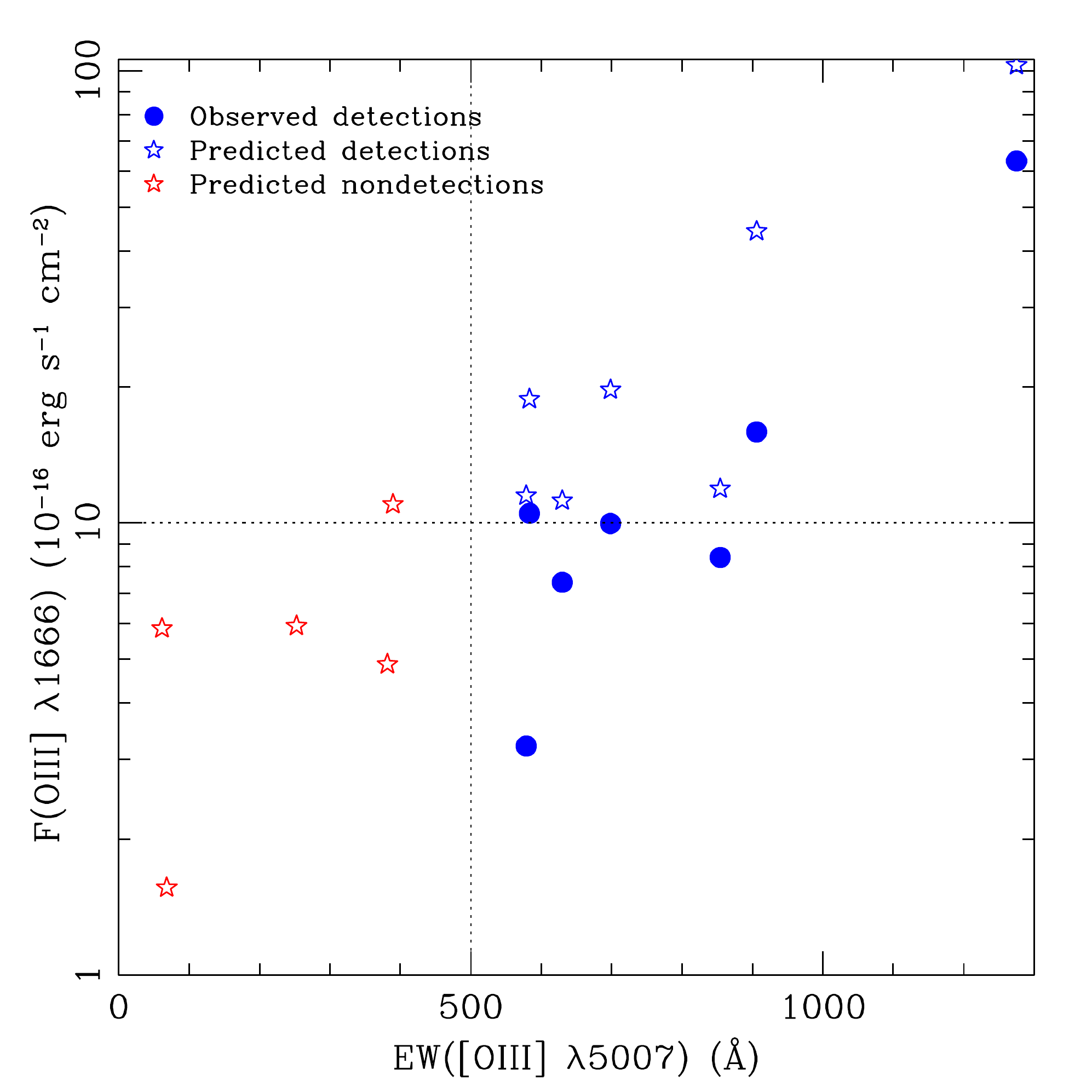}\label{fig9b}}
\caption{BPT diagram comparing the locations of targets with and without C/O detections.
As before, blue symbols represent the HST/COS observations presented here and 
purple symbols indicate line measurements take from other studies.
3$\sigma$ C/O detections are show as filled circles, while non-detections
are shown as open circles. 
The yellow shaded box highlights BPT real-estate that has generally produced 
successful C/O detections.
(b) Flux in \ion{O}{3}] $\lambda1666$ versus the [\ion{O}{3}] $\lambda5007$ 
equivalent width plotted for predicted \ion{O}{3}] $\lambda1666$ fluxes based 
on the SDSS optical spectra (stars) and actual F \ion{O}{3}] $\lambda1666$ 
measurements (filled circles).
C/O detections (3$\sigma$) are blue and non-detections are red.
This plot can be used to predict the success of future C/O detections (upper right had corner).}
\end{center}
\end{figure}

The importance of high excitation is depicted in Figure~\ref{fig9a}
where we have plotted our C/O sample on the BPT diagram.
Once again, significant C/O measurements are depicted as closed circles,
whereas low S/N and non-detections are shown as open circles.
It is striking that all of our significant COS detections lie in the upper left hand corner,
which we highlight with a yellow box. 
Since the x-axis of the BPT diagram is correlated with increasing oxygen abundance,
we expect our low metallicity sample to lie at low values of [\ion{N}{2}] \W6584/H$\alpha$.
However, the distinct separation of detections and non-detections demonstrates that
the UV \ion{O}{3}] and \ion{C}{3}] lines are most easily detected at very low abundances and
become more difficult to observe at abundances typical of star-forming galaxies in 
the SDSS DR7 (gray dots). 

Our requirement of large [\ion{O}{3}] $\lambda5007$ equivalent width helped select 
more high excitation targets, but the cut was not severe enough.
The 5 non-detection targets in our sample have 54 $<$ EW(5007) $<$ 390, 
whereas our C/O detections have significantly larger [\ion{O}{3}] $\lambda5007$ 
equivalent widths of 579 $<$ EW(5007) $<$ 1275.
A similar, but less extreme division is found for H$\beta$ equivalent width.
In Figure~\ref{fig9b} we plot the flux in \ion{O}{3}] $\lambda1666$ versus the 
[\ion{O}{3}] $\lambda5007$ equivalent width, demonstrating the the dichotomy 
in EW(5007) for detections (blue) and non-detections (red).
The original estimates of the \ion{O}{3}] $\lambda1666$ flux based on the SDSS
optical spectra (stars) are greater by a factor of $\sim$2 compared to the actual 
measurements from the seven detections (filled circles).
The vertical and horizontal dotted lines in Figure~\ref{fig9b} divide up the
F(O\iii] $\lambda1666$) versus EW([O\iii] $\lambda5007$) parameter
space such that successful detections were produced by targets with predictions 
in the upper right hand quadrant. 
Thus, this diagram to useful to guide future sample selections.

We have argued that, given the currently available instruments/telescopes, 
 simultaneously observing the UV O$^{+2}$ and C$^{+2}$ CELs 
 with HST/COS is the best method to measure C/O abundances in nearby galaxies.
 Further, we now appreciate the importance of selecting the highest excitation targets
 that have EW([O\iii] $\lambda5007$) measurements and predicted F(O\iii] $\lambda1666$) 
 values that place them in the upper right hand quadrant of Figure~\ref{fig9b}. 
 Observing a larger sample of UV CELs constructed from these criteria 
 is needed to confirm and delineate the C/O trend more clearly in nearby dwarf galaxies. 
 

\clearpage

\bibliography{mybib}{}

\begin{thebibliography}{130}
\expandafter\ifx\csname natexlab\endcsname\relax\def\natexlab#1{#1}\fi
\expandafter\ifx\csname href\endcsname\relax
  \def\href#1#2{}\fi
\expandafter\ifx\csname urllinklabel\endcsname\relax
  \def\urllinklabel{[LINK]}\fi
\expandafter\ifx\csname adsurllinklabel\endcsname\relax
  \def\adsurllinklabel{[ADS]}\fi

\bibitem[{{Abazajian} {et~al.}(2005){Abazajian}, {Adelman-McCarthy},
  {Ag{\"u}eros}, \& {others}}]{abazajian05}
{Abazajian}, K., {Adelman-McCarthy}, J.~K., {Ag{\"u}eros}, M.~A., \& {others}.
  2005, \aj, 129, 1755


\bibitem[{{Abazajian} {et~al.}(2009){Abazajian}, {Adelman-McCarthy},
  {Ag{\"u}eros}, {et~al.}}]{abazajian09}
{Abazajian}, K.~N., {Adelman-McCarthy}, J.~K., {Ag{\"u}eros}, M.~A., {et~al.}
  2009, \apjs, 182, 543


\bibitem[{{Adelman-McCarthy} {et~al.}(2008){Adelman-McCarthy}, {Ag{\"u}eros},
  {Allam}, \& {others}}]{adelman-mccarthy08}
{Adelman-McCarthy}, J.~K., {Ag{\"u}eros}, M.~A., {Allam}, S.~S., \& {others}.
  2008, \apjs, 175, 297


\bibitem[{{Akerman} {et~al.}(2004){Akerman}, {Carigi}, {Nissen}, {Pettini}, \&
  {Asplund}}]{akerman04}
{Akerman}, C.~J., {Carigi}, L., {Nissen}, P.~E., {Pettini}, M., \& {Asplund},
  M. 2004, \aap, 414, 931


\bibitem[{{Alam} {et~al.}(2015){Alam}, {Albareti}, {Allende Prieto},
  {et~al.}}]{alam15}
{Alam}, S., {Albareti}, F.~D., {Allende Prieto}, C., {et~al.} 2015, \apjs, 219,
  12


\bibitem[{{Alexandroff} {et~al.}(2013){Alexandroff}, {Strauss}, {Greene},
  {et~al.}}]{alexandroff13}
{Alexandroff}, R., {Strauss}, M.~A., {Greene}, J.~E., {et~al.} 2013, \mnras,
  435, 3306


\bibitem[{{Aoki} {et~al.}(2007){Aoki}, {Beers}, {Christlieb}, {Norris}, {Ryan},
  \& {Tsangarides}}]{aoki07}
{Aoki}, W., {Beers}, T.~C., {Christlieb}, N., {Norris}, J.~E., {Ryan}, S.~G.,
  \& {Tsangarides}, S. 2007, \apj, 655, 492


\bibitem[{{Baldwin} {et~al.}(1981){Baldwin}, {Phillips}, \&
  {Terlevich}}]{baldwin81}
{Baldwin}, J.~A., {Phillips}, M.~M., \& {Terlevich}, R. 1981, \pasp, 93, 5


\bibitem[{{Bayliss} {et~al.}(2014){Bayliss}, {Rigby}, {Sharon},
  {et~al.}}]{bayliss14}
{Bayliss}, M.~B., {Rigby}, J.~R., {Sharon}, K., {et~al.} 2014, \apj, 790, 144


\bibitem[{{Beers} \& {Christlieb}(2005)}]{beers05}
{Beers}, T.~C. \& {Christlieb}, N. 2005, \araa, 43, 531


\bibitem[{{Bensby} \& {Feltzing}(2006)}]{bensby06}
{Bensby}, T. \& {Feltzing}, S. 2006, \mnras, 367, 1181


\bibitem[{{Berg} {et~al.}(2015){Berg}, {Skillman}, {Croxall},
  {et~al.}}]{berg15}
{Berg}, D.~A., {Skillman}, E.~D., {Croxall}, K.~V., {et~al.} 2015, \apj, 806,
  16


\bibitem[{{Berg} {et~al.}(2013){Berg}, {Skillman}, {Garnett},
  {et~al.}}]{berg13}
{Berg}, D.~A., {Skillman}, E.~D., {Garnett}, D.~R., {et~al.} 2013, \apj, 775,
  128


\bibitem[{Berg {et~al.}(2012)Berg, Skillman, Marble, {et~al.}}]{berg12}
Berg, D.~A., Skillman, E.~D., Marble, A.~R., {et~al.} 2012, \apj, 754, 98


\bibitem[{{Bianchi} {et~al.}(2014){Bianchi}, {Conti}, \& {Shiao}}]{bianchi14}
{Bianchi}, L., {Conti}, A., \& {Shiao}, B. 2014, Advances in Space Research,
  53, 900


\bibitem[{{Binette} {et~al.}(2012){Binette}, {Matadamas}, {H{\"a}gele},
  {et~al.}}]{binette12}
{Binette}, L., {Matadamas}, R., {H{\"a}gele}, G.~F., {et~al.} 2012, \aap, 547,
  A29


\bibitem[{{Bohlin}(2010)}]{bohlin10}
{Bohlin}, R.~C. 2010, \aj, 139, 1515


\bibitem[{{Bolton} {et~al.}(2012){Bolton}, {Schlegel}, {Aubourg}, \&
  {others}}]{bolton12}
{Bolton}, A.~S., {Schlegel}, D.~J., {Aubourg}, {\'E}., \& {others}. 2012, \aj,
  144, 144


\bibitem[{Bresolin {et~al.}(2009a)Bresolin, Gieren, Kudritzki,
  {et~al.}}]{bresolin09a}
Bresolin, F., Gieren, W., Kudritzki, R.-P., {et~al.} 2009a, \apj, 700, 309


\bibitem[{{Brinchmann} {et~al.}(2004){Brinchmann}, {Charlot}, {White},
  {Tremonti}, {Kauffmann}, \& {others}}]{brinchmann04}
{Brinchmann}, J., {Charlot}, S., {White}, S.~D.~M., {Tremonti}, C.,
  {Kauffmann}, G., \& {others}. 2004, \mnras, 351, 1151


\bibitem[{{Brinchmann} {et~al.}(2008){Brinchmann}, {Kunth}, \&
  {Durret}}]{brinchmann08}
{Brinchmann}, J., {Kunth}, D., \& {Durret}, F. 2008, \aap, 485, 657


\bibitem[{{Brown} {et~al.}(2012){Brown}, {Tumlinson}, {Geha}, {Kirby},
  {VandenBerg}, {Mu{\~n}oz}, {Kalirai}, {Simon}, {Avila}, {Guhathakurta},
  {Renzini}, \& {Ferguson}}]{brown12}
{Brown}, T.~M., {Tumlinson}, J., {Geha}, M., {Kirby}, E.~N., {VandenBerg},
  D.~A., {Mu{\~n}oz}, R.~R., {Kalirai}, J.~S., {Simon}, J.~D., {Avila}, R.~J.,
  {Guhathakurta}, P., {Renzini}, A., \& {Ferguson}, H.~C. 2012, \apjl, 753, L21


\bibitem[{Campbell {et~al.}(1986)Campbell, Terlevich, \& Melnick}]{campbell86}
Campbell, A., Terlevich, R., \& Melnick, J. 1986, \mnras, 223, 811


\bibitem[{{Cardamone} {et~al.}(2009){Cardamone}, {Schawinski}, {Sarzi},
  {et~al.}}]{cardamone09}
{Cardamone}, C., {Schawinski}, K., {Sarzi}, M., {et~al.} 2009, \mnras, 399,
  1191


\bibitem[{Cardelli {et~al.}(1989)Cardelli, Clayton, \& Mathis}]{cardelli89}
Cardelli, J.~A., Clayton, G.~C., \& Mathis, J.~S. 1989, \apj, 345, 245


\bibitem[{{Carigi}(2000)}]{carigi00}
{Carigi}, L. 2000, \rmxaa, 36, 171


\bibitem[{{Carigi} \& {Peimbert}(2011)}]{carigi11}
{Carigi}, L. \& {Peimbert}, M. 2011, \rmxaa, 47, 139


\bibitem[{{Carigi} {et~al.}(2005){Carigi}, {Peimbert}, {Esteban}, \&
  {Garc{\'{\i}}a-Rojas}}]{carigi05}
{Carigi}, L., {Peimbert}, M., {Esteban}, C., \& {Garc{\'{\i}}a-Rojas}, J. 2005,
  \apj, 623, 213


\bibitem[{{Chiappini} {et~al.}(2003){Chiappini}, {Romano}, \&
  {Matteucci}}]{chiappini03}
{Chiappini}, C., {Romano}, D., \& {Matteucci}, F. 2003, \mnras, 339, 63


\bibitem[{{Chieffi} \& {Limongi}(2004)}]{chieffi04}
{Chieffi}, A. \& {Limongi}, M. 2004, \apj, 608, 405


\bibitem[{{Christensen} {et~al.}(2012){Christensen}, {Laursen}, {Richard},
  {et~al.}}]{christensen12}
{Christensen}, L., {Laursen}, P., {Richard}, J., {et~al.} 2012, \mnras, 427,
  1973


\bibitem[{{Conroy} \& {van Dokkum}(2012)}]{conroy12}
{Conroy}, C. \& {van Dokkum}, P.~G. 2012, \apj, 760, 71


\bibitem[{{Cooke} {et~al.}(2011){Cooke}, {Pettini}, {Steidel}, {Rudie}, \&
  {Nissen}}]{cooke11}
{Cooke}, R., {Pettini}, M., {Steidel}, C.~C., {Rudie}, G.~C., \& {Nissen},
  P.~E. 2011, \mnras, 417, 1534


\bibitem[{{Dufour}(1984)}]{dufour84}
{Dufour}, R.~J. 1984, \pasp, 96, 787


\bibitem[{{Dufour} {et~al.}(1982){Dufour}, {Shields}, \& {Talbot}}]{dufour82}
{Dufour}, R.~J., {Shields}, G.~A., \& {Talbot}, Jr., R.~J. 1982, \apj, 252, 461


\bibitem[{{Eisenstein} {et~al.}(2011){Eisenstein}, {Weinberg}, {Agol},
  {et~al.}}]{eisenstein11}
{Eisenstein}, D.~J., {Weinberg}, D.~H., {Agol}, E., {et~al.} 2011, \aj, 142, 72


\bibitem[{Ekta \& Chengalur(2010)}]{ekta10}
Ekta, B. \& Chengalur, J.~N. 2010, \mnras, 406, 1238


\bibitem[{{Erb} {et~al.}(2010){Erb}, {Pettini}, {Shapley}, {et~al.}}]{erb10}
{Erb}, D.~K., {Pettini}, M., {Shapley}, A.~E., {et~al.} 2010, \apj, 719, 1168


\bibitem[{Esteban {et~al.}(2009)Esteban, Bresolin, Peimbert,
  {et~al.}}]{esteban09}
Esteban, C., Bresolin, F., Peimbert, M., {et~al.} 2009, \apj, 700, 654


\bibitem[{{Esteban} {et~al.}(2014){Esteban}, {Garc{\'{\i}}a-Rojas}, {Carigi},
  {et~al.}}]{esteban14}
{Esteban}, C., {Garc{\'{\i}}a-Rojas}, J., {Carigi}, L., {et~al.} 2014, \mnras,
  443, 624


\bibitem[{{Esteban} {et~al.}(2002){Esteban}, {Peimbert}, {Torres-Peimbert}, \&
  {Rodr{\'{\i}}guez}}]{esteban02}
{Esteban}, C., {Peimbert}, M., {Torres-Peimbert}, S., \& {Rodr{\'{\i}}guez}, M.
  2002, \apj, 581, 241


\bibitem[{{Fabbian} {et~al.}(2009){Fabbian}, {Nissen}, {Asplund}, {Pettini}, \&
  {Akerman}}]{fabbian09}
{Fabbian}, D., {Nissen}, P.~E., {Asplund}, M., {Pettini}, M., \& {Akerman}, C.
  2009, \aap, 500, 1143


\bibitem[{{Ferland} {et~al.}(2013){Ferland}, {Porter}, {van Hoof},
  {et~al.}}]{ferland13}
{Ferland}, G.~J., {Porter}, R.~L., {van Hoof}, P.~A.~M., {et~al.} 2013, \rmxaa,
  49, 137


\bibitem[{{Fumagalli} {et~al.}(2011){Fumagalli}, {da Silva}, \&
  {Krumholz}}]{fumagalli11}
{Fumagalli}, M., {da Silva}, R.~L., \& {Krumholz}, M.~R. 2011, \apjl, 741, L26


\bibitem[{{Garc{\'{\i}}a-Rojas} \& {Esteban}(2007)}]{garcia-rojas07}
{Garc{\'{\i}}a-Rojas}, J. \& {Esteban}, C. 2007, \apj, 670, 457


\bibitem[{Garnett(1990)}]{garnett90}
Garnett, D.~R. 1990, \apj, 363, 142


\bibitem[{{Garnett}(1992)}]{garnett92}
{Garnett}, D.~R. 1992, \aj, 103, 1330


\bibitem[{{Garnett} {et~al.}(1999){Garnett}, {Shields}, {Peimbert},
  {et~al.}}]{garnett99}
{Garnett}, D.~R., {Shields}, G.~A., {Peimbert}, M., {et~al.} 1999, \apj, 513,
  168


\bibitem[{{Garnett} {et~al.}(1997){Garnett}, {Shields}, {Skillman}, {Sagan}, \&
  {Dufour}}]{garnett97}
{Garnett}, D.~R., {Shields}, G.~A., {Skillman}, E.~D., {Sagan}, S.~P., \&
  {Dufour}, R.~J. 1997, \apj, 489, 63


\bibitem[{{Garnett} {et~al.}(1995){Garnett}, {Skillman}, {Dufour},
  {et~al.}}]{garnett95}
{Garnett}, D.~R., {Skillman}, E.~D., {Dufour}, R.~J., {et~al.} 1995, \apj, 443,
  64


\bibitem[{{Gavil{\'a}n} {et~al.}(2005){Gavil{\'a}n}, {Buell}, \&
  {Moll{\'a}}}]{gavilan05}
{Gavil{\'a}n}, M., {Buell}, J.~F., \& {Moll{\'a}}, M. 2005, \aap, 432, 861


\bibitem[{{Gavil{\'a}n} {et~al.}(2006){Gavil{\'a}n}, {Moll{\'a}}, \&
  {Buell}}]{gavilan06}
{Gavil{\'a}n}, M., {Moll{\'a}}, M., \& {Buell}, J.~F. 2006, \aap, 450, 509


\bibitem[{{Green} {et~al.}(2012){Green}, {Froning}, {Osterman}, \&
  {others}}]{green12}
{Green}, J.~C., {Froning}, C.~S., {Osterman}, S., \& {others}. 2012, \apj, 744,
  60


\bibitem[{{Gunn} {et~al.}(1998){Gunn}, {Carr}, {Rockosi}, \& {others}}]{gunn98}
{Gunn}, J.~E., {Carr}, M., {Rockosi}, C., \& {others}. 1998, \aj, 116, 3040


\bibitem[{{Guseva} {et~al.}(2011){Guseva}, {Izotov}, {Stasi{\'n}ska}, {Fricke},
  {Henkel}, \& {Papaderos}}]{guseva11}
{Guseva}, N.~G., {Izotov}, Y.~I., {Stasi{\'n}ska}, G., {Fricke}, K.~J.,
  {Henkel}, C., \& {Papaderos}, P. 2011, \aap, 529, A149


\bibitem[{{Guseva} {et~al.}(2009){Guseva}, {Papaderos}, {Meyer}, {Izotov}, \&
  {Fricke}}]{guseva09}
{Guseva}, N.~G., {Papaderos}, P., {Meyer}, H.~T., {Izotov}, Y.~I., \& {Fricke},
  K.~J. 2009, \aap, 505, 63


\bibitem[{{Gustafsson} {et~al.}(1999){Gustafsson}, {Karlsson}, {Olsson},
  {Edvardsson}, \& {Ryde}}]{gustafsson99}
{Gustafsson}, B., {Karlsson}, T., {Olsson}, E., {Edvardsson}, B., \& {Ryde}, N.
  1999, \aap, 342, 426


\bibitem[{{Hainline} {et~al.}(2011){Hainline}, {Shapley}, {Greene}, \&
  {Steidel}}]{hainline11}
{Hainline}, K.~N., {Shapley}, A.~E., {Greene}, J.~E., \& {Steidel}, C.~C. 2011,
  \apj, 733, 31


\bibitem[{{Hainline} {et~al.}(2009){Hainline}, {Shapley}, {Kornei}, {Pettini},
  {Buckley-Geer}, {et~al.}}]{hainline09}
{Hainline}, K.~N., {Shapley}, A.~E., {Kornei}, K.~A., {Pettini}, M.,
  {Buckley-Geer}, E., {et~al.} 2009, \apj, 701, 52


\bibitem[{{Henry} {et~al.}(2015){Henry}, {Scarlata}, {Martin}, \&
  {Erb}}]{henry15}
{Henry}, A., {Scarlata}, C., {Martin}, C.~L., \& {Erb}, D. 2015, \apj, 809, 19


\bibitem[{Henry {et~al.}(2000)Henry, Edmunds, \& K\"oppen}]{henry00}
Henry, R. B.~C., Edmunds, M.~G., \& K\"oppen, J. 2000, \apj, 541, 660


\bibitem[{{Hirschi}(2007)}]{hirschi07}
{Hirschi}, R. 2007, \aap, 461, 571


\bibitem[{{Hirschi} \& {et al.}(2006)}]{hirschi06}
{Hirschi}, R. \& {et al.} 2006, in Reviews in Modern Astronomy, Vol.~19,
  Reviews in Modern Astronomy, ed. S.~{Roeser}, 101


\bibitem[{{Izotov} {et~al.}(2009){Izotov}, {Guseva}, {Fricke}, \&
  {Papaderos}}]{izotov09}
{Izotov}, Y.~I., {Guseva}, N.~G., {Fricke}, K.~J., \& {Papaderos}, P. 2009,
  \aap, 503, 61


\bibitem[{Izotov \& Thuan(1999)}]{izotov99}
Izotov, Y.~I. \& Thuan, T.~X. 1999, \apj, 511, 639


\bibitem[{{Izotov} {et~al.}(2012){Izotov}, {Thuan}, \& {Guseva}}]{izotov12}
{Izotov}, Y.~I., {Thuan}, T.~X., \& {Guseva}, N.~G. 2012, \aap, 546, A122


\bibitem[{{James} {et~al.}(2014){James}, {Pettini}, {Christensen},
  {et~al.}}]{james14}
{James}, B.~L., {Pettini}, M., {Christensen}, L., {et~al.} 2014, \mnras, 440,
  1794


\bibitem[{{Jenkins}(2014)}]{jenkins14}
{Jenkins}, E.~B. 2014, ArXiv e-prints


\bibitem[{{Jester} {et~al.}(2005){Jester}, {Schneider}, {Richards},
  {et~al.}}]{jester05}
{Jester}, S., {Schneider}, D.~P., {Richards}, G.~T., {et~al.} 2005, \aj, 130,
  873


\bibitem[{{Kalirai} {et~al.}(2013){Kalirai}, {Anderson}, {Dotter},
  {et~al.}}]{kalirai13}
{Kalirai}, J.~S., {Anderson}, J., {Dotter}, A., {et~al.} 2013, \apj, 763, 110


\bibitem[{{Kauffmann} {et~al.}(2003){Kauffmann}, {Heckman}, {White}, \&
  {others}}]{kauffmann03b}
{Kauffmann}, G., {Heckman}, T.~M., {White}, S.~D.~M., \& {others}. 2003,
  \mnras, 341, 33


\bibitem[{{Kennicutt}(1998)}]{kennicutt98}
{Kennicutt}, Jr., R.~C. 1998, \apj, 498, 541


\bibitem[{{Kennicutt} {et~al.}(2003){Kennicutt}, {Bresolin}, \&
  {Garnett}}]{kennicutt03a}
{Kennicutt}, Jr., R.~C., {Bresolin}, F., \& {Garnett}, D.~R. 2003, \apj, 591,
  801


\bibitem[{{Kewley} {et~al.}(2006){Kewley}, {Groves}, {Kauffmann}, \&
  {Heckman}}]{kewley06}
{Kewley}, L.~J., {Groves}, B., {Kauffmann}, G., \& {Heckman}, T. 2006, \mnras,
  372, 961


\bibitem[{Kniazev {et~al.}(2004)Kniazev, Pustilnik, Grebel, Lee, \&
  Pramskij}]{kniazev04}
Kniazev, A.~Y., Pustilnik, S.~A., Grebel, E.~K., Lee, H., \& Pramskij, A.~G.
  2004, \apjs, 153, 429


\bibitem[{Kobulnicky \& Skillman(1997)}]{kobulnicky97}
Kobulnicky, H.~A. \& Skillman, E.~D. 1997, \apj, 489, 636


\bibitem[{{Kobulnicky} \& {Skillman}(1998)}]{kobulnicky98}
{Kobulnicky}, H.~A. \& {Skillman}, E.~D. 1998, \apj, 497, 601


\bibitem[{{Kroupa}(2001)}]{kroupa01}
{Kroupa}, P. 2001, \mnras, 322, 231


\bibitem[{{Kroupa} {et~al.}(1993){Kroupa}, {Tout}, \& {Gilmore}}]{kroupa93}
{Kroupa}, P., {Tout}, C.~A., \& {Gilmore}, G. 1993, \mnras, 262, 545


\bibitem[{{Kurt} {et~al.}(1999){Kurt}, {Dufour}, {Garnett}, {et~al.}}]{kurt99}
{Kurt}, C.~M., {Dufour}, R.~J., {Garnett}, D.~R., {et~al.} 1999, \apj, 518, 246


\bibitem[{Lee {et~al.}(2009)Lee, Gil~de Paz, Tremonti, {et~al.}}]{jlee09}
Lee, J.~C., Gil~de Paz, A., Tremonti, C., {et~al.} 2009, \apj, 706, 599


\bibitem[{{Leitherer} {et~al.}(1999){Leitherer}, {Schaerer}, {Goldader},
  {et~al.}}]{leitherer99}
{Leitherer}, C., {Schaerer}, D., {Goldader}, J.~D., {et~al.} 1999, \apjs, 123,
  3


\bibitem[{{Limongi} \& {Chieffi}(2003)}]{limongi03}
{Limongi}, M. \& {Chieffi}, A. 2003, \apj, 592, 404


\bibitem[{{L{\'o}pez-S{\'a}nchez} {et~al.}(2007){L{\'o}pez-S{\'a}nchez},
  {Esteban}, {Garc{\'{\i}}a-Rojas}, {Peimbert}, \&
  {Rodr{\'{\i}}guez}}]{lopez-sanchez07}
{L{\'o}pez-S{\'a}nchez}, {\'A}.~R., {Esteban}, C., {Garc{\'{\i}}a-Rojas}, J.,
  {Peimbert}, M., \& {Rodr{\'{\i}}guez}, M. 2007, \apj, 656, 168


\bibitem[{{Maeder}(1990)}]{maeder90}
{Maeder}, A. 1990, \aaps, 84, 139


\bibitem[{{Maeder}(1992)}]{maeder92}
---. 1992, \aap, 264, 105


\bibitem[{{Maeder} {et~al.}(2015){Maeder}, {Meynet}, \& {Chiappini}}]{maeder15}
{Maeder}, A., {Meynet}, G., \& {Chiappini}, C. 2015, \aap, 576, A56


\bibitem[{{Marassi} {et~al.}(2014){Marassi}, {Chiaki}, {Schneider},
  {et~al.}}]{marassi14}
{Marassi}, S., {Chiaki}, G., {Schneider}, R., {et~al.} 2014, \apj, 794, 100


\bibitem[{{Marigo} {et~al.}(1996){Marigo}, {Bressan}, \& {Chiosi}}]{marigo96}
{Marigo}, P., {Bressan}, A., \& {Chiosi}, C. 1996, \aap, 313, 545


\bibitem[{{Marigo} {et~al.}(1998){Marigo}, {Bressan}, \& {Chiosi}}]{marigo98}
---. 1998, \aap, 331, 564


\bibitem[{{Mateo}(1998)}]{mateo98}
{Mateo}, M.~L. 1998, \araa, 36, 435


\bibitem[{{Mattsson}(2010)}]{mattsson10}
{Mattsson}, L. 2010, \aap, 515, A68


\bibitem[{{Meynet} {et~al.}(2006){Meynet}, {Ekstr{\"o}m}, \&
  {Maeder}}]{meynet06}
{Meynet}, G., {Ekstr{\"o}m}, S., \& {Maeder}, A. 2006, \aap, 447, 623


\bibitem[{{Meynet} {et~al.}(2010){Meynet}, {Hirschi}, {Ekstrom}, {Maeder},
  {Georgy}, {Eggenberger}, \& {Chiappini}}]{meynet10}
{Meynet}, G., {Hirschi}, R., {Ekstrom}, S., {Maeder}, A., {Georgy}, C.,
  {Eggenberger}, P., \& {Chiappini}, C. 2010, \aap, 521, A30


\bibitem[{{Meynet} \& {Maeder}(2002)}]{meynet02}
{Meynet}, G. \& {Maeder}, A. 2002, \aap, 390, 561


\bibitem[{{Meynet} {et~al.}(2003){Meynet}, {Maeder}, \& {Hirschi}}]{meynet03}
{Meynet}, G., {Maeder}, A., \& {Hirschi}, R. 2003, ArXiv Astrophysics e-prints


\bibitem[{{Moll{\'a}} {et~al.}(2015){Moll{\'a}}, {Cavichia}, {Gavil{\'a}n}, \&
  {Gibson}}]{molla15}
{Moll{\'a}}, M., {Cavichia}, O., {Gavil{\'a}n}, M., \& {Gibson}, B.~K. 2015,
  \mnras, 451, 3693


\bibitem[{Nava {et~al.}(2006)Nava, Casebeer, Henry, \& Jevremovic}]{nava06}
Nava, A., Casebeer, D., Henry, R. B.~C., \& Jevremovic, D. 2006, \apj, 645,
  1076


\bibitem[{{Nomoto} {et~al.}(2013){Nomoto}, {Kobayashi}, \&
  {Tominaga}}]{nomoto13}
{Nomoto}, K., {Kobayashi}, C., \& {Tominaga}, N. 2013, \araa, 51, 457


\bibitem[{{Peimbert}(2003)}]{peimbert03}
{Peimbert}, A. 2003, \apj, 584, 735


\bibitem[{{Peimbert} \& {Peimbert}(2010)}]{peimbert10}
{Peimbert}, A. \& {Peimbert}, M. 2010, \apj, 724, 791


\bibitem[{{Peimbert} {et~al.}(2005){Peimbert}, {Peimbert}, \&
  {Ruiz}}]{peimbert05}
{Peimbert}, A., {Peimbert}, M., \& {Ruiz}, M.~T. 2005, \apj, 634, 1056


\bibitem[{{Peimbert}(1967)}]{peimbert67}
{Peimbert}, M. 1967, \apj, 150, 825


\bibitem[{{Peimbert} {et~al.}(1986){Peimbert}, {Pena}, \&
  {Torres-Peimbert}}]{peimbert86}
{Peimbert}, M., {Pena}, M., \& {Torres-Peimbert}, S. 1986, \aap, 158, 266


\bibitem[{{Pettini} {et~al.}(2008){Pettini}, {Zych}, {Steidel}, \&
  {Chaffee}}]{pettini08}
{Pettini}, M., {Zych}, B.~J., {Steidel}, C.~C., \& {Chaffee}, F.~H. 2008,
  \mnras, 385, 2011


\bibitem[{Pilyugin \& Thuan(2005)}]{pilyugin05}
Pilyugin, L.~S. \& Thuan, T.~X. 2005, \apj, 631, 231


\bibitem[{{Portinari} {et~al.}(1998){Portinari}, {Chiosi}, \&
  {Bressan}}]{portinari98}
{Portinari}, L., {Chiosi}, C., \& {Bressan}, A. 1998, \aap, 334, 505


\bibitem[{{Quider} {et~al.}(2009){Quider}, {Pettini}, {Shapley}, \&
  {Steidel}}]{quider09}
{Quider}, A.~M., {Pettini}, M., {Shapley}, A.~E., \& {Steidel}, C.~C. 2009,
  \mnras, 398, 1263


\bibitem[{{Salpeter}(1955)}]{salpeter55}
{Salpeter}, E.~E. 1955, \apj, 121, 161


\bibitem[{{Salvadori} {et~al.}(2015){Salvadori}, {Sk{\'u}lad{\'o}ttir}, \&
  {Tolstoy}}]{salvadori15}
{Salvadori}, S., {Sk{\'u}lad{\'o}ttir}, {\'A}., \& {Tolstoy}, E. 2015, \mnras,
  454, 1320


\bibitem[{{Shapley} {et~al.}(2003){Shapley}, {Steidel}, {Pettini}, \&
  {Adelberger}}]{shapley03}
{Shapley}, A.~E., {Steidel}, C.~C., {Pettini}, M., \& {Adelberger}, K.~L. 2003,
  \apj, 588, 65


\bibitem[{{Sim{\'o}n-D{\'{\i}}az} \& {Stasi{\'n}ska}(2011)}]{simon-diaz11}
{Sim{\'o}n-D{\'{\i}}az}, S. \& {Stasi{\'n}ska}, G. 2011, \aap, 526, A48


\bibitem[{{Skillman} \& {Kennicutt}(1993)}]{skillman93}
{Skillman}, E.~D. \& {Kennicutt}, Jr., R.~C. 1993, \apj, 411, 655


\bibitem[{{Spite} {et~al.}(2013){Spite}, {Caffau}, {Bonifacio},
  {et~al.}}]{spite13}
{Spite}, M., {Caffau}, E., {Bonifacio}, P., {et~al.} 2013, \aap, 552, A107


\bibitem[{{Spite} {et~al.}(2005){Spite}, {Cayrel}, {Plez}, {et~al.}}]{spite05}
{Spite}, M., {Cayrel}, R., {Plez}, B., {et~al.} 2005, \aap, 430, 655


\bibitem[{{Stark} {et~al.}(2014){Stark}, {Richard}, {Siana},
  {et~al.}}]{stark14}
{Stark}, D.~P., {Richard}, J., {Siana}, B., {et~al.} 2014, \mnras, 445, 3200


\bibitem[{{Stasi{\'n}ska}(1982)}]{stasinska82}
{Stasi{\'n}ska}, G. 1982, \aaps, 48, 299


\bibitem[{{Terlevich} \& {Melnick}(1985)}]{terlevich85}
{Terlevich}, R. \& {Melnick}, J. 1985, \mnras, 213, 841


\bibitem[{Thuan {et~al.}(1995)Thuan, Izotov, \& Lipovetsky}]{thuan95}
Thuan, T.~X., Izotov, Y.~I., \& Lipovetsky, V.~A. 1995, \apj, 445, 108


\bibitem[{{Thuan} {et~al.}(1996){Thuan}, {Izotov}, \& {Lipovetsky}}]{thuan96}
{Thuan}, T.~X., {Izotov}, Y.~I., \& {Lipovetsky}, V.~A. 1996, \apj, 463, 120


\bibitem[{{Tolstoy} {et~al.}(2009){Tolstoy}, {Hill}, \& {Tosi}}]{tolstoy09}
{Tolstoy}, E., {Hill}, V., \& {Tosi}, M. 2009, \araa, 47, 371


\bibitem[{{Tominaga} {et~al.}(2014){Tominaga}, {Iwamoto}, \&
  {Nomoto}}]{tominaga14}
{Tominaga}, N., {Iwamoto}, N., \& {Nomoto}, K. 2014, \apj, 785, 98


\bibitem[{{Torres-Peimbert} {et~al.}(1980){Torres-Peimbert}, {Peimbert}, \&
  {Daltabuit}}]{torres-peimbert80}
{Torres-Peimbert}, S., {Peimbert}, M., \& {Daltabuit}, E. 1980, \apj, 238, 133


\bibitem[{Tremonti {et~al.}(2004)Tremonti, Heckman, Kauffmann,
  {et~al.}}]{tremonti04}
Tremonti, C.~A., Heckman, T.~M., Kauffmann, G., {et~al.} 2004, \apj, 613, 898


\bibitem[{{Tsujimoto} \& {Bekki}(2011)}]{tsujimoto11}
{Tsujimoto}, T. \& {Bekki}, K. 2011, \aap, 530, A78


\bibitem[{{van den Hoek} \& {Groenewegen}(1997)}]{vg97}
{van den Hoek}, L.~B. \& {Groenewegen}, M.~A.~T. 1997, \aaps, 123


\bibitem[{{van Zee} {et~al.}(1998){van Zee}, {Salzer}, \& {Haynes}}]{vanzee98a}
{van Zee}, L., {Salzer}, J.~J., \& {Haynes}, M.~P. 1998, \apjl, 497, L1


\bibitem[{{Venn}(1995)}]{venn95}
{Venn}, K.~A. 1995, \apj, 449, 839


\bibitem[{{Wheeler} {et~al.}(1989){Wheeler}, {Sneden}, \& {Truran}}]{wheeler89}
{Wheeler}, J.~C., {Sneden}, C., \& {Truran}, Jr., J.~W. 1989, \araa, 27, 279


\bibitem[{{York} {et~al.}(2000){York}, {Adelman}, {Anderson},
  {et~al.}}]{york00}
{York}, D.~G., {Adelman}, J., {Anderson}, Jr., J.~E., {et~al.} 2000, \aj, 120,
  1579


\end{thebibliography}

\clearpage

\end{document}